\documentclass{aa} 
\usepackage{graphicx}
\usepackage{txfonts}
\usepackage{amsmath}
\usepackage{graphicx}
\usepackage{txfonts}
\usepackage{natbib, twoopt}
\usepackage{subfig}
\usepackage{color}
\usepackage{mathrsfs}
\usepackage{xspace}
\usepackage{rotating}
\usepackage{lscape}
\usepackage{amssymb}
\usepackage{longtable}
\usepackage{pifont}
\usepackage{multirow}
\usepackage{adjustbox}
\usepackage{lineno}
\usepackage{gensymb}
\usepackage[plainpages=False,bookmarks=True,colorlinks=True,citecolor=blue,linkcolor=blue,urlcolor=magenta,breaklinks=True,pagebackref=False]{hyperref}
\usepackage[mathscr]{euscript}

\newcommand{\snf}{SNfactory\xspace}
\newcommand{\sne}{SNe~Ia\xspace}
\newcommand{\snia}{SN~Ia\xspace}
\newcommand{\sneia}{SNe~Ia\xspace}
\newcommand{\sn}{supernova\xspace}

\newcommand{\eg}{\emph{e.g.}\xspace}     
\newcommand{\ie}{\emph{i.e.}\xspace}     

\newcommand{\eqcoma}{\ ,}
\newcommand{\eqdot}{\ .}

\newcommand{\R}{\ensuremath{\mathrm{\cal{R}}}\xspace}

\newcommand{\SiIIawlength}{4131}
\newcommand{\Sia}{Si~\textsc{ii}~$\lambda$\SiIIawlength\xspace}
\newcommand{\Sib}{Si~\textsc{ii}~$\lambda$5972\xspace}
\newcommand{\Sic}{Si~\textsc{ii}~$\lambda$6355\xspace}
\newcommand{\CaHK}{Ca~\textsc{ii}~H\&K\xspace}
\newcommand{\Mg}{Mg~\textsc{ii}\xspace}
\newcommand{\FE}{Fe~$\lambda$4800\xspace}
\newcommand{\SW}{S~\textsc{ii}~W\xspace}
\newcommand{\SWl}{S~\textsc{ii}~$\lambda$5454\xspace}
\newcommand{\SWr}{S~\textsc{ii}~$\lambda$5640\xspace}
\newcommand{\OI}{O~\textsc{i}~$\lambda$7773\xspace}
\newcommand{\CaIR}{Ca~\textsc{ii}~IR\xspace}
\newcommand{\Rsjb}{\ensuremath{\R_{642/443}}\xspace}

\newcommand{\SWa}{S~\textsc{ii}~W~$\lambda$5454\xspace}
\newcommand{\SWb}{S~\textsc{ii}~W~$\lambda$5640\xspace}

\newcommand{\NsnTotal}{171\xspace}
\newcommand{\NsnValidation}{58\xspace}
\newcommand{\NsnTraining}{113\xspace}
\newcommand{\NsnSigmaClipping}{8\xspace}
\newcommand{\NSnTrainingFilterEMFA}{105\xspace}
\newcommand{\sugarMinWavelength}{3254\xspace}
\newcommand{\sugarMaxWavelength}{8649\xspace}
\newcommand{\sugarMedianWavelength}{$5305\AA$ \xspace}
\newcommand{\sugarMinPhase}{$-12$ \xspace}
\newcommand{\sugarMaxPhase}{$+48$ \xspace}
\newcommand{\numberbinsugar}{197 \xspace}

\newcommand{\noiselevelemfa}{18\% }
\newcommand{\AlphaOneImpactMax}{$0.17$ }
\newcommand{\AlphaTwoImpactMax}{$0.05$ }
\newcommand{\AlphaThreeImpactMax}{$0.05$}
\newcommand{\RvValue}{2.6 \xspace}
\newcommand{\RvValueError}{0.5 \xspace}

\newcommand{\sugar}{SUGAR\xspace}

\begin{document} 


   \title{\sugar: An improved empirical model of Type Ia Supernovae based on spectral features} 
   \titlerunning{\sugar model}
   \authorrunning{P.-F. L\'eget \& SNfactory}
   \date{Received 24 December 2018 / Accepted 19 September 2019}
\author{P.-F.~L\'eget \inst{\ref{lpc}, \ref{kipac}, \ref{lpnhe}}
\and   E.~Gangler \inst{\ref{lpc}}
\and   F.~Mondon \inst{\ref{lpc}}
\and   G.~Aldering \inst{\ref{lbnl}}
\and   P.~Antilogus \inst{\ref{lpnhe}}
\and   C.~Aragon \inst{\ref{lbnl}}
\and   S.~Bailey \inst{\ref{lbnl}}
\and   C.~Baltay \inst{\ref{yale}} 
\and   K.~Barbary \inst{\ref{lbnl}} 
\and   S.~Bongard \inst{\ref{lpnhe}}
\and   K.~Boone \inst{\ref{lbnl},\ref{ucb}}
\and   C.~Buton \inst{\ref{ipnl}}
\and   N.~Chotard \inst{\ref{ipnl}}
\and   Y.~Copin \inst{\ref{ipnl}}
\and   S.~Dixon \inst{\ref{lbnl}}
\and   P.~Fagrelius \inst{\ref{lbnl},\ref{ucb}}
\and   U.~Feindt \inst{\ref{okc}}
\and   D.~Fouchez \inst{\ref{cppm}}
\and   B.~Hayden \inst{\ref{lbnl}}
\and   W.~Hillebrandt \inst{\ref{garching}}
\and   A.~Kim \inst{\ref{lbnl}}
\and   M.~Kowalski \inst{\ref{berlin},\ref{desy}}
\and   D.~Kuesters \inst{\ref{berlin}}
\and   S.~Lombardo \inst{\ref{berlin}}
\and   Q.~Lin \inst{\ref{china}}
\and   J.~Nordin \inst{\ref{berlin}}
\and   R.~Pain \inst{\ref{lpnhe}}
\and   E.~Pecontal \inst{\ref{cral}}
\and   R.~Pereira \inst{\ref{ipnl}}
\and   S.~Perlmutter \inst{\ref{lbnl},\ref{ucb}}
\and   M.~V. Pruzhinskaya\inst{\ref{lpc},\ref{msu}}
\and   D.~Rabinowitz \inst{\ref{yale}} 
\and   M.~Rigault \inst{\ref{lpc}} 
\and   K.~Runge \inst{\ref{lbnl}} 
\and   D.~Rubin \inst{\ref{lbnl},\ref{stsci}}
\and   C.~Saunders \inst{\ref{lpnhe}}
\and   L.-P.~Says \inst{\ref{lpc}}
\and   G.~Smadja \inst{\ref{ipnl}} 
\and   C.~Sofiatti \inst{\ref{lbnl},\ref{ucb}}
\and   N.~Suzuki \inst{\ref{lbnl},\ref{ipmu}}
\and   S.~Taubenberger \inst{\ref{garching},\ref{eso}}
\and   C.~Tao \inst{\ref{cppm},\ref{china}}
\and   R.~C.~Thomas \inst{\ref{nersc}}\\
\textsc{The Nearby Supernova Factory}
}


\institute{\tiny Universit\'e Clermont Auvergne, CNRS/IN2P3, Laboratoire de Physique de Clermont, F-63000 Clermont-Ferrand, France.  
\label{lpc}
\and
 Kavli Institute for Particle Astrophysics and Cosmology,
    Department of Physics, Stanford University, 
    Stanford, CA 94305 \label{kipac}
\and 
 LPNHE, CNRS/IN2P3, Sorbonne Universit\'e, Paris Diderot, Laboratoire de
Physique Nucl\'eaire et de Hautes \'Energies, F-75005, Paris, France \label{lpnhe}
\and 
    Physics Division, Lawrence Berkeley National Laboratory, 
    1 Cyclotron Road, Berkeley, CA, 94720 \label{lbnl}
\and 
    Department of Physics, Yale University, 
    New Haven, CT, 06250-8121 \label{yale}
\and 
    Department of Physics, University of California Berkeley,
    366 LeConte Hall MC 7300, Berkeley, CA, 94720-7300 \label{ucb}
\and 
    Universit\'e de Lyon, F-69622, Lyon, France ; Universit\'e de Lyon 1, Villeurbanne ; 
    CNRS/IN2P3, Institut de Physique Nucl\'eaire de Lyon. \label{ipnl}
\and 
    The Oskar Klein Centre, Department of Physics, AlbaNova, Stockholm
    University, SE-106 91 Stockholm, Sweden \label{okc}
\and 
    Aix Marseille Universit\'e, CNRS/IN2P3, CPPM UMR 7346, 13288,
    Marseille, France \label{cppm}
\and
    Max-Planck Institut f\"ur Astrophysik, Karl-Schwarzschild-Str. 1,
    85748 Garching, Germany \label{garching}
\and
    Institut fur Physik, Humboldt-Universitat zu Berlin,
    Newtonstr. 15, 12489 Berlin \label{berlin} 
\and
    Deutsches Elektronen-Synchrotron, D-15735 Zeuthen, Germany \label{desy}
\and 
    Tsinghua Center for Astrophysics, Tsinghua University, Beijing
    100084, China \label{china} 
\and 
    Centre de Recherche Astronomique de Lyon, Universit\'e Lyon 1,
    9 Avenue Charles Andr\'e, 69561 Saint Genis Laval,
    France \label{cral}
\and    
    Lomonosov Moscow State University, Sternberg Astronomical Institute, Universitetsky pr.~13, Moscow 119234, Russia \label{msu}
\and
    Space Telescope Science Institute, 3700 San Martin Drive,
    Baltimore, MD 21218 \label{stsci}
\and
    European Southern Observatory, Karl-Schwarzschild-Str. 2, 85748
    Garching, Germany \label{eso}
\and
    Computational Cosmology Center, Computational Research Division, Lawrence Berkeley National Laboratory, 
    1 Cyclotron Road MS 50B-4206, Berkeley, CA, 94720 \label{nersc}
\and
    Kavli Institute for the Physics and Mathematics of the Universe,
    University of Tokyo, 5-1-5 Kashiwanoha, Kashiwa, Chiba, 277-8583, Japan \label{ipmu}
}
  \abstract
   {Type Ia Supernovae 
   (\sne) are widely used to measure the expansion of the Universe. Improving distance measurements  of  
\sne is one technique to better constrain the acceleration of expansion and determine its physical nature.}
   {This document develops a new \sne spectral energy distribution (SED) model, called the SUpernova Generator And Reconstructor (\sugar), which improves the spectral description of \sne, and consequently could improve the distance measurements.}
   {This model is constructed from \sne spectral properties and spectrophotometric data from The Nearby Supernova Factory collaboration. In a first step, a PCA-like method is used on spectral features measured at maximum light, which allows us to extract the intrinsic properties of \sne. Next, the intrinsic properties are used to extract the average extinction curve. Third, an interpolation using Gaussian Processes facilitates using data taken at different epochs during the lifetime of a \snia and then projecting the data on a fixed time grid. Finally, the three steps are combined to build the SED model as a function of time and wavelength. This is the \sugar model.}
   {The main advancement in \sugar is the addition of two additional parameters to characterize \sne variability. The first is tied to the properties of \sne ejecta velocity, the second is correlated with their calcium lines. The addition of these parameters, as well as the high quality the Nearby Supernova Factory data, makes \sugar an accurate and efficient model for describing the spectra of normal \sne as they brighten and fade.}
   {The performance of this model makes it an excellent SED model for experiments like ZTF, LSST or WFIRST.}

   \maketitle
%
 
 \section{Introduction}

Type Ia supernovae (\sne) are excellent cosmological probes: they are very
luminous objects, visible up to a redshift of $z\sim2$  
\citep{Guillochon17}, and their luminosity dispersion is naturally low 
and can be further reduced by an appropriate standardization process. As 
precise distance indicators, comparing their luminosity and their redshift
allowed \cite{1998Natur.391...51P,Perlmutter99}, \cite{Riess98}, and \cite{Schmidt98}  to demonstrate that 
the expansion of the Universe is accelerating, a feature that gave rise to the 
dark energy paradigm. This result, obtained with a small sample of \sne, has since been repeatedly		
confirmed with larger samples \citep{Astier06,Guy10,Suzuki2012,Rest13,
Betoule14,Scolnic17}, and in combination with other cosmological probes 
like the Cosmological Microwave Background (CMB) \citep{Planck2015},
Baryon Acoustic Oscillations (BAO) \citep{Delubac15}, or cosmic shear 
\citep{2017arXiv170801538T} gave birth to the so-called concordance model:
the flat-$\Lambda$CDM model. \\
\indent Improving our description of \sne as a cosmological probe is needed, 
not only to discriminate among alternate dark energy models \citep{Copeland06}, but also in order
to address existing tensions such as the 3.4 $\sigma$ difference between the values of the 
Hubble Constant $H_0$ from the local measurement by \cite{Riess16} and the cosmological fit from 
\cite{Planck2015}. The current uncertainty budget due to limited 
\snia statistics will be greatly improved by current surveys like ZTF \citep{Bellm2014}
or next generation surveys like LSST or WFIRST \citep{DESC2012,WFIRST}. 
Since, systematic and statistical uncertainties are already of the same order of magnitude 
\citep{Betoule14, Scolnic17}, a better understanding of systematics will be needed to improve 
the accuracy of \sne as a cosmology probe. \\
\indent Part of the systematic uncertainty comes from the standardization process: the
observed flux of the \snia has to be corrected for variations observed from one object to 
another. Two main contributions to this variation have been observed. The first one, seen by \cite{1974PhDT.........7R} and
\cite{1977SvA....21..675P,1984SvA....28..658P}, is the correlation between the peak luminosity of a \sne 
and the light curve decrease time, the so-called brighter-slower effect. Various parametrizations have been proposed
for this effect, the most commonly used being $\Delta m_{15}$ \citep{Phillips93}, stretch \citep{Perlmutter97} or
$X_1$ \citep{Guy07}. The second contribution, observed first by \cite{Hamuy95} and \cite{Riess96}, is 
that the peak luminosity depends on color, the so-called brighter-bluer effect. This effect can be explained by the 
presence of dust in varying quantities along the line of sight, possibly combined with an intrinsic 
color-brightness correlation after the stretch effect has been taken into account.
Most standardization techniques in photometry thus rely on a stretch and color relation 
known as the \cite{Tripp98} relation. The SALT2 model \citep{Guy07} is one such standardization method, 
and has over the years become the reference in cosmological analysis.
Despite many attempts, no consensus has yet emerged on how
to go beyond a 2-component model to describe \sne light curves. 
However, the observation of a step in the standardized luminosity with
respect to the host mass by \cite{Kelly10} and \cite{Sullivan10} has led to the inclusion of a
corrective term in the subsequent cosmological analysis. This corrective term
linked to the environment is a hint that more fundamental properties of \sne 
physics are not captured by the stretch-color standardization scheme, and that 
there is room for improvement. \\
\indent However, the most obvious indication that the parameterization used is insufficient comes
from the observation of a residual dispersion of \sne luminosity around the Hubble diagram 
after standardization.
In order to obtain statistically coherent results, 
\cite{Perlmutter99} and \cite{Astier06} introduced an intrinsic dispersion in luminosity as an additional 
uncertainty in the fit. \cite{Betoule14} estimates the dispersion value at 0.11
mag using a SALT2 standardization, for an observed total dispersion of 0.16 mag. 
Since the intrinsic dispersion is due to unmodeled \snia 
variations that may depend on the redshift, this may result in a significant error on 
the extraction of cosmological parameters \citep{Rigault18}. 
The precision of \sne as cosmological probes therefore depends 
on this intrinsic dispersion in luminosity. However, the intrinsic dispersion depends 
strongly on the assumptions about measurement uncertainties. A way to characterize the overall accuracy 
obtained by a given standardization method is to use the weighted Root Mean Square
(wRMS) metric on the standardized magnitude (e.g. \citealt{Blondin11}). 
Thus, in order to reduce the effects of unmodeled 
variations of \sne, an improved standardization procedure must also involve 
reducing the wRMS.\\
\indent Most efforts to improve the standardization of \sne involve the search 
for a new parameter correlated with the intrinsic luminosity. After the discovery 
of the mass-step, many efforts focused on describing the environmental effects 
of the \sne \citep{Kelly10,Sullivan10,Rigault13, Rigault14, Roman17, Rigault18}. A complementary approach 
consists of directly looking for this new parameter from the analysis of the light curves
or observed spectra. For example, \cite{Mandel17} proposed that the color of \sne 
is a mixture of intrinsic color and extinction by dust, and proposed a Bayesian model 
to separate these two components. For their part, \cite{Chotard11} showed that once the \CaHK variability 
is taken into account, the color law is compatible with a \cite{Cardelli89} extinction curve. Moreover, \cite{Mandel14} 
highlighted the dependence of the extinction curve on the minima of P-Cygni profiles. Thus the use of additional variables 
related to spectral indicators seems to be a promising avenue to describe the variability of \sne. \\
\indent The use of spectral indicators to standardize \sne has a long history
\citep{Nugent95,Arsenijevic08,Bailey09,Wang09b,Foley11,Chotard11}. 
As example, using only the flux ratio \Rsjb, \cite{Bailey09} is 
able to get a significatively better standardization in B-band  in comparison to the classical \cite{Tripp98} relation.
The method of using spectral information is also suggested by the analysis of \cite{Boone15}, which shows that the best method 
to measure distance with \sne is with spectroscopic twins \sne. Another recent study by \cite{Nordin18} has shown that the use 
of spectral information from the UV part of the spectra improves distance measurement compared to the \cite{Tripp98} relation. However,
those methods do not currently lead to a spectral energy distribution (SED) 
model, which is necessary for cosmological analyses on 
purely photometric data, such as LSST \citep{DESC2012}. Thus we propose here a full SED model 
which will be based on spectral indicators generalizing the 
procedure originally developed in \cite{Chotard11}. This method uses spectral features, which allows for the 
addition of more than one intrinsic parameter and they offer a possible way to separate intrinsic properties from extrinsic properties. 
From it, we build a full SED model which may be used for purely photometric surveys. 

The aim of this study is to revisit the parametrization of \sne SED in light of the \snf 
spectrophotometric dataset \citep{Aldering02}, and to seek new sources of variability by statistical analysis. Our model is 
trained using spectral indicators, derived around maximum light in B-band, as features to describe the model. They 
provide both a reduced dimensionality description of spectra, and a description which is linked to the physics of 
the explosion. In addition, we select indicators insensitive to reddening in order to decouple the characterization of the 
reddening from effects purely linked to the intrinsic part of the explosion. This new \sne SED model is 
named the SUpernova Generator And Reconstructor (SUGAR) model. 

Another approach to develop a new SED model was undertaken in parallel by \cite{Saunders18}, using the same dataset and based on a "SALT2 like" framework. 
\cite{Saunders18} generalized a strategy that was originally proposed by \cite{Guy07} and which was first performed on broad band photometry from the 
SNfactory dataset in \cite{Kim13}. In brief, \cite{Saunders18} did a Principal Component Analysis-like study on interpolated spectral time series in order 
to go beyond the classical \cite{Tripp98} relation. \cite{Saunders18} is able to significatively improve the SED description with respect to the SALT2 model. 
The SNEMO model developed in \cite{Saunders18} differs from the \sugar model in that SNEMO attempts to find principal components purely based on spectral 
time series variability in the relative luminosities at each wavelength and as a function of time, whereas \sugar uses spectral features to try to find a compact description of such 
spectral time series. However, both models share some technical details in the model training and were developed in common. 

This paper is organized as follows: Section~\ref{descriptiondatset} presents 
how the spectrophotometric time series of the \snf were obtained, Section~\ref{deriveddata}
focuses on the intermediary data used in building the \sugar model (spectral indicators at maximum 
light, dimensionality reduction through factor analysis, derivation of extinction parameters and time 
interpolation). Section~\ref{sugarmodeltrainingsection} describes the \sugar model: 
its formalism, training and components. Section~\ref{sugarperfs} presents the performance 
of the model for fitting spectral time series using \sugar model. These results are
compared to the performance achieved by SALT2 for the same data. Finally, Section~\ref{discussionsugarsection} discusses adding 
additional components to the
\sugar model and some technical choices that were made in the training of the \sugar model. The appendices 
describe details of the mathematical implementation of the \sugar model. The SUGAR template is available online at \url{http://supernovae.in2p3.fr/sugar}, while the 
data used to train SUGAR are available at \url{https://snfactory.lbl.gov/sugar}\footnote{The data link will become active upon journal publication.}.

\section{Spectrophotometric time series from \snf}

\label{descriptiondatset}

This analysis is based on \NsnTotal SNe Ia obtained by the \snf collaboration 
beginning in 2004 \citep{Aldering02} with the SuperNova Integral Field Spectrograph (SNIFS,
\cite{Lantz04}) installed on the University of Hawaii 2.2-m telescope (Mauna Kea).
SNIFS is a fully integrated instrument optimized for semi-automated observations of
point sources on a structured background over an extended optical window at moderate
spectral resolution. SNIFS has a fully-filled $6.4''\times6.4''$ spectroscopic field-of-view
subdivided into a grid of $15\times15$ contiguous square spatial elements (spaxels).
The dual-channel spectrograph simultaneously covers 3200-5200 \AA{} (B-channel)
and 5100-10000 \AA{} (R-channel) with 2.8 and 3.2 \AA~resolution, respectively.
The data reduction of the x, y, $\lambda$ data cubes is 
summarized by \cite{Aldering06} and updated in Sect.~2.1 of  \cite{Scalzo10}. 
The flux calibration is developed in Sect. 2.2 of \cite{Pereira13} based on the atmospheric 
extinction derived in \cite{Buton13}. In addition, observations 
are obtained at the \snia location at least one year after the explosion to serve as a final 
reference to enable subtraction of the underlying host and the host subtraction, as described in 
\cite{Bongard11}. For every \snia followed, the \snf creates a spectrophotometric time 
series, typically composed of $\sim$14 epochs, with the first spectrum taken on average three 
days before maximum light in $B$-band \citep{Bailey09,Chotard11}. The sample of \NsnTotal SNe Ia contains 
the objects with good final references, objects that passed quality cuts suggested by \cite{Guy10},
and is restricted to objects with at least one observation in a time window of $\pm$2.5 days around maximum light in B-band.
The size of this window is kept identical with respect to the study of \cite{Chotard11} and 
is discussed in \cite{LegetPhD}. After flux calibration, host-galaxy subtraction and correction
for Milky Way extinction, the flux of the \NsnTotal spectra is integrated in synthetic top-hat filters defined in 
\cite{Pereira13} and for reference a SALT2 fit is applied with the model from \cite{Betoule14} in order to obtain 
the $X_1, C$ and $m_B$ parameters. The spectra of the \NsnTotal \sne are transformed to the rest 
frame with a fiducial cosmology and an arbitrary Hubble constant. It is allowing to respect blinding of future cosmological analysis using this dataset. For the spectral 
analysis here, the spectra are rebinned  at 1500 $\text{km s}^{-1}$ 
between \sugarMinWavelength and \sugarMaxWavelength \AA\  (\numberbinsugar bins per spectra) for computational efficiency while still resolving spectral 
features and converted to the absolute AB magnitude system. The training is done on \NsnTraining \sne observed before 2010 and 
the validation is done on \NsnValidation \sne observed after 2010. 

\section{Derived data}
\label{deriveddata}

Any empirical SN~Ia modeling must solve three problems\footnote{In general, empirical SNIa modeling must 
also handle spectroscopic and photometric data that are not observed on 
the same day or by the same instruments. However, this is not a problem with the SNfactory dataset.}:
choosing what features to model, accounting for color, and dealing with the data sampling.

For the features modeling of SUGAR, we used spectral features at maximum light to describe the intrinsic part of the SED. 
This is described successively in 
Section~\ref{si_derivation_section} and Section~\ref{emfa_section}. Section~\ref{si_derivation_section} describes 
how the spectral features are selected and measured. Section~\ref{emfa_section} explains how the spectral features 
are projected onto a new basis that allows us to work in an orthogonal basis for the \sugar training.

To estimate the average color curve of \sne, we will generalize the method 
of \cite{Chotard11}. This is described in Section~\ref{extinction_law_section}.

To deal with the data sampling, we project the observed spectra onto a common time grid so that the SED model 
can be calculated on the same time grid. This is done in Section~\ref{gp_interp} using the Gaussian Process method.

The following sections describe those intermediate steps that will determine the full SUGAR SED model in 
Section~\ref{sugarmodeltrainingsection}. We will use the spectral features derived and color curve parameters 
derived at maximum light combined with the interpolated spectra to infer the full SED at all epochs.

\subsection{Spectral indicators at maximum light}
\label{si_derivation_section}

Spectral indicators are metrics of empirical features 
of the input spectrum such as equivalent widths or line velocities. They offer an efficient characterization of spectral 
variability by representing the underlying spectral complexity with a few key numbers, and play a role in 
non-linear dimensionality reduction of the original data. These two characteristics, 
interpretability and simplicity, make them ideal for describing the intrinsic part of \sne SED, 
and consequently derive the extrinsic part in the same spirit as \cite{Chotard11}.
This decoupling restricts the set of spectral indicators to 
pseudo-equivalent widths and the wavelengths of P-Cygni profile minima. 
As the spectral indicators evolve with phase, we select the spectrum closest to 
maximum light in B-band, if it is within a time window of $\pm$ 2.5 days. 
This window size was chosen to optimize the trade-off between the total number of 
\sne in the sample and the potential loss in precision due to time evolution.

\begin{figure}
	\centering
	\includegraphics[scale=0.35]{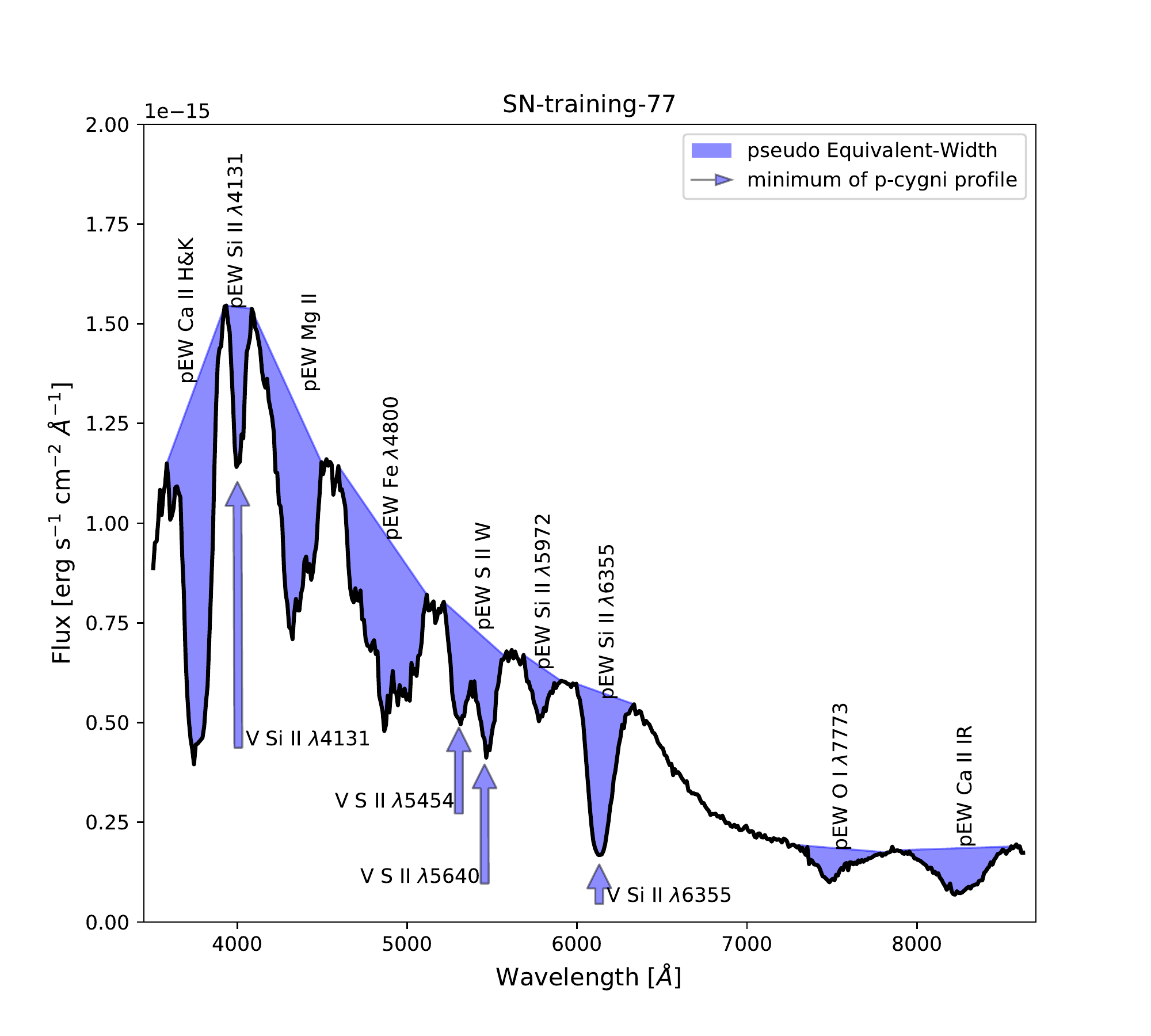}
	\caption{\small The nine pseudo-equivalent widths of absorption lines at maximum light
		       as well as the four minima of P-Cygni profiles at maximum of light which
		       are used in our analysis. They are represented on the spectrum of SN-training-77 at 
		       a phase of 1 day after maximum brightness.}
	\label{Spectral_indicator_used_by_sugar}
\end{figure}

\indent Near maximum light and within the spectral range covered by the 
\snf, it is possible to systematically obtain the 13 spectral indicators represented 
in Figure~\ref{Spectral_indicator_used_by_sugar}. They are the  
9 pseudo-equivalent widths of \CaHK, \Sia, \Mg, \FE, \SW, \Sib, \Sic, \OI \footnote{We will refer to this as \OI but note it overlaps with Mg~\textsc{ii}~$\lambda$7812.}, and \CaIR features, 
as well as 4 minima of P-Cygni profiles (\Sia, \SWl, \SWr, and \Sic). For the pseudo-equivalent 
widths, we rely only on well-defined troughs that are present in all \sne in the sample, and 
consider line blends as a whole. Some of the possible minima were also discarded either 
because the corresponding lines form a complex mixture or because 
the corresponding trough is too shallow to accurately define a minimum for some 
\sne. This is why the \Sib feature velocity is rejected. The spectral indicators and
their uncertainties were automatically derived from the spectra at maximum light following the procedure of 
\cite{Chotard11}, which is described in detail in Appendix~\ref{app:measurements}.

\subsection{Factor analysis on spectral features}
\label{emfa_section}

\subsubsection{Factor analysis model}
\label{emfa_model}

The space defined by the 13 selected spectral indicators has too high dimensionality to  
efficiently train a model. Additionally, some of the spectral indicators are correlated, and 
therefore contain redundant information. Most of the model variation can be captured 
in a reduced number of dimensions.
Principal Component Analysis (PCA) \citep{Pearson1901} is 
one of the methods to implement such a dimensionality reduction. It consists of diagonalizing the covariance 
matrix of the sample and projecting the data into the resulting eigenvectors basis.
In this new basis, the variables are uncorrelated, and an approximation of the 
input data is found by neglecting the dimensions corresponding to the smallest eigenvalues.
This method has been employed in the case of \sne by \cite {Guy07}, 
\cite{Kim13}, and \cite{Sasdelli2015}. However, in the case considered here, some directions are 
dominated by noise, so their eigenvectors would align along the direction of measurement 
errors rather than the intrinsic sample variance. To solve this problem, we employ a variant of PCA, Factor Analysis
\citep{spearman1904general, spearman1927abilities}, which has the advantage of taking into account the variance 
caused by measurement uncertainties. This technique decomposes the observables 
into two terms, one representing the explanatory factors and one representing the noise  
affecting each variable. This is expressed by the following 
relation \citep{Ghahramani97theem}:

\begin{equation}
\textbf{x}_i = \boldsymbol{\Lambda} \textbf{q}_i + \textbf{u}_i \ ,
\end{equation}

\noindent where $\textbf{x}_i$ is the spectral feature vector with $s$ components 
for the $i^{\text{th}}$ \snia. $\textbf{q}_i$ is the explanatory factor vector
of dimension $ m \leq s$, and is linearly related to $ \textbf {x}_i $ by the 
matrix $\boldsymbol{\Lambda}$ of dimensions $ m \times s $.  
$\mathbf{u}_i$ is the noise with variance $\mathbf{\Psi}$ of dimensions $ s \times s $. In order 
to fix the normalization of $\boldsymbol{\Lambda}$ and $\textbf{q}_i$, 
we assume that the $\textbf{q}_i$ are drawn from a centered normal distribution:

\begin{equation}
P\left(\textbf{q}_i\right) \sim {\cal N}\left(0 ,\textbf{I} \right) \eqdot
\end{equation}

In the framework of the PCA, and up to a normalization, the matrix $\boldsymbol{\Lambda}$ and
the $\textbf{q}_i$ are respectively equivalent to the
eigenvector matrix and the projections into the new basis. Factor analysis 
thus consists of determining the matrices $\boldsymbol{\Lambda}$ and $\boldsymbol{\Psi}$ 
that maximize the likelihood, under the assumption that the  $\textbf{x}_i$ 
are distributed according to a normal distribution with both intrinsic scatter and measurement noise:

\begin{equation}
 P\left(\textbf{x}_i\right) \sim {\cal N}\left(0 ,\boldsymbol{\Lambda} \boldsymbol{\Lambda}^T +\boldsymbol{\Psi} \right) \eqdot
\end{equation}

To estimate $\boldsymbol{\Lambda}$, $\boldsymbol{\Psi}$, and the explanatory factors 
$\textbf{q}_i$, \cite{Ghahramani97theem} propose a solution based on an expectation-maximization 
algorithm where $\textbf{q}_i$ and $\boldsymbol{\Lambda}$ are estimated iteratively. 
Here, unlike in conventional factor analysis, a reliable estimate of the spectral indicator
measurement error is provided. For each \snia $i$, $\boldsymbol{\Psi}_i$ is the known diagonal matrix that 
contains the squared errors of the spectral indicators, hence, it is not necessary to fit for a global $\mathbf{\Psi}$.
We adapted the aforementioned expectation-maximization algorithm 
to take this change into account. Additional details can be 
found in Appendix~\ref{app:EMFA}. 

Finally, a prescription is needed to normalize each individual variable. 
The amplitude of the spectral indicator variation, expressed in \AA\ , is not a good indicator 
of their impact on the spectral shape. 
As an example, \cite{Branch06a} show that the weak \Sib line can play a significant role in 
subclassing \sne. As a consequence, the input data are normalized to unit variance prior 
to the factor analysis, so that no spectral indicator is favored a priori. 
Within this framework, the eigenvalues of $\mathbf{\Lambda\Lambda}^T$ represent the variance 
explained by each factor, and $s - \mbox{Tr}(\mathbf{\Lambda\Lambda}^T)$ is the variance coming 
from the noise, where here $s=13$ is the number of spectral indicators and Tr is the trace operator.

\subsubsection{Outlier rejection}
\label{sec:emfa_outlier}

Outliers affect the sample variance of any population. As we 
care most about a correct description of the bulk of \sne, it is desirable to identify and 
remove these outliers. Amongst them, \sne of type SN1991T \citep{Filippenko91T} or 
SN1991bg \citep{Filippenko92c} are known to have different spectral and photometric behavior 
from other \sne. However, basing an outlier rejection on this empirical identification has 
two issues. The first is that within the spectral indicator space, some of those subtypes may 
not appear as distinct subclasses, and do not offer objective grounds for rejection. The second issue is 
that the attribution of a given \snia to one of these subclasses by SNID \citep{Blondin07} may 
provide inconsistent results depending on the epoch considered for the identification.
In order to apply a self-contained criterion for defining an outlier, we thus resorted to a 
$\chi^2$-based definition for identifying outliers: this quantity can be interpreted as the squared distance  
to the center of the distribution normalized by the natural dispersion. For  
\snia $i$, it is given by

\begin{equation}
  \chi^2_i= \textbf{x}_i^T\left(\boldsymbol{\Lambda}
   \boldsymbol{\Lambda}^T  + \boldsymbol{\Psi_i} \right)  \textbf{x}_i \eqdot
\end{equation}

In the case of a Gaussian distribution, $\chi^2_i$ should follow a 
$\chi^2$-distribution with 13 degrees of freedom. An iterative cut at 
$3 \ \sigma$ on the value of $\chi^2_i$ rejects \NsnSigmaClipping \sne 
from the sample. A visual inspection carried out on each of these \NsnSigmaClipping 
\sne, as well as a sub-classification made using SNID, show that 4 of 
those objects exhibit shallow silicon features as defined by \cite{Branch06a}, 2 of them have high-velocity silicon features as defined by \cite{Wang09b}, one is within the broad line subcategory, 
and the remaining one has a failed estimation of one of the spectral indicators.

\indent We studied the influence of the $3 \ \sigma$ cut by carrying 
out the analysis successively with and without the cut. Even though this had
no major effect on the direction of the vectors, the cut is applied to the rest 
of the analysis to avoid training the model on outliers.

\subsubsection{Factor analysis results on spectral features}
\label{emfa_results}

\begin{figure}
	\centering
	\includegraphics[scale=0.47]{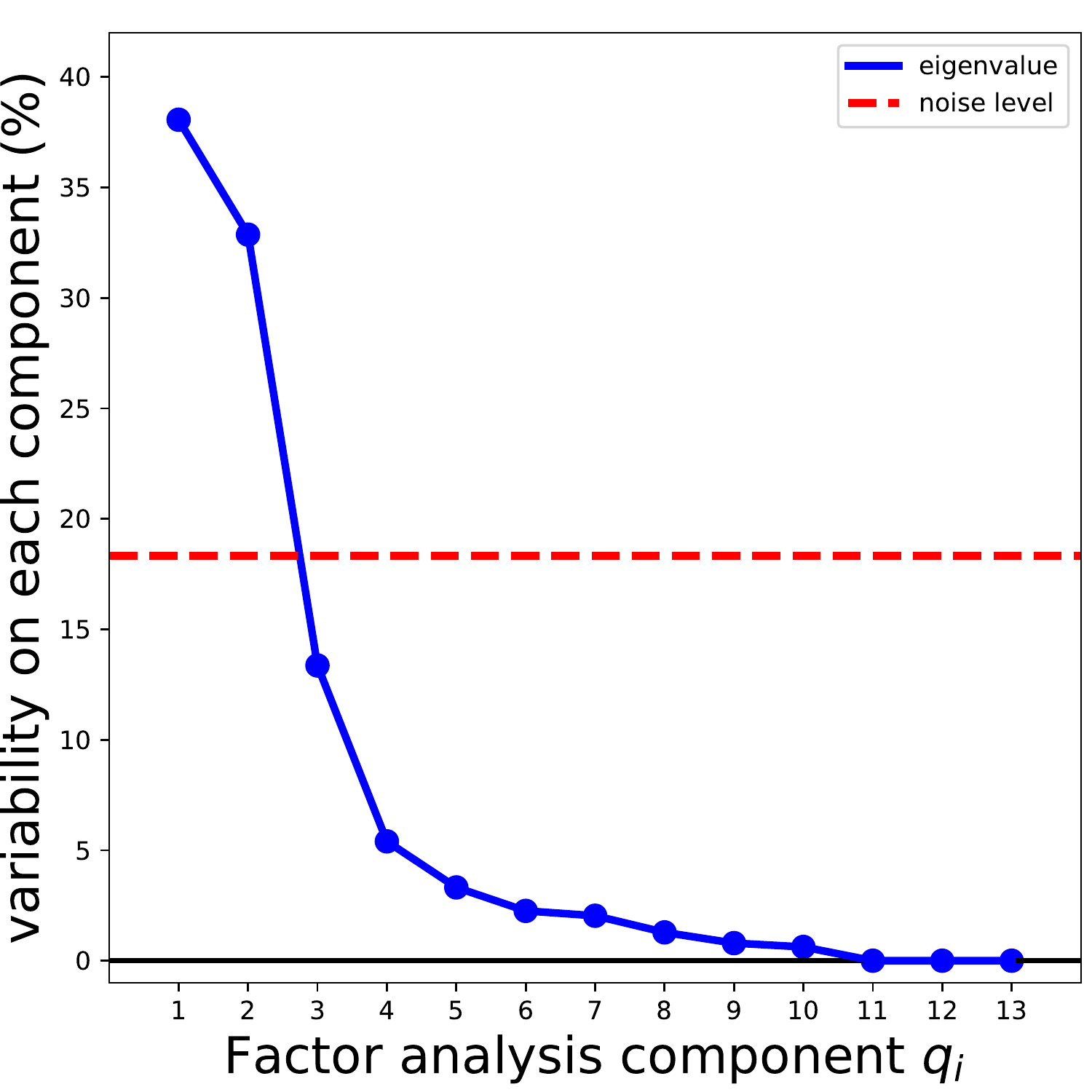}
	\caption{\small 
	Relative importance of the eigenvalues associated with the eigenvectors. They are 
	ordered in decreasing order of variance, and the total amounts to 100\% 
	once the contribution of the noise (indicated by a red line) is taken into account. We can 
	see that the first three vectors dominate the variability of this space.}
	\label{Table_valeur_propre}
\end{figure}

\begin{figure*}
	\centering
	\includegraphics[scale=0.6]{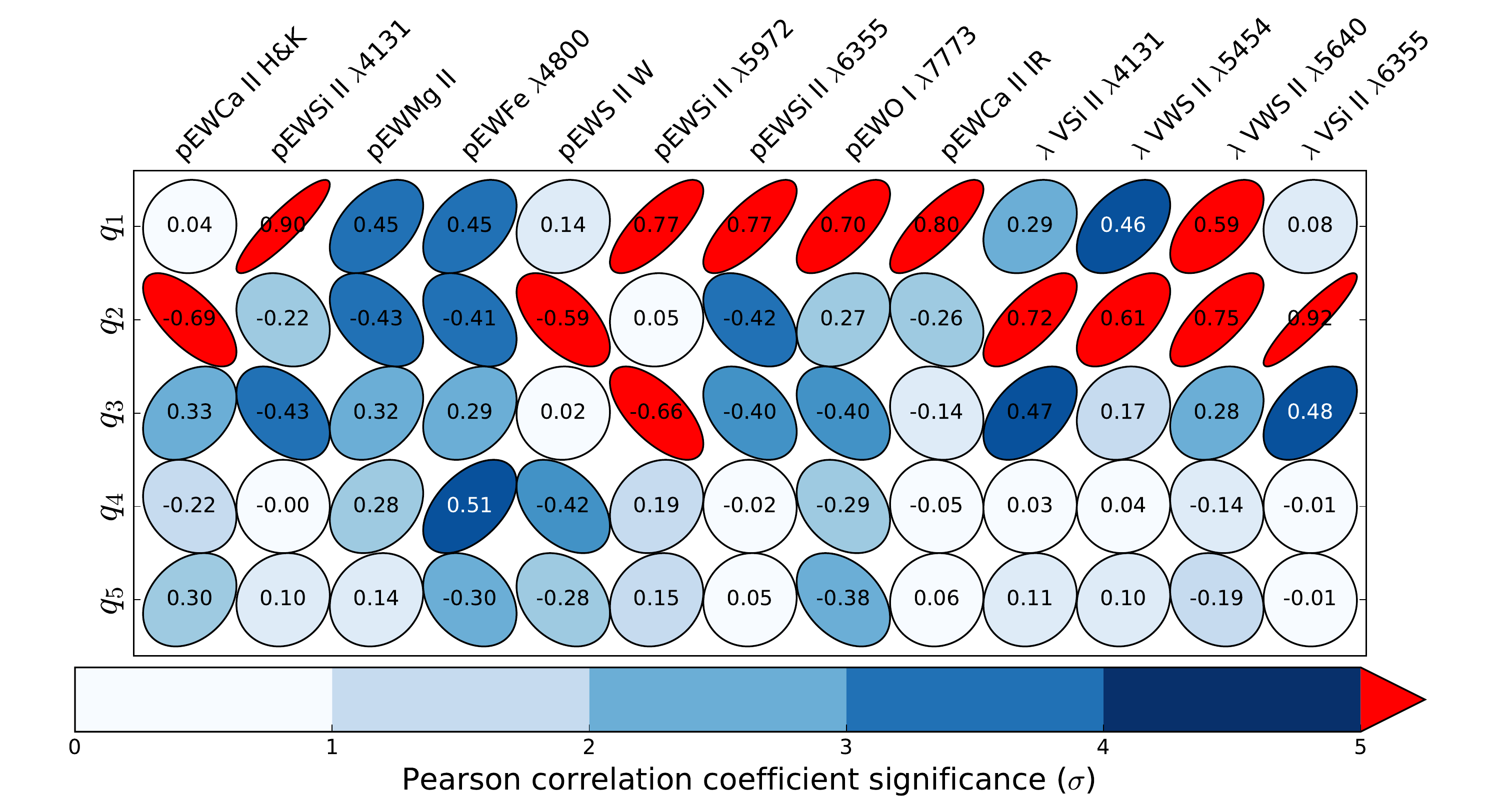}
	\caption{\small Pearson correlation coefficients between the spectral indicators and 
	the first five factors. The eccentricities of the ellipses indicate the magnitude of the 
	correlation, and the color the importance of the correlation in units of $\sigma $. 
	The vectors 1 and 2 correspond respectively to a global correlation of the 
	pseudo-equivalent widths and of the line velocities. Only the first 3 vectors 
	display correlations with a significance higher than $ 5 \sigma $. }
	\label{Corr_coeff_caballo}
\end{figure*}

\begin{figure}
	\centering
	\includegraphics[scale=0.7]{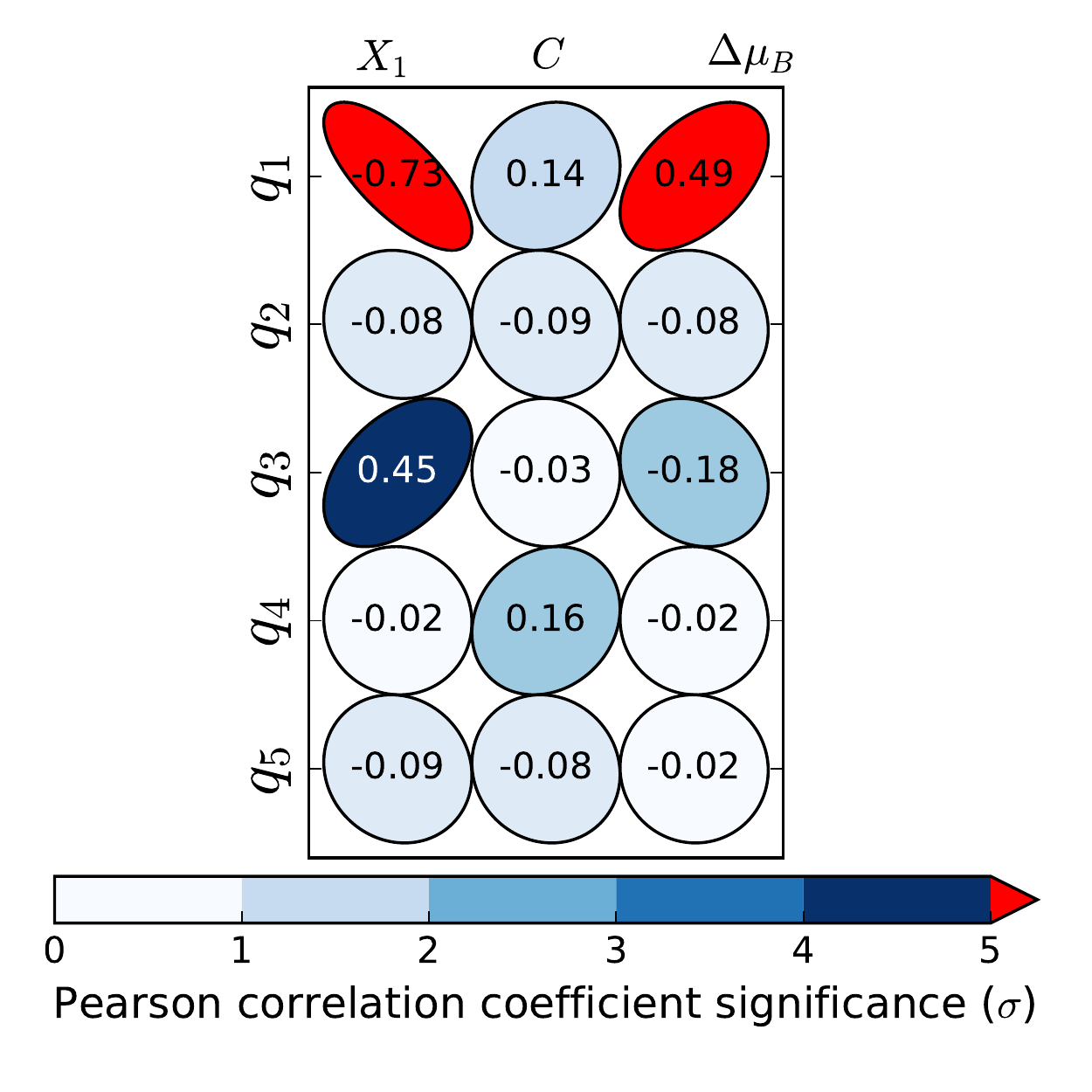}
	\caption{\small Pearson correlation coefficients between the SALT2 parameters and 
	the first five vectors. The eccentricities of the ellipses indicate the magnitude of the 
	correlation, and the color the importance of the correlation in units of $\sigma$.}
	\label{Corr_coeff_caballo_salt2}
\end{figure}

The EM-FA algorithm described in the previous sections is applied to the 13 spectral features measured at maximum light. 
The relative weight of the eigenvalues on the total variance is shown in Figure~\ref{Table_valeur_propre},
and the correlations between the eigenvectors and the spectral indicators are presented 
in Figure~\ref{Corr_coeff_caballo}. The significance of the correlation is
expressed in units of $\sigma$, meaning that the null hypothesis, i.e. 0 correlation
is rejected at $\sigma$ confidence level. Figure~\ref{Table_valeur_propre} shows 
that two vectors dominate the variability in the space of spectral indicators. 
The first vector is linked with coherent variations of the pseudo-equivalent widths, 
except the widths of the \CaHK and \SW lines as seen in Figure~\ref{Corr_coeff_caballo}. 
The second vector is anticorrelated with these same two features, and 
describes a coherent variation of the velocities. Since the first vector is very similar 
to the pseudo-equivalent width of \Sia, which is shown by \cite{Arsenijevic08} to be highly
correlated with the stretch parameter, it is therefore natural to assume that this vector represents 
the main source of intrinsic variability already known for \sne. The second vector is 
mostly driven by the line velocities and \mbox{pEW(\CaHK)} and \mbox{pEW(\SW)} lines, 
with smaller dependencies from the other pseudo-equivalent widths. Interestingly, while investigating the spectral diversity of \sne
beyond stretch and color, \cite{Chotard11} demonstrated the role of \mbox{pEW(\CaHK)}, while \cite{Wang09b} 
focused on the role of the speed of \Sic. Our second vector unifies these two approaches. 
The interpretation of the next vectors is less straightforward. 
As their rank increases, they describe less of the sample variance, and thus exhibit lower correlations 
with the original data. The third vector is the last of the eigenvectors to show strong correlation with one of the spectral indicators, 
\mbox{pEW \Sib}. The correlations with the SALT2 parameters are presented in 
Figure~\ref{Corr_coeff_caballo_salt2}. As expected from the correlations with the pseudo-equivalent 
widths, $q_1$ is strongly correlated with $X_1$. Also, $q_3$ exhibits a significant 
correlation with $X_1$, which may be an indication that the stretch is driven by two different parameters. 
Remaining factors are only weakly correlated with $X_1$. The correlations between SALT2 color 
and any of the factors are close to zero. The SALT2 absolute magnitude cosmology residual,
$\Delta \mu_B$, are mainly correlated with $q_1$ and $q_3$, presumably due to the correlation of both with $X_1$.

\indent Within this framework, the eigenvalues of $\mathbf{\Lambda\Lambda}^T$ represent the variance 
explained by each factor, and $s - \mbox{Tr}(\mathbf{\Lambda\Lambda}^T)$ is the variance coming 
from the noise, where here $s=13$ is the number of spectral indicators and Tr is the trace operator. 
Noise, in this situation, represents \noiselevelemfa of the variability observed, 
so only the first two vectors clearly outweigh the noise and it is legitimate to ask whether or not the 
other vectors are sensitive to statistical fluctuations. For the rest of the analysis, we keep 
the first three factors for the final training and discuss this choice in the Section~\ref{discussion_number_of_component}, by looking at 
the impact on the final model.

\subsection{Extinction curve estimation}
\label{extinction_law_section}

In this section, we estimate the average extinction curve by generalizing the procedure described in 
\cite{Chotard11}. For this, a model of the SED at maximum light is derived. This model will allow us to deduce the average extinction curve and the procedure to estimate it: 
the results are described in the following.

\subsubsection{Empirical description}

The SED model at maximum light presented here is a generalization of the model 
presented in \cite{Chotard11}, in which the authors removed intrinsic variability tied to spectral features and measured 
the remaining spectral variation, which was found to be consistent with an extinction curve like that for the Galaxy. 
The major difference is that the intrinsic description is 
here motivated by the factors derived in Section~\ref{emfa_section} rather than being described only by \mbox{pEW(\Sia)} and 
\mbox{pEW(\CaHK)} as was done in \cite{Chotard11}. To model the SED we define for the \snia $i$  
the vector $\textbf{x}_i \equiv \{h_{i,1}, h_{i,2}, h_{i,3}\}$, 
which is the true value of the measured factor $\textbf{q}_i = \{q_{i,1}, q_{i,2}, q_{i,3}\}$. We propose that these 
are related to the intrinsic absolute magnitude by:

\begin{equation}
\label{Model_SED_at_max}
     M_i(\lambda)=M_0(\lambda)+  \ \sum_{j=1}^{j=3} h_{i,j} \alpha_j(\lambda)  + 
     A_{\lambda_0,i} \ \gamma(\lambda) \eqcoma
\end{equation}

\noindent  where $M_0(\lambda)$ is the average spectrum in absolute magnitude,  $\alpha_j(\lambda)$
is the spectrum related to the factor $h_{i,j}$, where $j$ is the factor index running from 1 
to 3 (the choice of the number of components is discussed in Section~\ref{discussion_number_of_component}). 
$A_{\lambda_0,i}$ is a free parameter that can be interpreted as the absorption due 
to extinction at a reference wavelength $\lambda_0$ (set here at the median wavelength, 
\sugarMedianWavelength, without loss of generality). $\gamma(\lambda)$ is an arbitrary 
function that represents the effect of the extinction, and we have not set any prior on its shape. Once the function  $\gamma(\lambda)$ 
is fixed, it is possible to fit the total-to-selective extinction ratio of the \cite{Cardelli89} law, 
$R_V$, and the absorption in V-band, $A_V$ to it; this will be described in the next sub-section. The model parameters 
here are $M_0(\lambda)$, the $\alpha_j(\lambda)$, the $\gamma(\lambda)$, the $h_{i,j}$, 
and the $A_{\lambda_0,i}$. It is understood that we are modeling 
the SED at specific wavelengths where we have measurements. For notational efficiency we rewrite Equation~
\ref{Model_SED_at_max} as:

\begin{eqnarray}
     \textbf{M}_i &=& \textbf{A} \textbf{h}_{i}  \eqcoma \textrm{where}\\
\textbf{A}&=&\left(\textbf{M}_{0},\boldsymbol{\gamma}, \boldsymbol{\alpha}_{1},\boldsymbol{\alpha}_{2},\boldsymbol{\alpha}_{3} \right) \eqcoma \\
\textbf{h}_i^T&=&\left(1,A_{\lambda_0,i},h_{i,1},h_{i,2}, h_{i,3}\right) \ .
\end{eqnarray}

\noindent In \cite{Chotard11}, the parameters $\alpha_j(\lambda)$ were computed in sequential order, and the reddening 
curve $\gamma(\lambda)$ as well as the extinction $A_{\lambda_0,i}$ were determined in a second step. We improve on this 
procedure by applying a global fit for all parameters at once. This is described below, and allows us to derive a global average $R_V$ 
for all \sne, and an $A_V$ for each \snia.

\subsubsection{Fitting the model at maximum light and $R_V$}
\label{global_fit_SED_RV_max}

In this section we explain the framework for fitting 
the free parameters of Equation~\ref{Model_SED_at_max} and how we then determine the $A_V$ and $R_V$ 
parameters. Fitting the parameters of the absolute SED of Equation~\ref{Model_SED_at_max} is done using an orthogonal distance regression. 
Estimating these parameters within the framework of an orthogonal distance regression amounts to minimizing the following $\chi^2$:

\begin{multline}
     \label{chi2_max_0}
     \chi^2 = \sum_i \left( \textbf{M}_i^{obs} - \textbf{A} \textbf{h}_{i} \right) ^T 
     \textbf{W}_{\textbf{M}_i}  \left( \textbf{M}_i^{obs} - \textbf{A} \textbf{h}_{i} \right)
     + \\
      \left( \textbf{q}_i - \textbf{x}_{i} \right)^T \textbf{W}_{\textbf{q}_i}
      \left( \textbf{q}_i - \textbf{x}_{i} \right)  \eqcoma
\end{multline}

\noindent where $\textbf{M}_i^{obs}$ is the observed spectrum of the \snia $i$ in absolute AB magnitude and
 $\mathbf{W}_{\mathbf{M}_i}$ is the weight matrix of $\mathbf{M}_i^{obs}$, defined as:

\begin{equation}
    \textbf{W}_{\textbf{M}_i}= \left[ \begin{pmatrix}
    \ddots &   &0 \\ 
       &  \sigma_{\lambda i}^2&\\ 
    0  &  & \ddots
\end{pmatrix} \ + \ \left(\sigma_{cal}^2+\sigma_{z}^2\right)  \begin{pmatrix}
    1      & \cdots & 1 \\ 
    \vdots & \ddots & \vdots \\ 
    1      & \cdots & 1 
\end{pmatrix} \ + \ \textbf{D} \right]^{-1}\eqcoma
\end{equation}

\noindent where $\sigma_{\lambda i}$ is the uncertainty of the \snia $i$ at the
wavelength $\lambda$ derived from spectral error, $\sigma_{cal}$ the 
per-spectrum calibration uncertainty, taken to be 0.03 mag, $\sigma_{z}$ the combination of the redshift error and the 
uncertainty of 300 $\text{km s}^{-1}$ due to peculiar velocities, and $\textbf{D}$ 
is the dispersion matrix. The dispersion matrix is fitted to deal with the remaining variability
which is not described by the three factors and the extinction. It includes an 
estimate of the remaining chromatic and  achromatic (grey) dispersions. The estimation of $\textbf{D}$ 
is described in Appendix~\ref{app:disp_matrix}. $\mathbf{W}_{\mathbf{q}_i}$ is the weight 
matrix of $\textbf{q}_i$ that comes from projecting the spectral feature uncertainties into the factor sub-space, and is defined as:

\begin{equation}
     \textbf{W}_{\textbf{q}_i} = \boldsymbol{\Lambda}^T \boldsymbol{\Psi}_i^{-1} \boldsymbol{\Lambda}\eqcoma
\end{equation}

\noindent where $\boldsymbol{\Lambda}$ and $\boldsymbol{\Psi}_i$ are the same as in Section~\ref{emfa_section}. 
The unknown parameters of the model are the matrix $\textbf{A}$ and the vector $\textbf{h}_i$.
These are found by minimizing equation \ref{chi2_max_0} using an expectation-minimization algorithm 
described in Appendix~\ref{app:ODR}.

\indent Once the parameters that minimize equation \ref{chi2_max_0} have been found, we can compare $\gamma_{\lambda}$ with 
an extinction curve to see if it is compatible. It is straightforward to directly estimate the global mean $ R_V $ from $\gamma_{\lambda}$ by minimizing

\begin{equation}
     \chi_{R_V}^2 = \left[\boldsymbol{\gamma}-\left(\textbf{a}+\frac{1}{R_V}\textbf{b}\right) \right]^T \ \left[ \boldsymbol{\gamma}-\left(\textbf{a}+\frac{1}{R_V}\textbf{b}\right)\right]\eqcoma
\end{equation}

\noindent where the \textbf{a} and $\mathbf{b}$ vectors 
are the extinction curve coefficients defined in \cite{Cardelli89}.
Minimizing $\chi_{R_V}^2$ as a function of $R_V$ then gives:

\begin{equation}
R_{V}=\frac{\textbf{b}^T\textbf{b}}{\boldsymbol{\gamma}^T\textbf{b}-\textbf{a}^T\textbf{b}}
\eqdot
\end{equation}

Note that the value we quote for $R_V$ within the context of our model depends on the assumptions adopted
in the evaluation of the dispersion matrix (discussed in detail in Section~\ref{discussion_grey_disp_matrix}) and 
therefore does not necessarily correspond to the true mean properties of dust. This uncertainty associated with 
separating \snia color behavior from dust behavior is common to all \snia fitting methods. Once the value of the global mean $ R_V $ is fixed, 
we determine the absorption in the V-band, $A_V$ for each \snia. This amounts to minimizing

\begin{multline}
     \chi_{A_V}^2 = \sum_i \left(\boldsymbol{\delta}\tilde{\textbf{M}}_i -A_{V,i} \ \left(\textbf{a}+\frac{1}{R_V}\textbf{b}\right) \right)^T  \textbf{W}_{\textbf{M}_i} \\ \left(\boldsymbol{\delta}\tilde{\textbf{M}}_i -A_{V,i} \ \left(\textbf{a}+\frac{1}{R_V}\textbf{b}\right)\right) \eqcoma
\end{multline}
 
\noindent where $A_{V, i}$ is the absorption in the V-band for the \snia $i$, 
and $\boldsymbol{\delta}\tilde{\textbf{M}}_i$ are the residuals 
corrected for the intrinsic variabilities

\begin{equation}
   \boldsymbol{\delta}\tilde{\textbf{M}}_i =   \textbf{M}_{i}^{obs} -\textbf{M}_{0}-  \ \sum_j q_{i,j} \boldsymbol{\alpha}_{j} 
   \eqdot
\end{equation}

\noindent Taking the derivative of $\chi_{A_V}^2$ by $A_V$ we find: 

\begin{equation}
\label{equation_sur_AV}
A_{V,i}=\frac{\left(\textbf{a}+\frac{1}{R_V}\textbf{b}\right)^T \textbf{W}_{\textbf{M}_i}\boldsymbol{\delta}\tilde{\textbf{M}}_i}{\left(\textbf{a}+\frac{1}{R_V}\textbf{b}\right)^T\textbf{W}_{\textbf{M}_i}\left(\textbf{a}+\frac{1}{R_V}\textbf{b}\right)}\eqdot
\end{equation}

\subsubsection{Results}
\label{sec:result_at_max}

\begin{figure}
	\centering
	\includegraphics[scale=0.4]{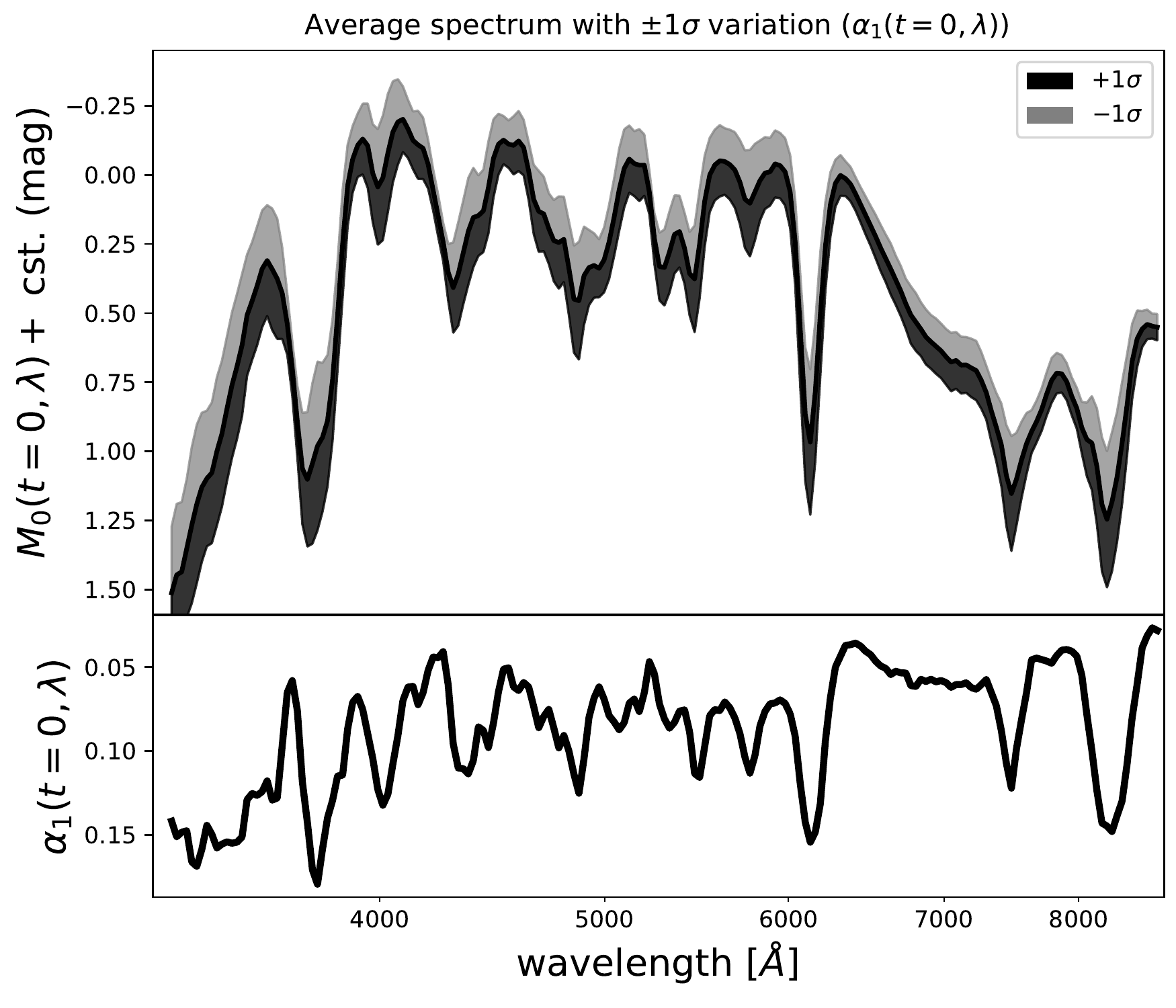}
	\includegraphics[scale=0.4]{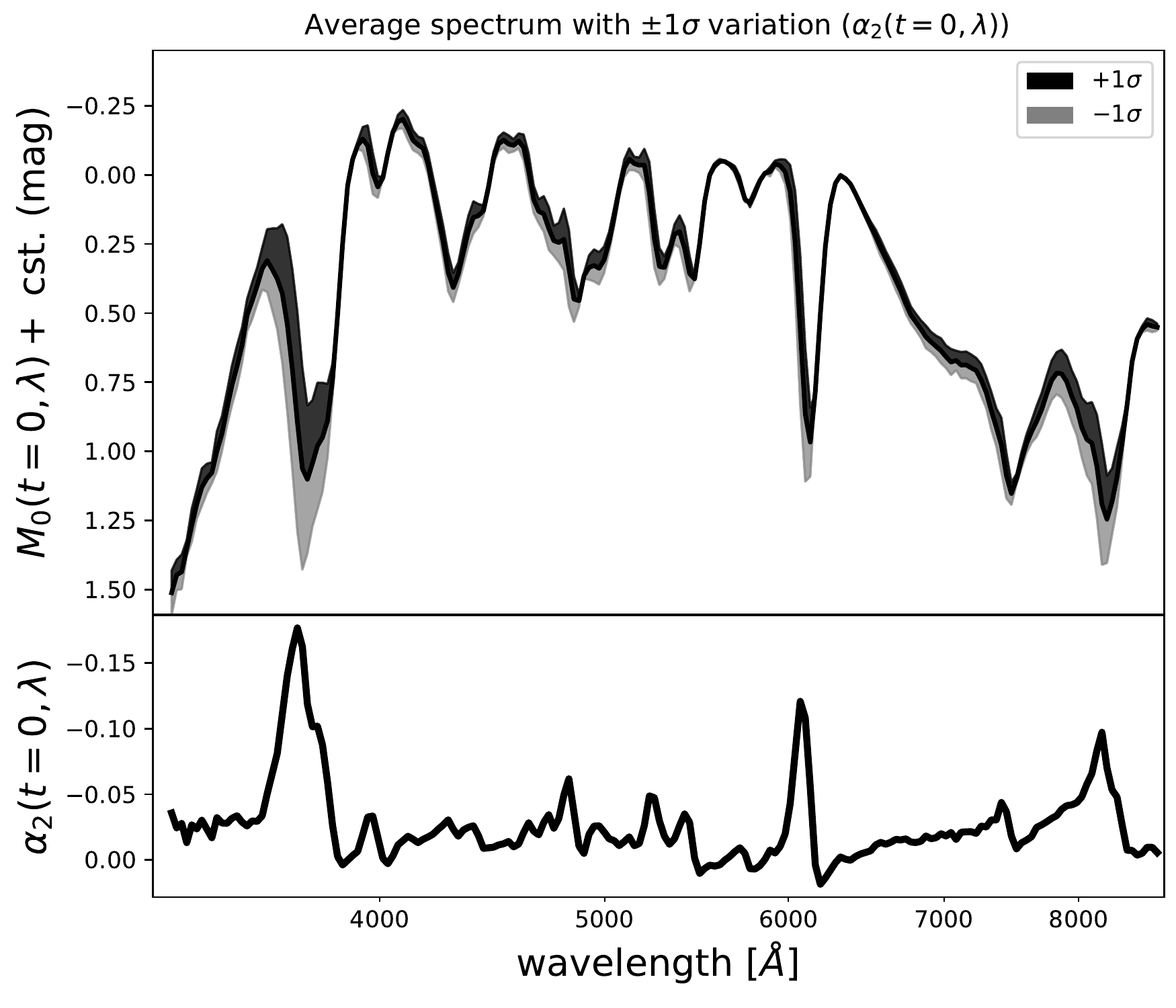}
	\includegraphics[scale=0.4]{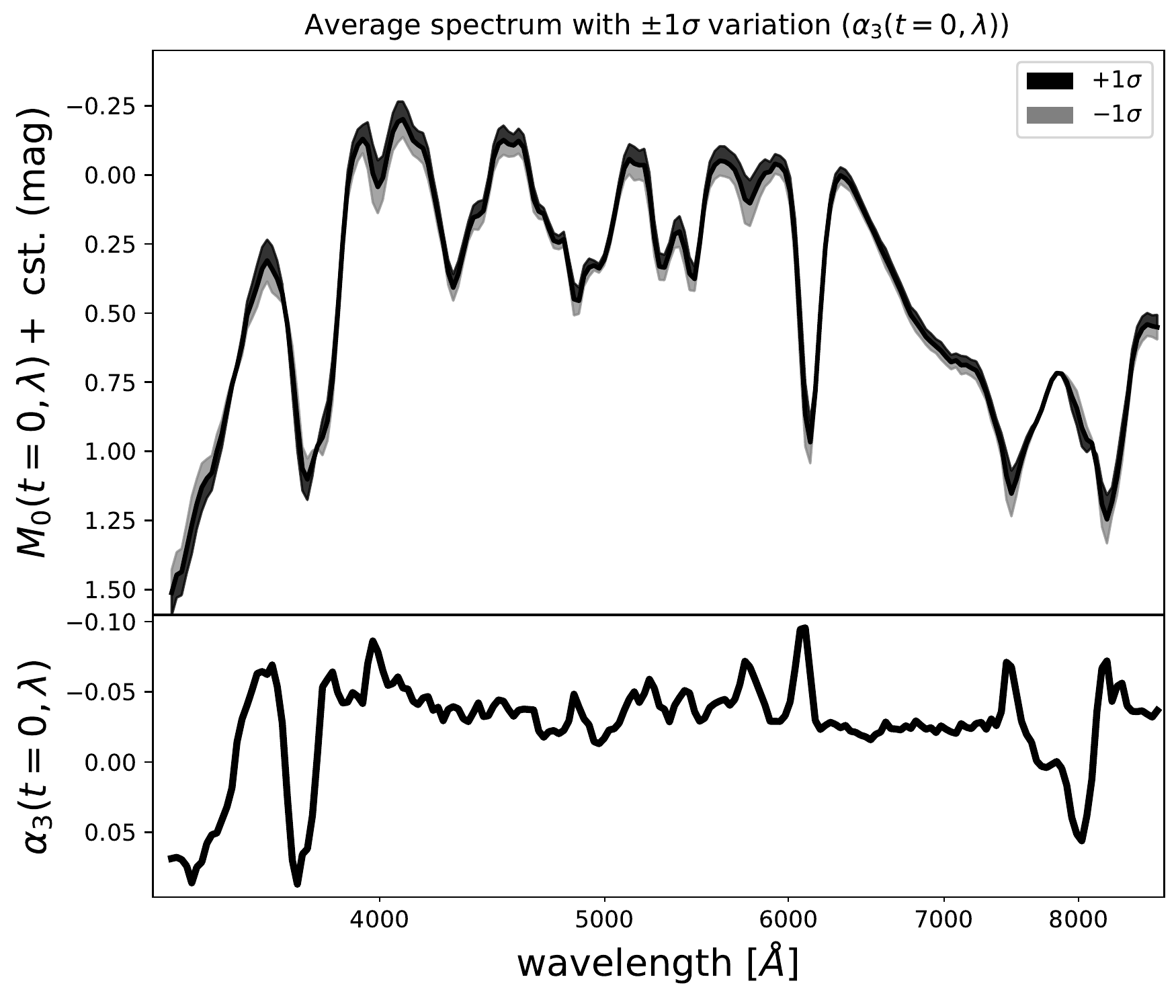} 
	\caption{\small \textbf{Top:} Top panel, average spectrum $\mathbf{M}_{0}$ and the 
	predicted effect of a variation of $q_1$ by $\pm 1 \sigma$. Bottom panel, the corresponding 
	$\mathbf{\alpha}_{1}$ vector. \textbf{Middle}: Top panel, average spectrum $\mathbf{M}_{0}$ and the predicted 
	effect of a variation of $q_2$ by $\pm 1 \sigma$. Bottom panel, the corresponding 
	$\mathbf{\alpha}_{2}$ vector. \textbf{Bottom:} Top panel, average spectrum $\mathbf{M}_{0}$ and the predicted 
	effect of a variation of $q_3$ by $\pm 1 \sigma$. Bottom panel, the corresponding 
	$\mathbf{\alpha}_{3}$ vector.}
	\label{Alpha123}
\end{figure}

The model above is trained on the \NSnTrainingFilterEMFA \sne  that remain in the training sample.
The vectors $\boldsymbol{\alpha}_{j}$ for $j = 1$, $j = 2$ and $j = 3,$ as well as their effect on the average
spectrum $\textbf{M}_{0}$, are presented in~Figure~\ref{Alpha123}. 
To quantify the overall impact of each vector in units of magnitude, we compute the average RMS of $h_j \boldsymbol{\alpha}_j$,
 the deviation from the average spectrum $\textbf{M}_{0}$. \\
\indent The vector $\boldsymbol{\alpha}_{1}$ presented in Figure~\ref{Alpha123}, has an average impact of \AlphaOneImpactMax  
mag when multiplied by the standard deviation of $q_1$. This is the expected amplitude for a stretch effect. 
The structure of this vector is associated with variations
of the line depths, consistent with the correlation of the factor $q_1$ with both the 
pseudo-equivalent widths and the stretch (cf. Figure~\ref{Corr_coeff_caballo} and Figure~\ref{Corr_coeff_caballo_salt2}). \\
\indent The vector  $\boldsymbol{\alpha}_{2}$ shown in Figure~\ref{Alpha123}, which is correlated 
with velocities, has a much weaker impact 
of \AlphaTwoImpactMax mag on the average spectrum $\textbf{M}_{0}$. This is consistent with the fact that this effect has not 
yet been detected on purely photometric data. Moreover, the associated variability is centered in 
localized structures such as the regions of the \CaHK, \Sic, and \CaIR lines. 
A closer scrutiny shows that the variability especially affects the bluer edge of the features, leading to 
an overall effect on the line velocities. This is as expected from the correlations of $q_2$ and the minima of P-Cygni profile
shown in Figure~\ref{Corr_coeff_caballo}. \\
\indent The vector  $\boldsymbol{\alpha}_{3}$ shown in Figures~\ref{Alpha123}, has an impact  
of \AlphaThreeImpactMax~mag on the average spectrum $\textbf{M}_{0}$. 
Like the first component $q_1$, $q_3$ is correlated with the stretch  (cf. Figure~\ref{Corr_coeff_caballo_salt2}), but 
the corresponding vector  $\boldsymbol{\alpha}_{3}$ is less structured than $\boldsymbol{\alpha}_{1}$ in the optical band. 
However, the expected correlations with the spectral features are consistent with the behavior of
$q_3$ (cf. Figure~\ref{Corr_coeff_caballo}). Moreover, the most prominent features
affect the extreme UV part of the spectrum as well as the 
\CaHK and \CaIR regions. For the latter, brighter \sne exhibit a stronger trough in the higher
velocity part of the blends, which could link this vector to the presence of high-velocity calcium structure in 
ejectas. \\
\indent In summary, the analysis of the SED components at maximum light confirms that the stretch 
has the dominant effect on magnitudes as expected, but one also has to take into account other
variabilities which are difficult to detect in photometric bands since they are linked to localized features such as velocities. 

\begin{figure}
	\centering
	\includegraphics[scale=0.3]{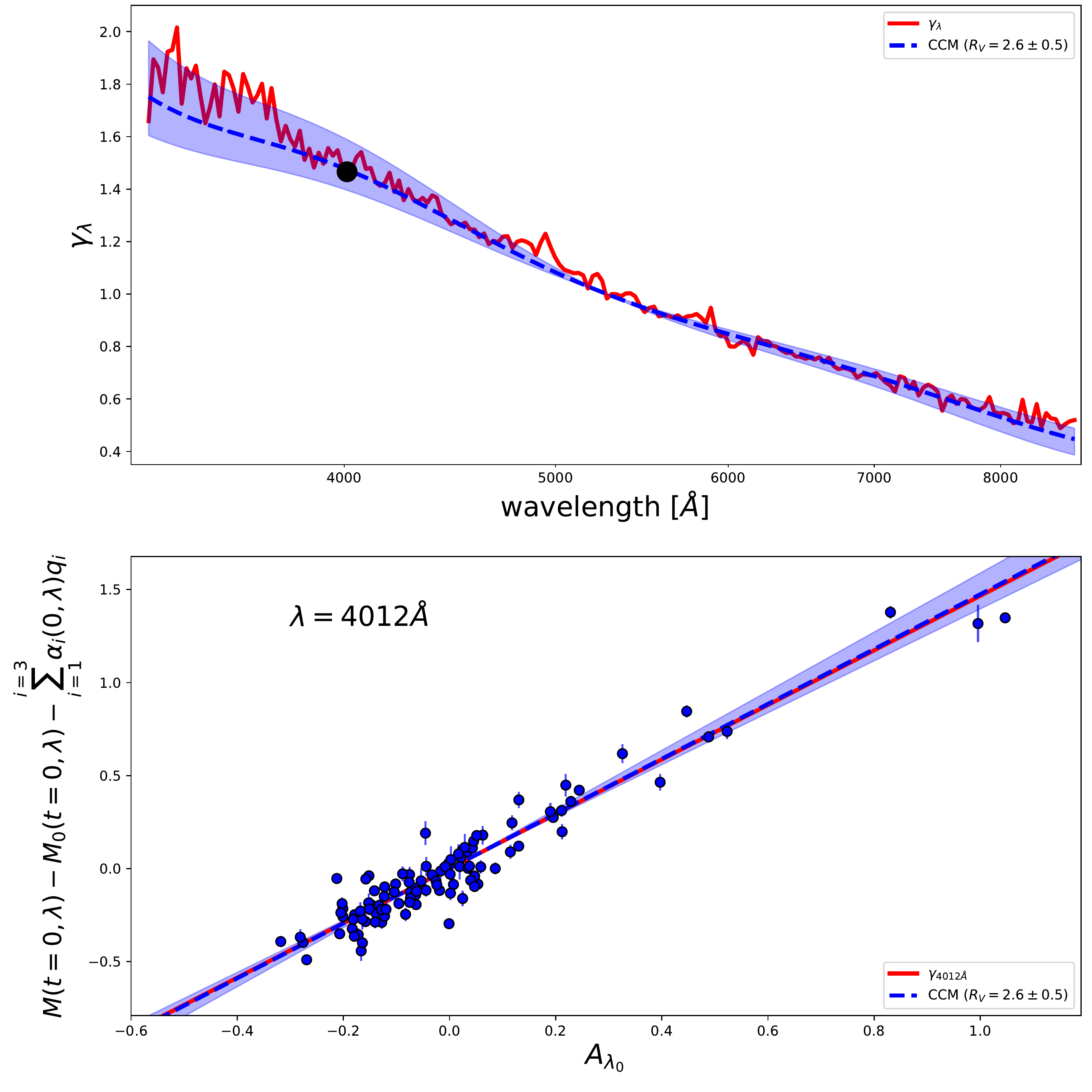}
	\caption{\small \textit{Top panel:} Empirical average extinction curve, $\gamma (\lambda)$,
	 compared to the best fit by a \cite{Cardelli89} law, which gives an 
	 $R_V = \RvValue$. The shaded blue area represent a  $\pm \RvValueError$ range on $R_V$,
	 in order to illustrate the typical range of value that are measured for \snia in the literature.
	 The black dot on this curve indicates the wavelength used on the lower panel graph. 
	 \textit{Lower panel:} Residuals after 
	 correction for the intrinsic behavior at $4012 \AA$ as a function of $ A_{\lambda_0}$. 
	 Each dot corresponds to a single \snia.
	 The $\gamma_{4012 \AA}$ slope between $ M $ and $ A_{\lambda_0} $ is indicated by the 
	 red continuous line. Dashed represent the \cite{Cardelli89} law for $R_V = \RvValue$. The shaded blue area represent a  
	 $\pm \RvValueError$ range on $R_V$, in order to illustrate the typical range of value that are measured for \snia in the literature.}
	\label{CCM_Law_bin42}
\end{figure}

The $\gamma(\lambda)$  curve is shown in Figure~\ref{CCM_Law_bin42} and is consistent  
with a dust extinction curve. 
This shows that three eigenvectors provide a sufficient description of the intrinsic variability from the purpose of deriving a color law.
The fit of  $\gamma(\lambda)$ to a \cite{Cardelli89} law gives a value of \mbox{$R_V = \RvValue$}.
The \cite{Cardelli89} law fit coincides remarkably well
with $\gamma(\lambda)$, except in the UV, where it differs slightly. 
The $R_V$ value is mainly driven by the reddest \sne. Indeed, as 
can be seen in the lower panel of Figure~\ref{CCM_Law_bin42}, three \sne have 
the dominant contribution on the final value of $R_V$ and removing any of these from the sample 
would significantly alter the result, hence the rather large value of the uncertainty. \\
\indent We made the choice of fitting an average extinction curve for the whole sample, but 
there are indications from observations that extinction curves exhibit some diversity \citep{Amanullah15,Kim18}. 
Moreover, any such extinction variation, or intrinsic color variation (e.g. \citealt{Foley11,Polin18}), that the model at maximum is 
unable to capture goes into the dispersion matrix.
However, in order to keep the model simple, we keep an average value of $R_V$ and 
we will leave $R_V$ variation analysis for the future. 

\begin{figure}
	\centering
	\includegraphics[scale=0.29]{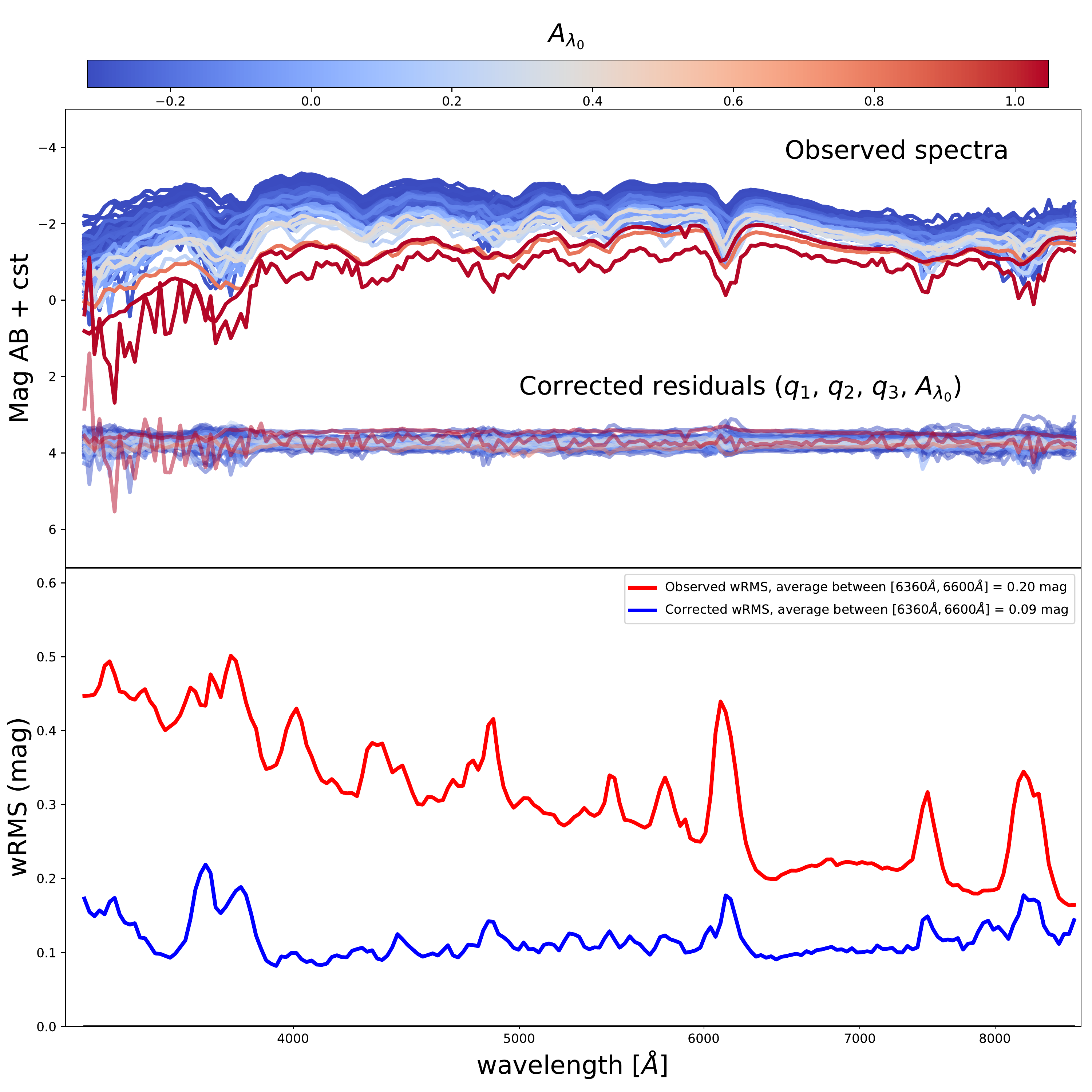}
	\caption{\small \textit{Top panel:} the \NSnTrainingFilterEMFA spectra in the absolute AB magnitude
	before correction (upper set of spectra) and after correction for intrinsic properties and color (lower set of spectra). 
	An arbitrary offset has been applied for each set. The color code indicates the value of $A_{\lambda_0}$.
	\textit{Lower panel:} the wRMS of the residuals for the uncorrected (in red) and 
	corrected (in blue) spectra.
	 }
	\label{3_vec_corrected}
\end{figure}

\begin{figure}
	\centering
	\includegraphics[scale=0.29]{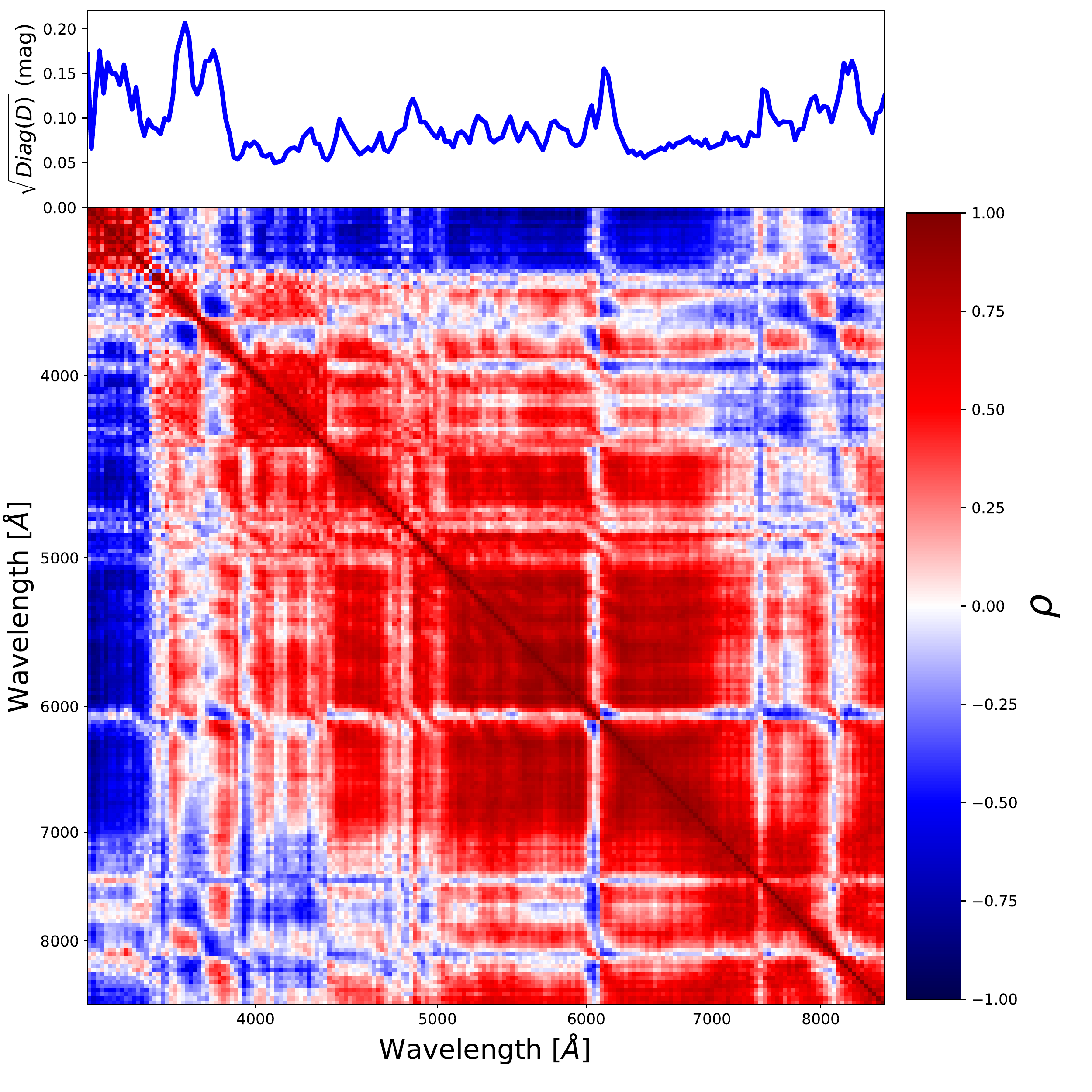}
	\caption{\small The square root of the diagonal of the dispersion matrix $\textbf{D}$ (top) and the correlation matrix corresponding to the 
	dispersion matrix $\textbf{D}$ (bottom). The shading represents the degree of correlation, as given by the Pearson correlation coefficient, $\rho$.}
	\label{corr_matrix}
\end{figure}

Figure~\ref{3_vec_corrected} presents the set of spectra before any correction 
and the residuals resulting from the difference between the observed spectrum and Equation~\ref{Model_SED_at_max}.
In the top panel, the effect of extinction is clearly visible for 
spectra before any correction. After correction, the residuals between the models and data
are mainly shifted in magnitude with 
respect to one another and no effect on color appears to remain. This is the signature expected for a mostly grey residual offset. \\
\indent The amplitude of the correction made by the three factors $q_1$, $q_2$, $q_3$, and the extinction curve,
given by the wRMS as a function of wavelength, is presented in the lower panel of 
Figure~\ref{3_vec_corrected}. Before any correction, the dispersion in the B-band is around 0.4 mag, as expected.
Spectra before any correction exhibit both localized structure coming from intrinsic variabilities and a slowly increasing 
dispersion in the bluer part of the spectrum, which is the signature of extinction. Once spectra are corrected by 
the three factors $q_1$, $q_2$, $q_3$, and the extinction curve, the wRMS of the residuals 
drops to a floor of 0.1 magnitude. However, localized structures remain, concentrated in UV,  \CaHK, \Sic, \OI and 
\CaIR regions. The amplitude of the baseline is compatible with the value of the grey intrinsic dispersion 
added in cosmological fits for nearby supernovae \citep{Betoule14}. The fact that a part of the residual is grey is also visible in the dispersion 
matrix as can be seen in Figure~\ref{corr_matrix}. Indeed, the high correlation across all wavelengths is consistent with grey fluctuation. 
However, some effects observed in the dispersion matrix are not due to the grey effect, but to the unmodeled variabilities not captured by SUGAR in certain spectral zones, which explains some features 
observed in the dispersion matrix and in the Figure~\ref{3_vec_corrected}.

\subsection{Time interpolation using Gaussian Process}
\label{gp_interp}

Now, we want to extend our model from maximum light to the full spectral time series. 
Spectra from the \snf are taken at different epochs in
the development of each \snia. 
The treatment of these observations taken at different times is an 
important step to handle before constructing the full SED model. 
In order to work with a fixed time grid, it is necessary to use a method of interpolation. One solution is to use Gaussian Processes, 
as was done for \sne in works like \cite{Kim13}, \cite{Boone15} and  \cite{Saunders18}. A complete review of Gaussian Processes
can be found in \cite{Rasmussen06}. The Gaussian Process implementation presented here is based
on this review. However, we have developed a specific 
implementation in order to take into account wavelength dependencies and to
accelerate computation of the interpolation. These are described below. \\
\indent A Gaussian Process is a generalization of the Gaussian probability distribution, and is a 
non-parametric way to interpolate data. The main assumption of Gaussian Processes is 
that observed data are the realization of Gaussian random fields that are characterized by an average function 
and a correlation function. Therefore, the distribution of \sne magnitude $m$ at a given phase $t$ and a 
given wavelength $\lambda$ follows this relation:

\begin{equation}
m\left(t,\lambda \right) \sim {\cal N}\left(m_0\left(t,\lambda \right),\textbf{K}_{\lambda}\right)\eqcoma
\end{equation}

\noindent where $m_0\left(t,\lambda \right)$ is the average function of the \sne SED, and
$K_{\lambda}$ corresponds to the correlation matrix (commonly called the kernel) that describes 
time correlations between different epochs. In the absence of an explosion model that would yield
the analytical form of $K_{\lambda}$, we choose a squared exponential kernel, 
to which the measurements and calibration uncertainties are added:

\begin{equation}
K_{\lambda} \left(t_i,t_j\right)=\phi_{\lambda}^2 \ \exp\left[-\frac{1}{2}
 \left(\frac{t_i-t_j}{l_\lambda}\right)^2\right] + \sigma_{grey}^2 \ \delta_{ij}'\eqcoma 
\end{equation}

\noindent where $\phi_{\lambda}^2$ represent the spectral variance around 
the average function for the wavelength $\lambda$, $l_{\lambda}$ corresponds
to the temporal correlation length for the wavelength $\lambda$, $t_i$ and $t_j$ correspond 
to the phase of observation $i$ $j$, and $\sigma_{grey}$ is the grey 
error taken to be 0.03 mag.  All wavelengths are treated independently of each other, i.e. the 
interpolation is performed on $\textbf{y}_{\lambda n}$, the $n^{\text{th}}$ \snia light curve for each wavelength $\lambda$
from which the average function has been subtracted. For a given set of global hyperparameters 
($\phi_{\lambda}$,  $l_{\lambda}$), the interpolation $\textbf{y}_{\lambda n}'$ and the covariance 
matrix of uncertainties on the interpolation, $\text{cov}\left(\textbf{y}_{\lambda n}'\right)$, are given by :

\begin{equation}
\textbf{y}_{\lambda n}' = \textbf{K}_{\lambda n}(\textbf{t}',\textbf{t})^T\left(\textbf{K}_{\lambda n}(\textbf{t},\textbf{t})+ \boldsymbol{\sigma}_{\lambda}^2 \textbf{\textit{I}}\right)^{-1}
\textbf{y}_{\lambda n}
\end{equation}
\begin{multline}
\text{cov}\left(\textbf{y}_{\lambda n}'\right) = \textbf{K}_{\lambda n}(\textbf{t}',\textbf{t}') - \\
\textbf{K}_{\lambda n}(\textbf{t}',\textbf{t})^T\left(\textbf{K}_{\lambda n}(\textbf{t},\textbf{t})+ \boldsymbol{\sigma}_{\lambda}^2 \textbf{\textit{I}}\right)^{-1}\textbf{K}_{\lambda n}(\textbf{t}',\textbf{t}) \eqcoma
\end{multline}

\noindent where the vector $\boldsymbol{\sigma}_{\lambda}^2$ is the error on the magnitude for 
the wavelength $\lambda$ and for all observed phases.  $\textbf{t}'$ is the new time grid and \textbf{t} 
the observed phases of the \snia.
Gaussian Processes assume that the observed data are distributed around an average function. 
This function is a necessary input, because in a region with no data points,
the Gaussian Process interpolation converges to this average function (after $\sim 3$ temporal correlation lengths).
The calculation of this average function is performed as follows. 
The training spectra are binned according to phase from maximum light in B-band, ranging from 
\sugarMinPhase to \sugarMaxPhase days. Each phase bin spans two days and contains at least 3 spectra. 
The weighted average spectrum is computed for each phase bin and given as a first approximation of the average function. 
However, due to some lower signal-to-noise spectra, this version of the average function is not smooth. A Savitsky-Gollay filter is used 
to smooth the average function, which is then used for the Gaussian Process interpolation.\\
\indent What remains is to determine the pair of hyperparameters $\theta_{\lambda}=(\phi_{\lambda}, l_{\lambda})$ 
 that represent, respectively, the amplitude of the fluctuation around the average function and the temporal correlation length. 
To estimate the hyperparameters for each wavelength, we maximize the product of all individual likelihoods:

\begin{multline}
{\cal L}_{\lambda}=  \prod_{n} \frac{1}{2\pi^{\frac{N}{2}}} \ 
\frac{1}{|\textbf{K}_{\lambda n}(\textbf{t},\textbf{t})+ \boldsymbol{\sigma}_{\lambda}^2 \textbf{\textit{I}}| ^{\frac{1}{2}}} \times  \\ 
\exp\left(-\frac{1}{2} \textbf{y}_{\lambda n}^T\left(\textbf{K}_{\lambda n}(\textbf{t},\textbf{t})+ \boldsymbol{\sigma}_{\lambda}^2 \textbf{\textit{I}}\right)^{-1} \textbf{y}_{\lambda n} \right) \eqdot
 \label{max_likelihood_GP_hyper}
 \end{multline}

\begin{figure}
	\centering
	\includegraphics[scale=0.27]{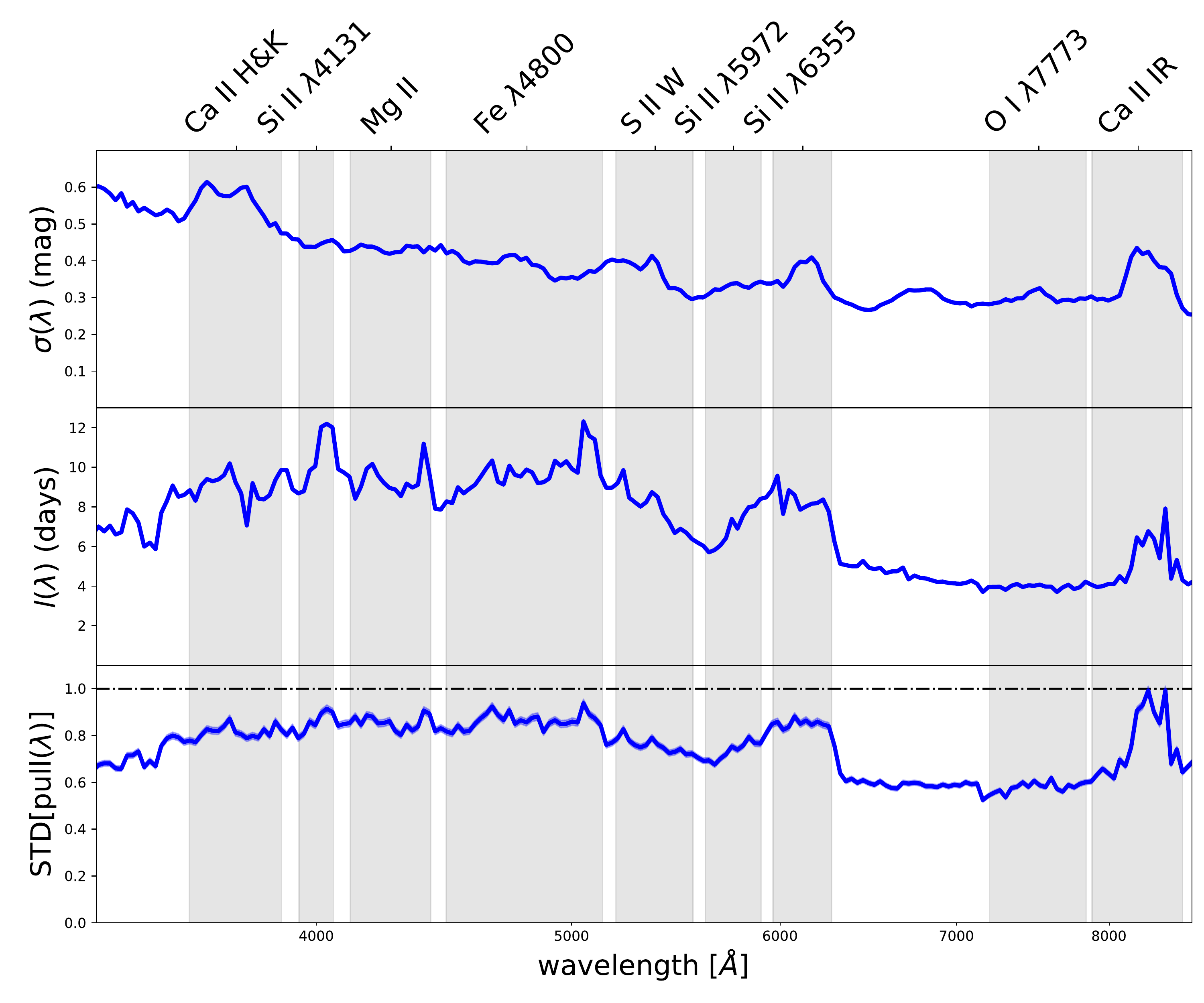}
	\caption{\small Results of the Gaussian Process interpolation. From the top to 
	the bottom are represented the amplitude of the kernel, the correlation length and the standard 
	deviation of the pull distribution in terms of wavelength. The grey areas represent 
	the presence of the absorption line that were used in our original factor analysis.}
	\label{gaussian_processes_adjustement}
\end{figure}

This differs from previous work for two main reasons: the wavelength dependence of the hyperparameters, 
and their estimation using all the \sne. These two new features are 
justified mainly by \sne physics and computation time. Since \sne are standardizable candles, we treat them as 
realizations of the same Gaussian random field, i.e. we assume that they share the same set 
of hyperparameters. In this case, we can use all available \sne to estimate these hyperparameters. The wavelength 
dependence of the hyperparameters is due to e.g. dust extinction, intrinsic variability, 
the second maximum in infrared, etc. It follows that the standard deviation around the average function
and the temporal correlation length should vary across wavelength. Moreover, the wavelength dependence allows us to be 
very efficient in terms 
of computation time. Indeed, it sped up matrix inversion from the classical 
{$\mathcal{O}(N_{\lambda}^3 \times N_{t}^3)$} to {$\mathcal{O}(N_{\lambda} \times N_{t}^3)$}, 
where $N_{\lambda}$ is the number of bin in wavelength (\numberbinsugar here), 
and $N_{t}$ the number of observed phases ($\sim$ 14 in average). On a Mac Book Pro with 
a 2.9 GHz Intel Core i5 processor and 8 GB of RAM, 
finding all hyperparameters ($2 \times 197$) and computing the interpolation and the pull distribution of residuals 
took under 6~minutes. \\
\indent Results of hyperparameter adjustment in terms of wavelength 
are shown in Figure~\ref{gaussian_processes_adjustement}. 
First, it can be seen that the structure of hyperparameters as a function of wavelength is not
random. Indeed, for the parameter $\phi_{\lambda}$, its value varies from 
0.3 magnitude to more than 0.6 magnitude: this number is difficult to interpret directly
because it is the amplitude around an average value at all phases of the data. 
Nevertheless, the increase in the ultraviolet and blue wavelengths can be explained by the 
dispersion caused by dust extinction of the host-galaxies 
and a larger intrinsic dispersion in this wavelength range \citep{Maguire12,Nordin18}. 
There are also structures in the peaks of this 
intrinsic dispersion that correspond to the regions of \CaHK, \Sic and \CaIR lines.  

\indent The correlation length, $l_{\lambda}$, has structure in the silicon and calcium regions and  
remains globally stable with no features between $6000\AA$ 
and $8000\AA$ (average value of 7 days). \\
\indent Once the hyperparameters have been estimated, the interpolations for each \snia are performed onto a 
new phase grid. The phase grid chosen covers a range
between \sugarMinPhase and \sugarMaxPhase days around maximum brightness in B-band, divided into 3 day bins. \\
\indent Finally, to check if both the predictions (interpolation) and associated errors are in agreement with what is
expected from a Gaussian Process, we compute the pull distribution of the residuals. Because we assume 
that our data are realizations of a Gaussian random fields, 
the pull distribution should follow a centered unit normal distribution.
For each wavelength we fit a Gaussian to the pull distribution and calculate its standard deviation.
The results are shown in the bottom of Fig.~\ref{gaussian_processes_adjustement}.
The pull standard deviation is on average lower than 1 with an average value of 0.8. \\

\indent These interpolation results will be used to establish the SED model that is described in the
following sections.

\section{The \sugar model}
\label{sugarmodeltrainingsection}

\subsection{The model}
\label{presentation_sugar_model}

The \sugar model assumes that 
SED variation at any epoch may be described by spectral features measured at maximum. The \sugar SED model is based 
on a linear combination of factors derived in Section~\ref{emfa_results} and an extinction curve taken 
as a \cite{Cardelli89} law derived in Section~\ref{extinction_law_section}.
To model the \sugar SED, we propose a model similar to the one we constructed
for maximum light. This \sugar model is based on the three factors 
for each \snia $i$, combined into the vector \mbox{$\textbf{x}_i \equiv \{h_{i,1}, h_{i,2}, h_{i,3}\}$}, which in the model is the true 
value of the measured factor $\textbf{q}_i = \{q_{i,1}, q_{i,2}, q_{i,3}\}$, and are related to the intrinsic magnitude by:

\begin{equation}
\label{Model_SED_SUGAR_equation}
     M_i(t,\lambda)=M_0(t,\lambda)+  \ \sum_{j=1}^{j=3} h_{i,j} \alpha_j(t,\lambda) 
      + A_{V,i} \ f\left(\lambda,R_V\right) +\Delta M_{grey \ i} \eqcoma
\end{equation}

\noindent where $M_0(t,\lambda)$ is the magnitude of the average spectral 
time series, $\alpha_j(t,\lambda)$ is the intrinsic variation related 
to the factor $h_{i,j}$; as discussed in Section~\ref{emfa_results}, the numbers of factors is set to three. As at maximum, 
$h_{i,j}$ represents the true value of the measured factor $q_{i,j}$ derived in Section~\ref{emfa_section}.
$A_{V,i}$ is the extinction in the V-band, and $f\left(\lambda, R_V\right)$ 
is the \cite{Cardelli89} law where $R_V$ is the extinction ratio that is derived in 
Section~\ref{extinction_law_section}. Finally, the term $\Delta M_{grey \ i}$ is a 
grey and time independent term comparable to the parameter $X_0$ of SALT2, which  
allows us to work independently of the distance.
Indeed, the effect of the distance 
on magnitude is equivalent to an additive constant independent of wavelength. 
Moreover, the parameter 
$\Delta M_{grey \ i}$ also contains the average of the spectral time series residuals with the \sugar model. 
The addition of $\Delta M_{grey \ i}$ differs from what was done 
when determining the extinction curve in Section~\ref{extinction_law_section}, and this choice is discussed in Section~\ref{discussion_grey_disp_matrix}.
We model the SED at the same wavelengths and epochs where the Gaussian Process 
interpolation is done. For notational efficiency we rewrite Equation~\ref{Model_SED_SUGAR_equation} as:

\begin{eqnarray}
     \textbf{M}_i &=& \textbf{A} \textbf{h}_{i}  \eqcoma \textrm{where} \\
\textbf{A}&=&\left(\textbf{M}_{0}, \textbf{f}\left(R_V\right), 1,\boldsymbol{\alpha}_{1},\boldsymbol{\alpha}_{2},\boldsymbol{\alpha}_{3} \right) \eqcoma \\
\textbf{h}_i^T&=&\left(1,A_{V,i}, \Delta M_{grey \ i},h_{i,1},h_{i,2}, h_{i,3}\right) \ .
\end{eqnarray}

\indent The unknown parameters of the \sugar model are 
therefore the average spectrum $M_0(t,\lambda)$, the intrinsic coefficients 
$\alpha_j(t,\lambda)$, and the grey offset $\Delta M_{grey \ i}$. Each \snia  
is parametrized during the training by the $q_{i,j}$ determined in Section~\ref{emfa_results} and $A_{V,i}$ derived in 
Section~\ref{extinction_law_section}. The average spectral time series $M_0(t,\lambda)$, the intrinsic coefficients 
$\alpha_j(t,\lambda)$, and the grey offset $\Delta M_{grey \ i}$ are determined 
in the same way as the modeled described at maximum light in the 
Section~\ref{extinction_law_section}. The main differences with Section~\ref{extinction_law_section} 
are that the model now depends explicitly on the phases of observation, the extinction curve is fixed at the maximum light value (same $R_V$),
and the model is trained on the Gaussian process interpolation of spectra derived in Section~\ref{gp_interp} 
instead of measured spectra. The procedure is described below.

\subsection{Fitting the SUGAR model}
\label{sugar_fitting_model}

We use an orthogonal distance regression method to estimate the parameters of the \sugar model. This is a 
version of the method described in Section~\ref{extinction_law_section}, modified to include the Gaussian Process 
interpolation, the known extinction parameters, and the grey offset. To estimate the \sugar parameters, we minimize the 
following $\chi^2$:

\begin{multline}
     \label{chi2_sugar_0}
     \chi^2 = \sum_i  \left( \textbf{M}_i^{gp} - \textbf{A} \textbf{h}_{i} \right) ^T 
     \textbf{W}_{\textbf{M}_i}  \left( \textbf{M}_i^{gp} - \textbf{A} \textbf{h}_{i} \right)
     +  \\ 
     \left( \textbf{q}_i - \textbf{x}_{i} \right)^T \textbf{W}_{\textbf{q}_i}
      \left( \textbf{q}_i - \textbf{x}_{i} \right) \eqcoma
\end{multline}

\noindent where the vector $\mathbf{M}_i^{gp}$ are the Gaussian 
Process interpolations of the observed spectra (Section~\ref{gp_interp}) which is in the following vector format:

\begin{equation}
\textbf{M}_{i}^{gp} =\left(\textbf{y}_{3340\AA \ i}'   \cdots \ \textbf{y}_{\lambda \ i}'  \cdots \ 
\textbf{y}_{8580\AA \ i}' \right)\eqcoma
\end{equation}

\noindent where $ \textbf {y}_{\lambda \ i} '$ is the light curve at wavelength $\lambda$
interpolated by Gaussian Processes onto new time phase as described in Section~\ref{gp_interp}. 
As in Section~\ref{extinction_law_section}, the uncertainties affecting the $\mathbf{x}_i$ are propagated 
through the weight matrix $\mathbf{W}_{\mathbf{x}_i}$. The $\mathbf{h}_i$ plays a role analogous to 
the $\mathbf{q}_i$ of Section~\ref{emfa_section} and are estimated like the other free parameters
of the model. Finally, the matrix $\textbf{W}_{\textbf{M}_i}$ is the weight matrix of 
the \snia $i$ and follows this matrix format: 

\begin{equation}
\label{Big_matrice_for_SUGAR_from_GP}
    \textbf{W}_{\textbf{M}_i}= \begin{pmatrix}
    \text{cov}\left(\textbf{y}_{3340\AA \ i}'\right) & &  & &0\\ 
       &\ddots& & &\\ 
            &&\text{cov}\left(\textbf{y}_{\lambda \ i}'\right) & &\\ 
             && & \ddots&\\ 
    0  &  &  &&\text{cov}\left(\textbf{y}_{8580\AA \ i}'\right)
\end{pmatrix}^{-1}\eqcoma
\end{equation}

\noindent where $\text{cov}\left(\textbf{y}_{\lambda \ i} '\right)$ is the covariance of interpolation
uncertainties from the Gaussian Process interpolations determined in Section~\ref{gp_interp}. The 
interpolation is done only in terms of the phase relative to B-band maximum, 
this makes the weight matrix block diagonal for each \snia. In this analysis,  
we do not add any covariance in terms of wavelength coming from the calibration uncertainties, for reasons of computation 
time, and because they are expected to be small. These block diagonal matrices allowed us to accelerate the minimization algorithm. 
To estimate the 16653 free parameters of the \sugar model, it took just less than 1 hour on 
a Mac Book Pro with a 2.9 GHz Intel Core i5 processor and 8 GB of RAM, thanks to sparse linear algebra.
Note that while the covariance terms were neglected, the uncertainty at each wavelength was 
taken into account during the training of the Gaussian Processes (these are the blocks 
$\text{cov} \left (\textbf{y}_{\lambda \ i} '\right)$). The unknown parameters of the model (the matrix  $\textbf{A}$ and 
vectors $\textbf{h}_i$) are computed by minimizing equation~\ref{chi2_sugar_0} using an expectation and 
minimization algorithm.  This algorithm, described in Appendix~\ref{app:ODR}, avoids degeneracies between parameters.

\subsection{The \sugar components}
\label{result_sugar}

\begin{figure}[!ht]
	\centering
	\includegraphics[scale=0.35]{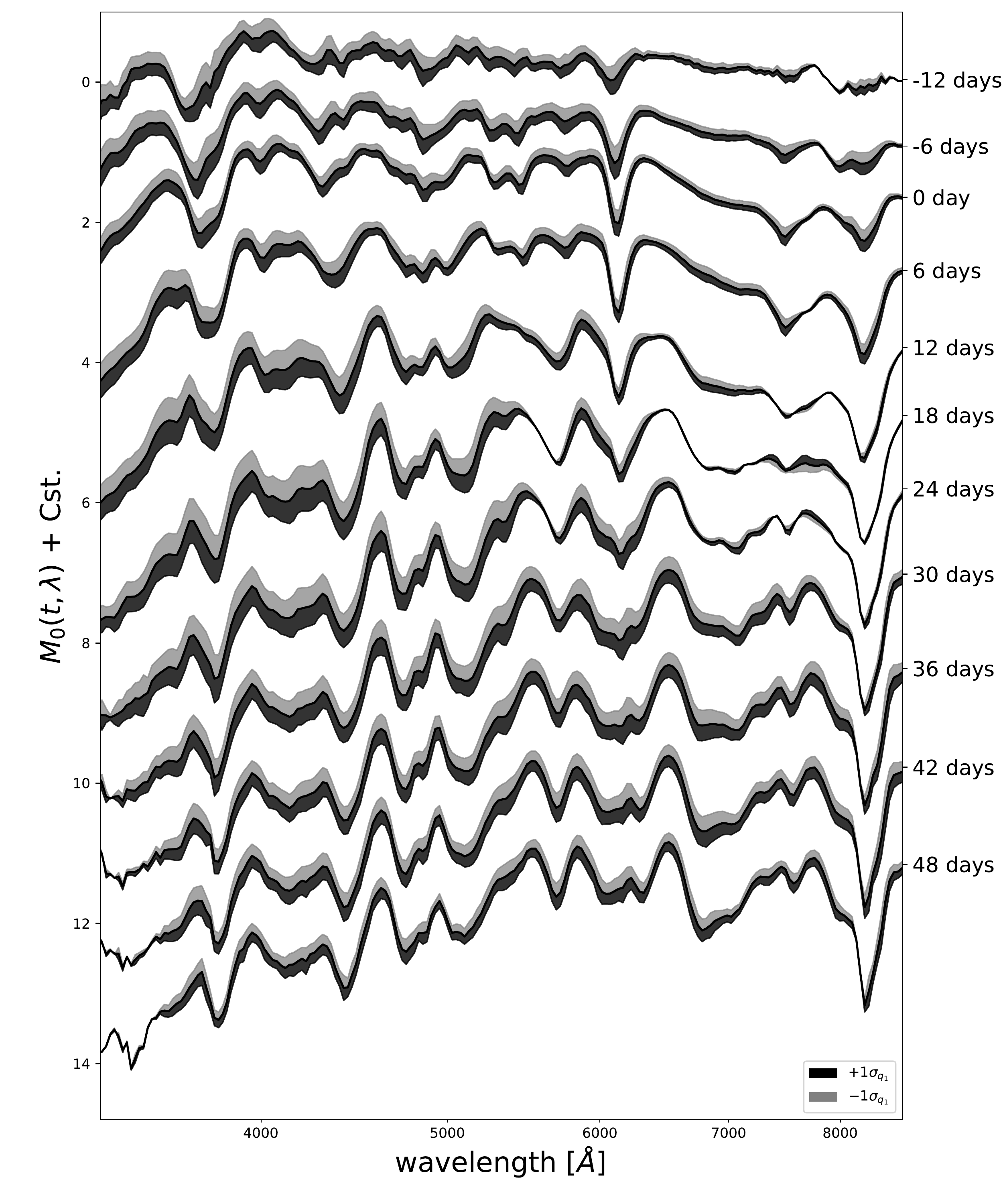}
	\caption{\small 
	Average spectral time series and the effect of a variation of $q_1$
	by $\pm 1 \sigma$. Each phase is separated by a constant magnitude offset.
	 }
	\label{alpha1_TIME}
\end{figure}

\begin{figure}
	\centering
	\includegraphics[scale=0.35]{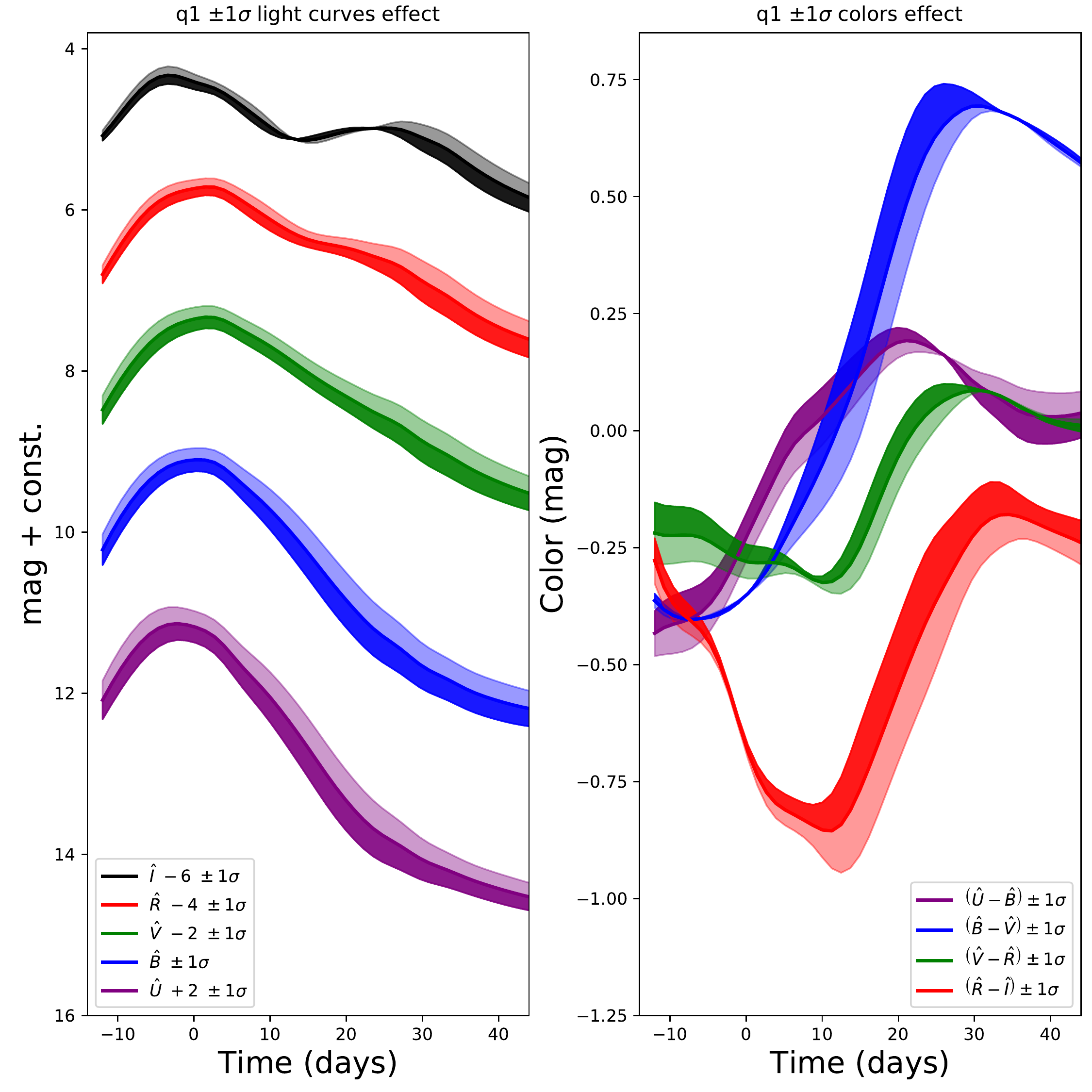}
	\caption{\small 
	Average light curves and color evolution with phase in synthetic bands, and the predicted effect of a variation of $q_1$ by $\pm 1 \sigma$.}
	\label{alpha1_LC}
\end{figure}

\begin{figure}
	\centering
	\includegraphics[scale=0.35]{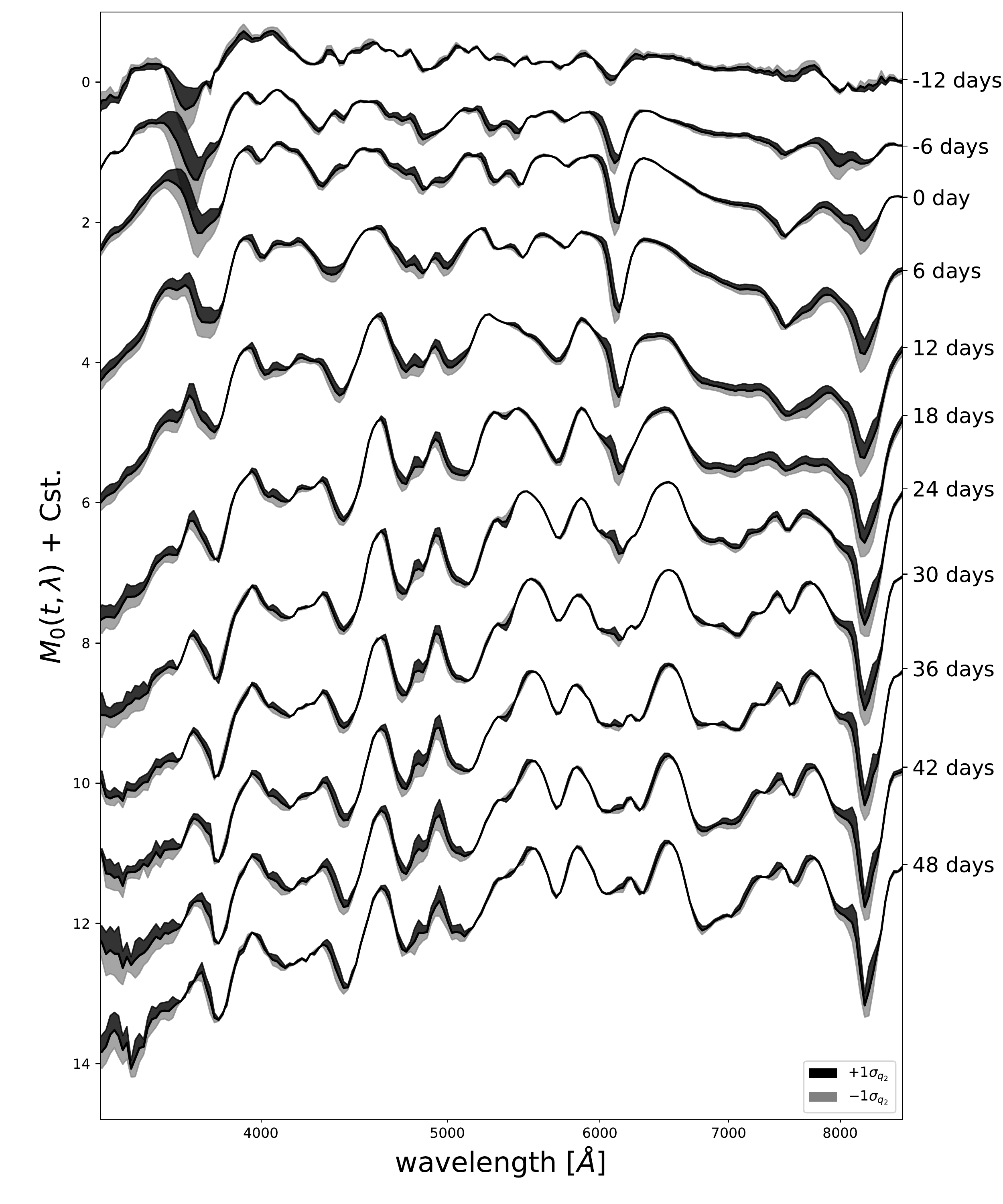}
	\caption{\small 
	Average spectral time series and the effect of a variation of $q_2$
	by $\pm 1 \sigma$. Each phase is separated by a constant magnitude offset.}
	\label{alpha2_TIME}
\end{figure}

\begin{figure}
	\centering
	\includegraphics[scale=0.35]{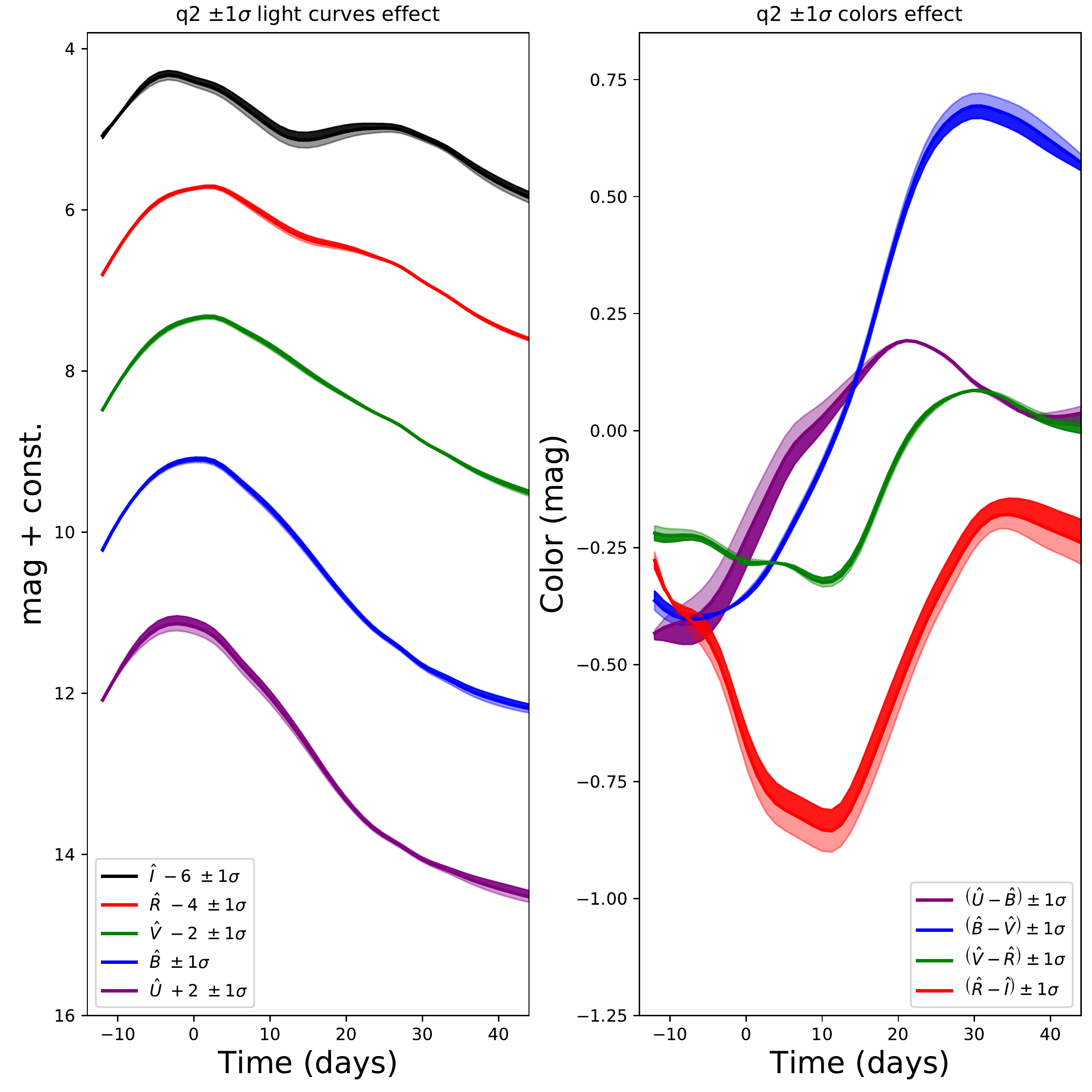}
	\caption{\small 
		Average light curves and color evolution with phase in synthetic bands, and the predicted effect of a variation of $q_2$ by $\pm 1 \sigma$.}
	\label{alpha2_LC}
\end{figure}

\begin{figure}
	\centering
	\includegraphics[scale=0.35]{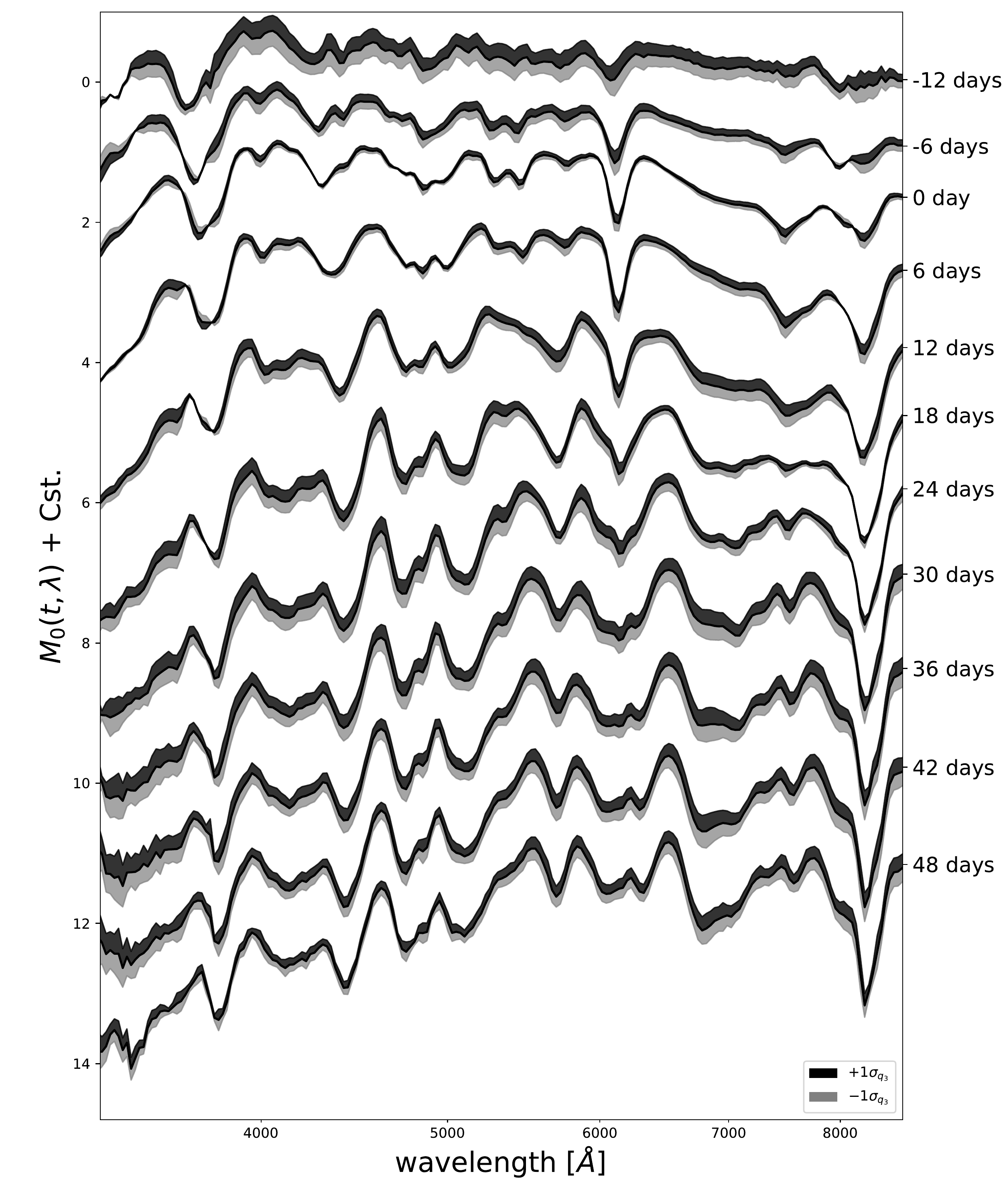}
	\caption{\small 
	Average spectral time series and the effect of a variation of $q_3$
	by $\pm 1 \sigma$. Each phase is separated by a constant magnitude offset.}
	\label{alpha3_TIME}
\end{figure}

\begin{figure}
	\centering
	\includegraphics[scale=0.35]{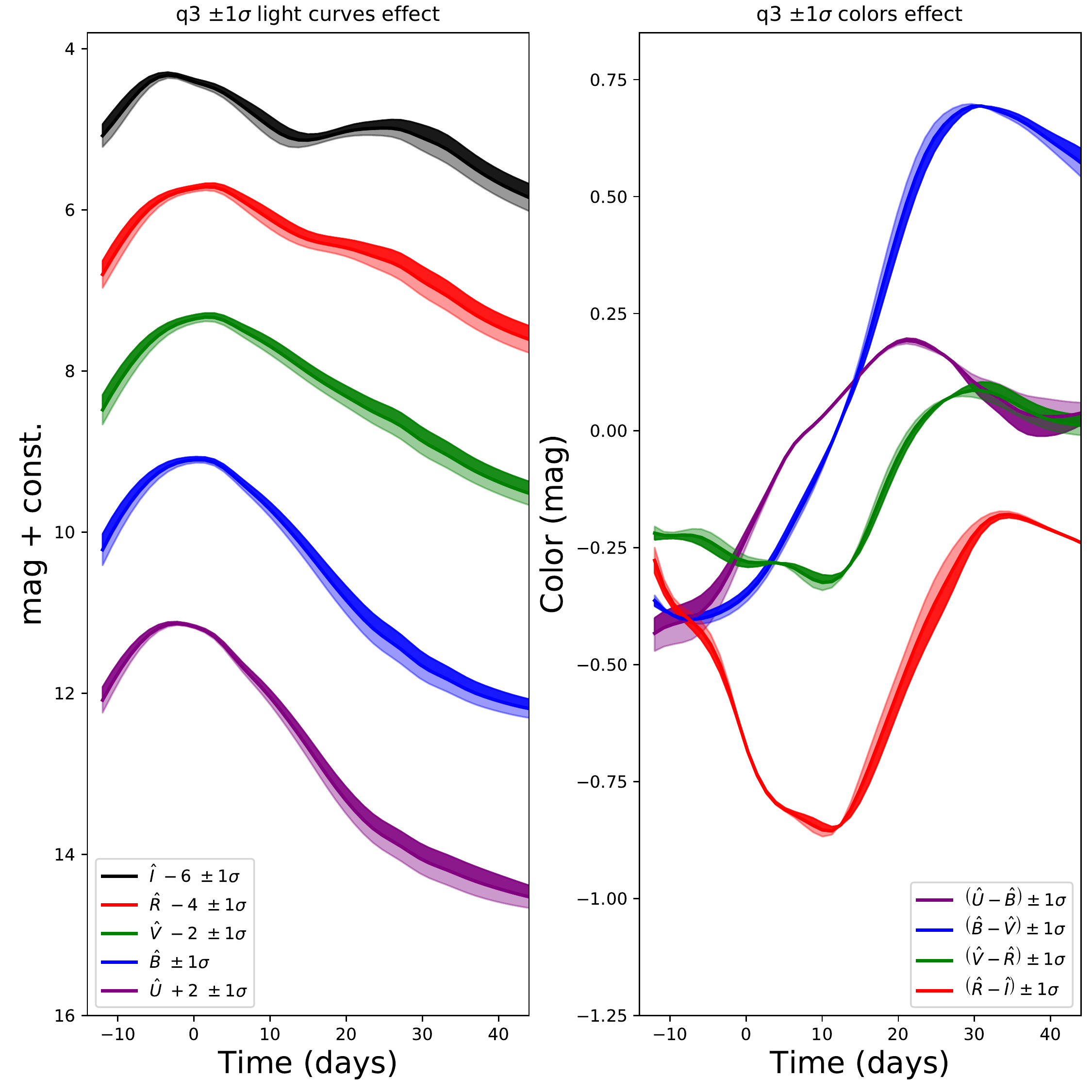}
	\caption{\small 
		Average light curves and color evolution with phase in synthetic bands, and the predicted effect of a variation of $q_3$ by $\pm 1 \sigma$.}
	\label{alpha3_LC}
\end{figure}

The \sugar model is described by a set of spectral time series components $(M_0, \alpha_1,\alpha_2,\alpha_3)$ at
3 day intervals. The SED of a given \snia is then obtained by a linear combination of those components employing the 
$(q_1,q_2,q_3)$ factors describing the supernova. For illustrative purposes, we present a spectral time series in 6 
day intervals as well as their integration in synthetic top-hat filter system comprised of five bands 
with the following wavelength ranges: $\hat{U}\ [3300.0 - 3978.0]\ \AA ; \hat{B}\ [3978.0 - 4795.3]\ \AA ; 
\hat{V}\ [4795.3 - 5780.6 ]\ \AA ; \hat{R}\  [5780.6 - 6968.3]\ \AA ; \hat{I}\ [6968.3 - 8400.0 ]\ \AA$. The diacritic hat serves as a reminder that these are not standard Johnson-Cousins filters. 
The effect of the vectors $\mathbf{\alpha}_{j}$ on the average 
spectrum $M_0$, on the average light curves (obtained by integration of $M_0$ in the $\hat{U}$, $\hat{B}$, $
\hat{V}$, $\hat{R}$, $\hat{I}$ bands) and on the average colors $\hat{U}-\hat{B}$, $\hat{B}-\hat{V}$, $\hat{V}-
\hat{R}$, $\hat{R}-\hat{I}$ is presented in Figures~\ref{alpha1_TIME} and~\ref{alpha1_LC} for $\mathbf{\alpha}
_{1}$, Figures~\ref{alpha2_TIME} and~\ref{alpha2_LC} for $\mathbf{\alpha}_{2}$, and Figures~\ref{alpha3_TIME} and~
\ref{alpha3_LC} for $\mathbf{\alpha}_{3}$. The upper and lower contours correspond to a variation of $\pm 1 \sigma$ of 
the associated $q_j$ parameter. The $\alpha_j(t=0)$ components obtained when training the color law with only data at 
maximum or with the full time series are almost identical: all conclusions from Section~\ref{sec:result_at_max} 
remain valid and we focus here on the temporal behavior.

Of the three factors, $q_1$ has the strongest impact: this can be seen from the time series presented in Figure~\ref{alpha1_TIME} 
and on the broad band light curves presented in Figure~\ref{alpha1_LC}. This is associated with a stretch 
effect, visible in $\hat{U}$, $\hat{B}$, and $\hat{V}$ bands as an enlargement of the variation band as 
we move away from maximum. Between $-12$ days and $+30$ days, the $q_1$ factor has a strong influence on spectroscopic 
details. The brightest \sne exhibit shallower troughs in their spectral features: this is especially visible 
for \Sia, \Sib and \CaIR, but also valid for most of the lines. This effect is more pronounced for 
\sne at earlier phases, and fades over time. After +30 days, the factor 
$q_1$ is less sensitive to localized features, and shows a relative enhancement of the optical $\hat{B}$ to $\hat{R}$ 
bands with respect to $\hat{U}$ and $\hat{I}$. Interestingly, between +12 days and +24 days, the brightest \sne 
are bluer than at maximum or at later phases. This is true in absolute value, but also relative to the average 
\snia. This shows that the \sne color space is driven by a variable intrinsic component in addition to reddening by dust.
This intrinsic color variation is linked with the position of the second peak in $\hat{I}$, 
which appears later for brighter \sne. \\
\indent As at maximum, the effect of the $q_2$ factor for the whole time series (Figure~
\ref{alpha2_TIME}) is mainly localized around specific spectral features, with little impact on broad band light 
curves (Figure~\ref{alpha2_LC}), with the exception of $\hat{I}$-band. This factor associates higher line velocities around 
maximum with deeper absorption troughs, strongly visible before maximum and up to +18 days in the \CaHK 
region, and at all phases in the \CaIR region. The net effect described by $q_2$ is that higher velocities are associated with 
slightly dimmer \sne. At later phases velocity effects are still observable, and at all phases after $-6$ days, high-velocity
\sne have localized differences in the $4700-5100\AA$ spectral region. In addition, they are bluer in $\hat{V}$, $
\hat{R}$, and $\hat{I}$ after maximum, with a maximal effect on the $\hat{I}$-band at around +12 days. A slight stretch 
effect is visible in $\hat{U}$ and $\hat{B}$, and also results in a later phase for the second maximum 
in $\hat{I}$ for the brightest \sne in $\hat{B}$, as was mentioned for $ q_1 $.  
The shape of variations of the light curve in $\hat{I}$ are very different for each factor, indicating that this band 
can help reconstruct the $q_2$ factor when using photometric data only. Furthermore, while the $q_2$ factor 
has a small impact on the variations of individual light curves, it has a sizable influence on colors, especially 
$\hat{R} -\hat{I}$, $\hat{U} -\hat{B}$, and after +25 days, on $\hat{B} - \hat{V}$. In addition to the 
$\hat{I}$-band, the different color variation pattern in $\hat{B} -\hat{V}$ at late phases offers a way of 
disentangling the $q_1$ and $q_2$ factors when dealing with photometric data. \\
\indent As seen in Figure~\ref{alpha3_TIME}, the influence of $q_3$ is minimal around maximum light, similar 
to the results from Section~\ref{extinction_law_section}. It grows at other phases and appears as a stretch effect on the light 
curves presented in Figure~\ref{alpha3_LC}. Unlike $q_1$, the stretch described by $q_3$ shows little correlation with the 
magnitude at maximum. While $q_3$ has a larger impact on the light curves than $q_2$, the reverse is true for 
individual colors. At late phases, $q_3$ exhibits a brighter-redder correlation that might be employed to distinguish this 
vector from $q_1$ when using photometric data only. The influence of $q_3$ on spectral structures is mainly visible in calcium regions, although 
variations can also be observed in other regions. Around \CaIR, between $-6$ days and maximum, 
brighter \sne exhibit a deeper absorption trough at high calcium velocity. This high-velocity calcium feature is visible for all 
$q_3$ values at $-12$ days, but fades faster for dimmer \sne. Meanwhile the lower velocity counterpart is only 
visible in $q_3$ for the dimmest supernovae at $-12$ days, and appears later for brighter \snia.

\section{\sugar Performance}
\label{sugarperfs}

\subsection{Fitting spectra with \sugar}
\label{sugarfit}

The measurement of spectral indicators is very sensitive to the signal-to-noise ratio of the spectrum, 
and at high redshift, measuring the indicators in the observer's reference frame is not possible. 
However, once the model has been trained, 
the parameters $q_1$, $q_2$, $q_3$, $A_V$, and $ \Delta M_ {grey} $ for a given \snia can be 
directly estimated from its spectral time series. Estimating the parameters 
$q_1$, $q_2$, $q_3$, $A_V$, and $\Delta M_{grey}$ is done directly on spectra by 
minimizing the following $ \chi_i^2$:

\begin{equation}
     \chi_i^2 = \left( \textbf{M}'_{i} -\textbf{A}'\textbf{x}_i   \right)^T 
     \textbf{W}_{\textbf{M}'_i} \ \left(\textbf{M}'_{i} -\textbf{A}'\textbf{x}_i \right) \eqcoma
\end{equation}

\noindent where $ \textbf{M}'_{i}$ is the vector containing the entire spectral 
time series of \snia $i$, ordered in the same way as for the \sugar training:

\begin{equation}
\textbf{M}'_{i}=\left(\textbf{y}_{3340\AA \ i}   \cdots \ 
\textbf{y}_{\lambda \ i}  \cdots \ \textbf{y}_{8580\AA \ i} \right)\eqdot
\end{equation}

\noindent where $\textbf{y}_{\lambda \ i}$ corresponds to the light curve
at wavelength $\lambda$ for the observed phases. The model is projected 
onto observation phases using cubic spline interpolation. For notation efficiency, it is thus ordered as:

\begin{equation}
\textbf{A}'=\left(\textbf{M}_{0}^{'},1, f\left(\lambda,R_V\right), 
\boldsymbol{\alpha}_{1'},\boldsymbol{\alpha}_{2'},\boldsymbol{\alpha}_{3'}\right)\eqdot
\end{equation}

\noindent The vector $\textbf {x}_i$ contains the parameters that describe
 the \snia $i$ according to the \sugar model:

\begin{equation}
\textbf{x}_i^T=\left(1,\Delta M_{grey \ i},A_{V\ i},q_{i,1},q_{i,2}, q_{i,3} \right)\eqdot
\end{equation}

\noindent Finally, $\textbf{W}_{\textbf{M}'_i} $ is the weight matrix that comes from
the $\textbf{M}'_i$ errors. Estimating $\textbf{x}_i$ amounts to minimizing the $\chi_i^2$
with respect to $\textbf{x}_i$, giving the solution: 

\begin{equation}
\textbf{x}_i=\left( \textbf{A}'^T \textbf{W}_{\textbf{M}'_i}\textbf{A}' 
\right)^{-1} \left( \textbf{A}'^T \textbf{W}_{\textbf{M}'_i}\textbf{M}'_i \right)
\end{equation}

Therefore the covariance on the $\textbf{x}_i$ is given by:

\begin{equation}
\label{equation_covariance_SUGAR_param}
\text{cov}\left(\textbf{x}_i\right)=\left(\textbf{A}'^{T} 
\textbf{W}_{\textbf{M}'_i}\textbf{A}'\right)^{-1}\eqdot
\end{equation}

\begin{figure}
	\centering
	\includegraphics[scale=0.5]{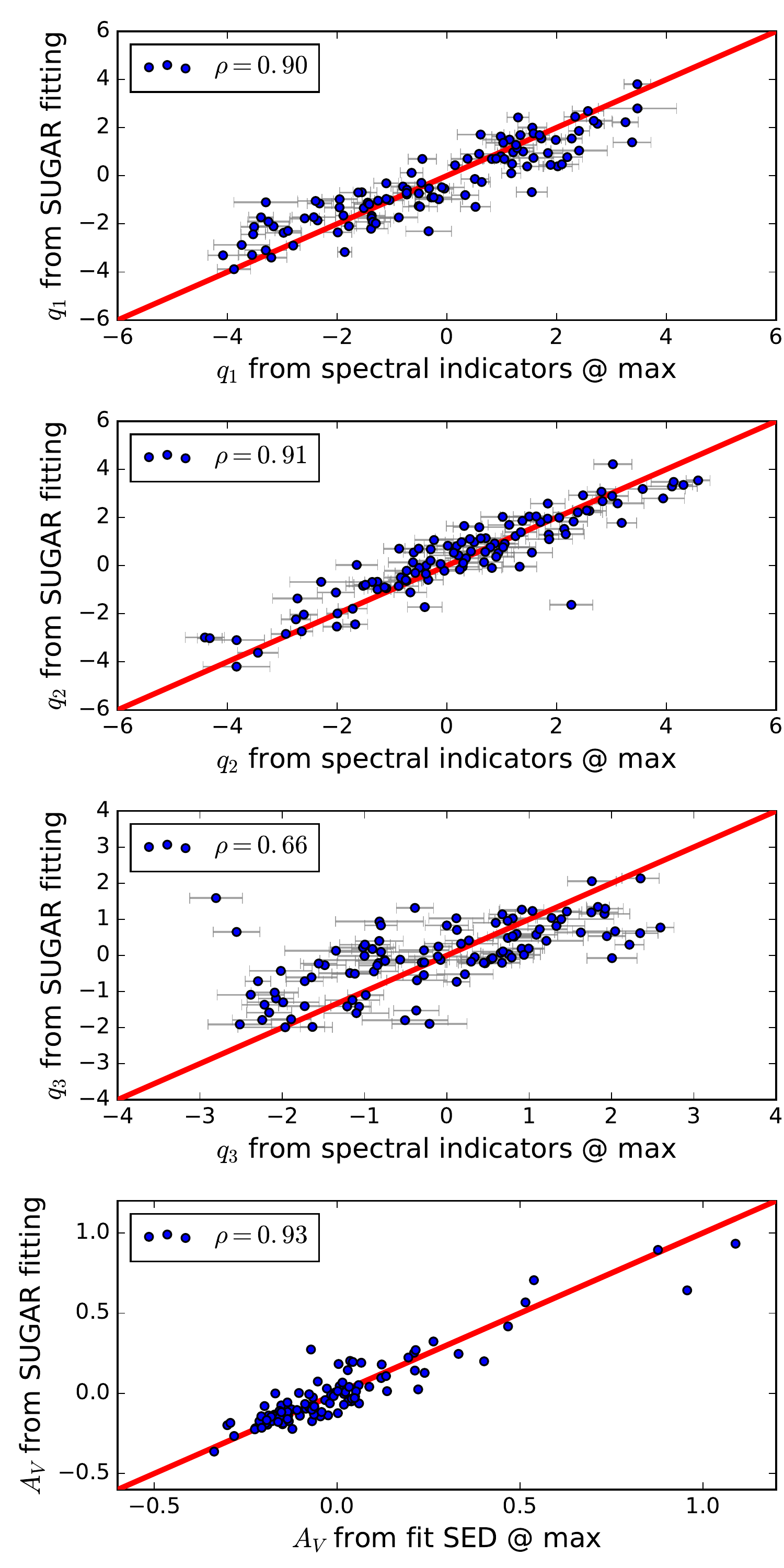}
	\caption{\small The factors estimated directly from the spectral indicators,
	 as a function of the factors estimated only from the \sugar model, and 
	 the $A_V$ estimated at maximum as a function of those 
	 estimated with SUGAR. The red lines represent the  
	 equation $x = y$.}
	\label{qiFA_vs_qiSUGAR}
\end{figure}

\begin{figure*}
	\centering
	\includegraphics[scale=0.55]{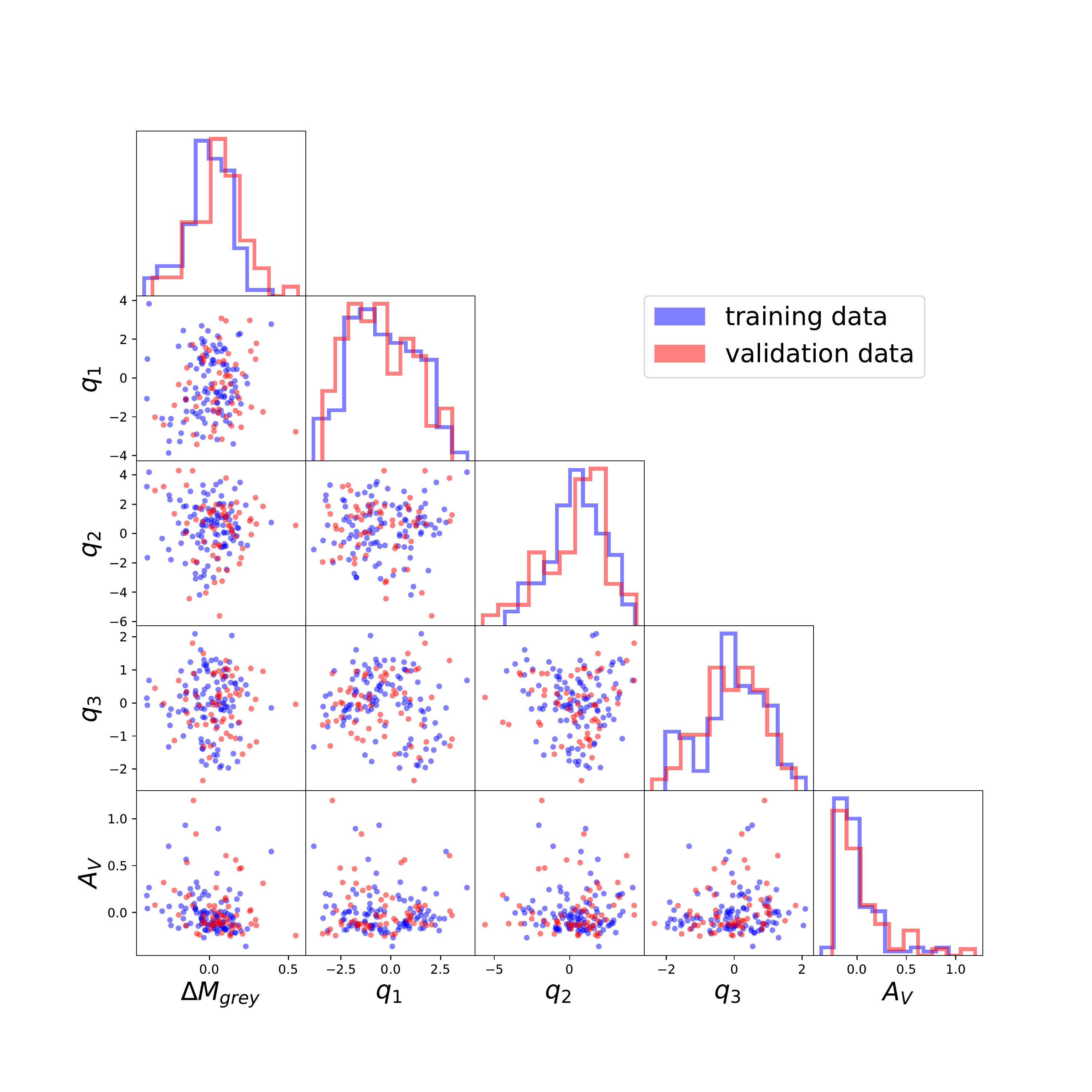}
	\caption{\small SUGAR parameters fitted directly from spectral data for the \NSnTrainingFilterEMFA \sne used to train \sugar (blue) and for the \NsnValidation \sne from the validation sample (red).}
	\label{train_vs_valid}
\end{figure*}

The term $\textbf{x}_i$ is computed for the \NSnTrainingFilterEMFA \sne that were used to 
train SUGAR. The values obtained are compared with the factors and $A_V$ measured at 
maximum light in Figure~\ref{qiFA_vs_qiSUGAR}. Both quantities are 
highly correlated. However, this correlation decreases with decreasing order of the
factor index. This is because the factor uncertainties increase with index due to the noise 
being greater for higher-order components. The increase in noise therefore reduces the correlation. The 
extinction term is quite compatible between its maximum estimate and its estimation with SUGAR.\\
\indent The term $\textbf{x}_i$ is also computed for the \NsnValidation \sne kept for validation. The distribution of the 
$\textbf{x}_i$ parameters from the validation sample is compared to the $\textbf{x}_i$ parameters from the training in Figure~\ref{train_vs_valid}.
From this, one can see that the training and validation samples are compatible, as will be confirmed in following sections.

\subsection{Comparison with SALT2}

\subsubsection{Difference between \sugar and SALT2 models}

Throughout this article, we use the SALT2 model as a benchmark to evaluate the properties of the \sugar model, 
because it has become a reference for cosmological analyses. SALT2 describes the SED of \sne in the following 
functional form \citep{Guy07}:

\begin{equation}
F(t,\lambda)=X_0 [F_0(t,\lambda)+X_1 F_1(t,\lambda)]\ \exp[C\, C_L(\lambda)]\eqcoma
\end{equation}

\noindent where the observed flux $F$ depends on the individual \snia parameters $(X_0,X_1,C)$ and on the 
model average flux $F_0$ (defined up to a multiplicative factor), the model flux variation $F_1$, 
and an empirical color law $C_L$. In the regime where $X_1 F_1$ is small with respect to $F_0$, this model, 
once expressed in magnitudes, can be identified with the \sugar model: ($F_0, F_1, C_L$) and $(X_1,C)$ take 
the roles of ($M_0, \alpha_1, f(R_V)$), and ($q_1, A_V$), respectively.  $\Delta M_{grey}$ is the 
equivalent of a linear combination of $(\log X_0, X_1,C)$. \\
\indent Although the major differences between the models come from the inclusion of 2 additional intrinsic 
components in \sugar, its color law, and the data on which they were trained, there are many other differing characteristics, 
which are summarized Table~\ref{Table_diff_salt_sugar}. 
The inclusion of two additional intrinsic components is enabled by 
the level of detail available in our spectrophotometric time series. We therefore expect the \sugar model to provide a more 
faithful representation of the spectral details than SALT2. Though analogous in spirit, $X_1$ and $q_1$ are derived employing a quite 
different paradigm: $F_1$ comes from a PCA analysis performed in flux space, while the $\alpha_i$ are trained in  
magnitude space and driven to reproduce the prominent spectral indicators. A direct generalization of the SALT2 
approach with more components is presented in \cite{Saunders18}. While the empirical SALT2 $C_L$ function models the average 
color variation, our color curve was shown to be accurately described by a CCM extinction law.
The SALT2 color parameter $c$ thus mixes information about reddening by interstellar media (described by $A_V$) 
with intrinsic \sne properties that were not directly matched with one of the $q_i$ parameters. These differences in 
the treatment of colors would manifest in long-range spectral variation (because the extinction curve is smooth). 
In regions where the phase coverage is sufficient, the different phase interpolation methods used are not expected to contribute 
significantly to differences between models. Otherwise, and for early phases in particular, we expect the 
differences to be driven by the training samples rather than interpolation technique. \\
\indent The wavelength and phase coverage of \sugar are restricted compared to 
SALT2. This is directly related to the training data: \sugar is trained on low redshift spectrophotometric time series, 
while SALT2 is trained on a wide redshift range of photometric data, employing a handful of spectra to help refine the SED 
details. As SALT2 uses low and high redshift data it offers a better coverage of the UV domain. It also has a 
larger phase coverage thanks to the rolling cadence strategy, while \snf \sne have to be specifically targeted. 
While SALT2 could be easily trained with \snf data added to their sample, the reverse is not true: major adaptation 
would be needed to allow \sugar to incorporate photometric data in its training. However, this does not prevent fitting 
photometric light curves with \sugar in the same way as with SALT2, provided they fall within the phase and spectral 
coverage of SUGAR. In contrast, SALT2 was not designed to fit spectrophotometric data.

\begin{table*}
\begin{adjustbox}{max width=\textwidth} 
\begin{tabular}{llllp{1cm}p{1cm}lllp{1cm}rrr}
   \hline \hline
Model properties & SALT2 & SUGAR  \\
  \hline  
     Rest frame wavelength coverage & 2000 -- 9200 $\AA$ & \sugarMinWavelength -- \sugarMaxWavelength \AA\\
     Rest frame phase coverage & $-20$ to +50 days & \sugarMinPhase to \sugarMaxPhase days \\
     Training Data & Spectroscopy and photometry & Spectrophotometric time series  \\
     Redshift range & Low and hight redshift & Low redshift\\
     Interpolation & Cubic spline & Gaussian Process with squared exponential kernel \\
     Intrinsic properties from & Weighted PCA like on SED & EM-FA on spectral features \\
     Number of intrinsic component & 1 & 3 \\
     Color & Color law (third order polynomial)& \cite{Cardelli89} law \\
     Error on the model & Yes & No \\
  \hline
  \end{tabular}
      \end{adjustbox}
  \caption{Table of main differences between the SALT2.4 model and the \sugar model.}
  \label{Table_diff_salt_sugar}
\end{table*}

\subsubsection{Spectral residuals}
 
\begin{figure*}
	\centering
	\includegraphics[scale=0.35]{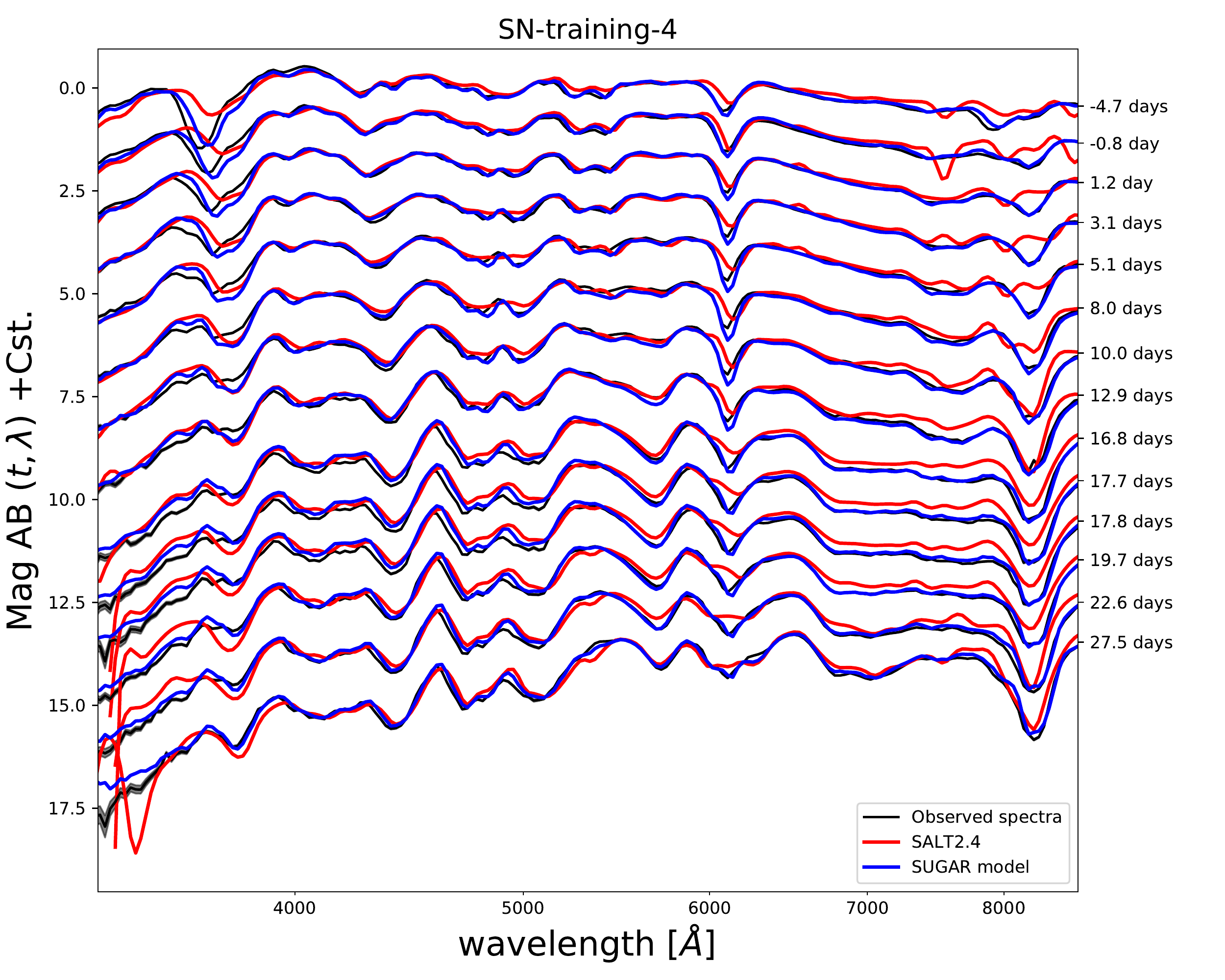} 
	\includegraphics[scale=0.35]{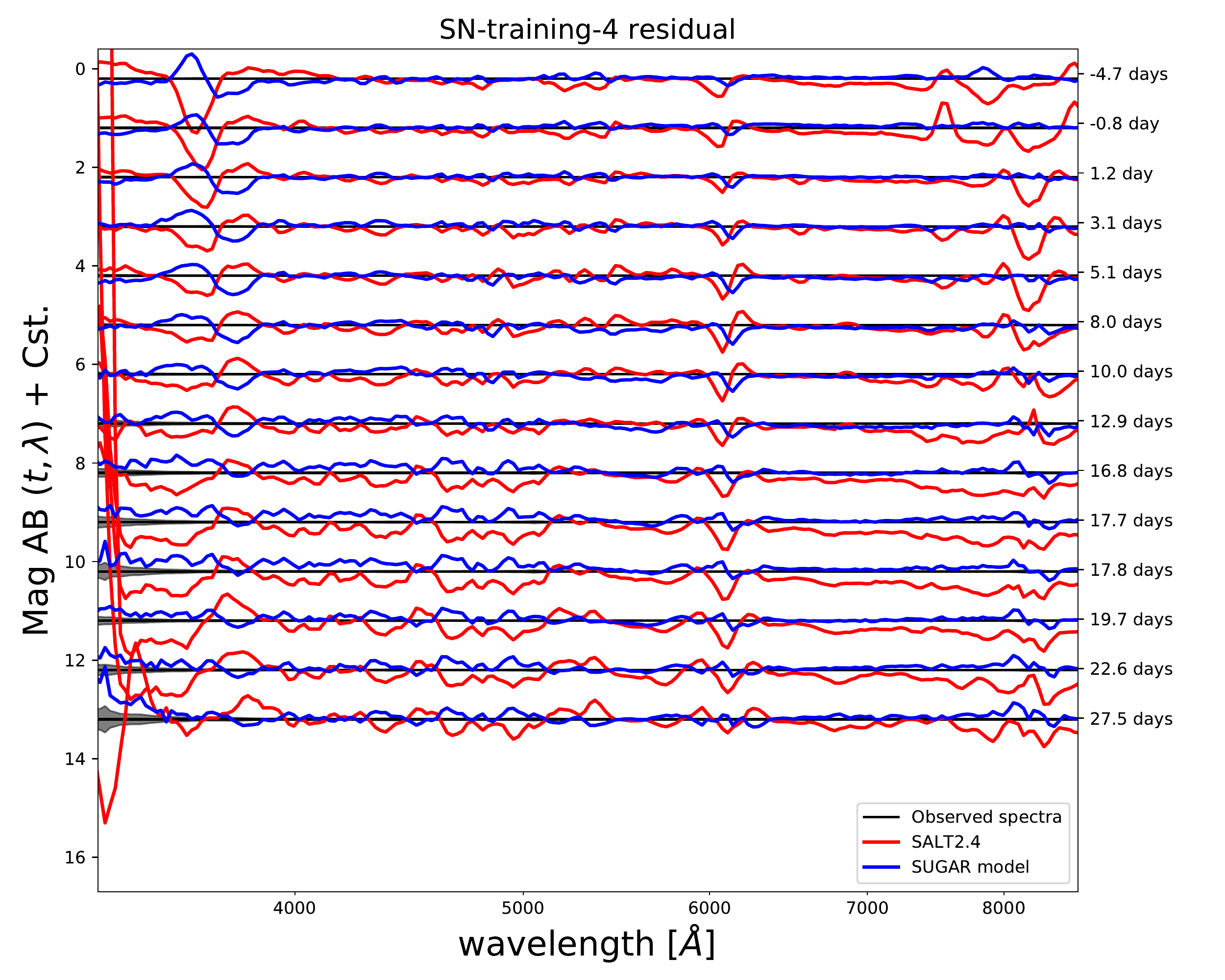}
	\caption{\small Left: observed spectral series of SN-training-4 (in black), compared
	 to SALT2 (in red) and \sugar (blue). Right: residuals
	 of SALT2 (red) and \sugar (blue) models to the observations. The black line 
	 represents the zero value of the residuals.}
	\label{PTF09fox_time_series}
\end{figure*}

The \sugar model is designed to improve the spectral description of \sne by adding components beyond stretch 
and color. While the ideal model would perfectly match the data, the level of mismatch can be quantified by the residuals, 
i.e. the difference between the left and right term of Equation~\ref{Model_SED_SUGAR_equation}. The improvement with 
respect to SALT2 is then studied by comparing the data to the spectral reconstruction predicted by both models.
One example of these comparisons is presented Figure~\ref{PTF09fox_time_series}, which shows the spectra and 
residuals for SN-training-4. For this supernova, \sugar gives a better 
description than SALT2. The improvement is especially clear in the UV after +17 days and in the 
infrared at any phase, and particularly in the \OI and \CaIR regions. This is not surprising insofar as SALT2 is 
essentially trained to reproduce the bluer part of the spectrum. 
The \sugar description is also better in the regions around lines, even if this description is not totally satisfactory in some areas: \CaHK, \Sic, and the bluer part of the spectrum at late phases.
In the B-band, the reconstruction of the spectral details is quite similar 
between \sugar and SALT2, although \sugar seems to follow more faithfully the spectral details, as can be seen in the \Mg 
area.

\begin{figure*}
	\centering
	\includegraphics[scale=0.45]{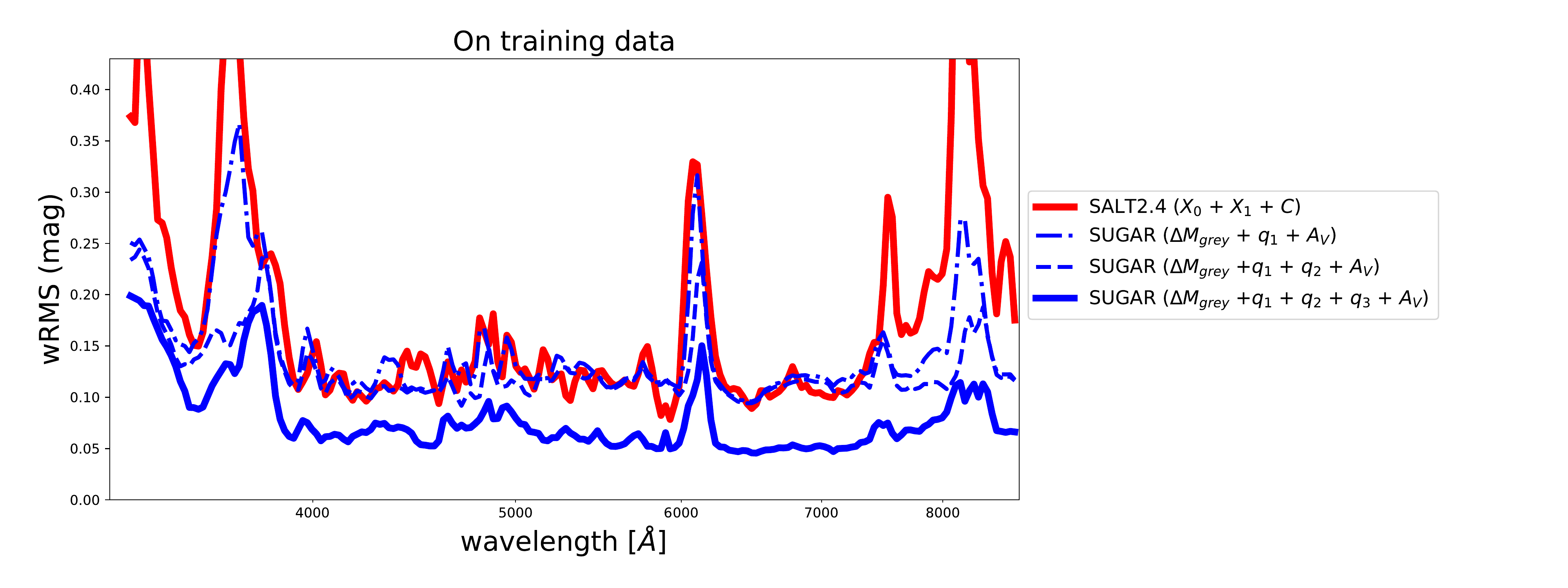}
	\includegraphics[scale=0.45]{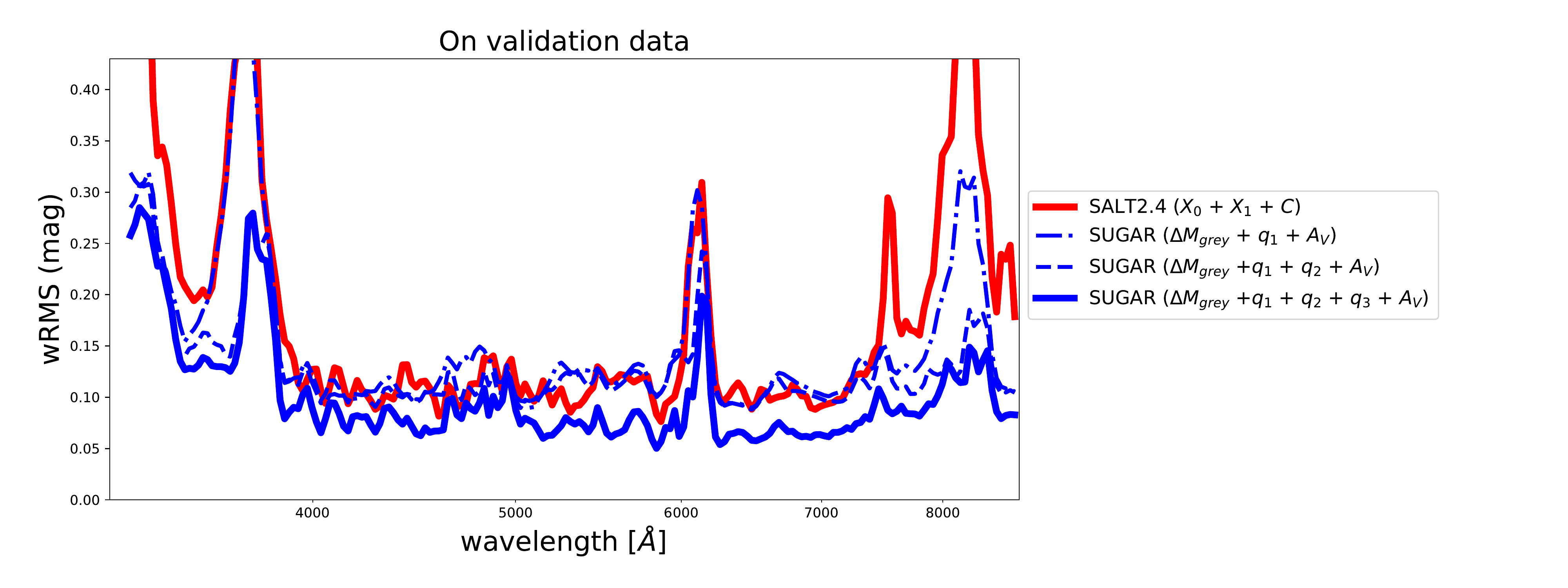}
	\caption{\small  wRMS of the residuals as a function of the wavelength for 
	the full \sugar model (blue line), for the \sugar model only corrected by 
	$q_1$ and $A_V$ (blue dashed dotted line), for the \sugar model only correct by $q_1$, $q_2$ 
	and $A_V$ (blue dashed lines), and for the SALT2 model (red line). \textbf{Top}: Training data. \textbf{Bottom}: Validation data.}
	\label{Residual_model_plot}
\end{figure*}

\begin{figure*}
	\centering
	\includegraphics[scale=0.45]{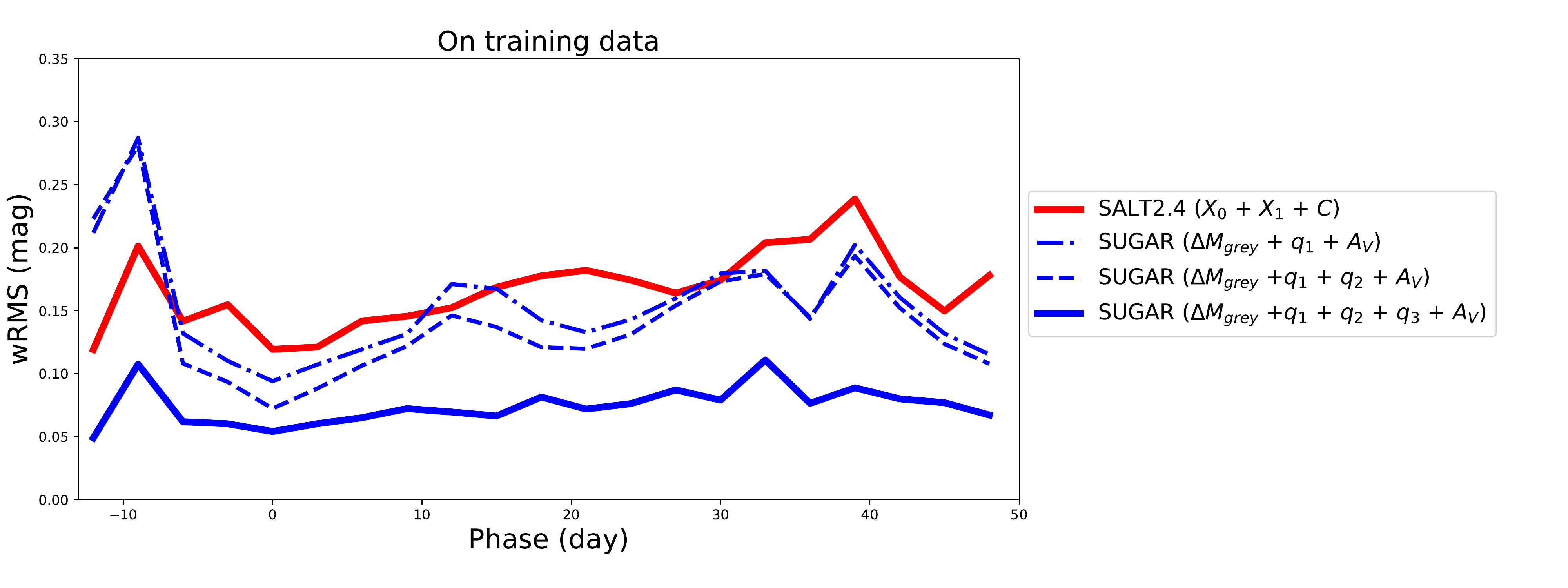}
	\includegraphics[scale=0.45]{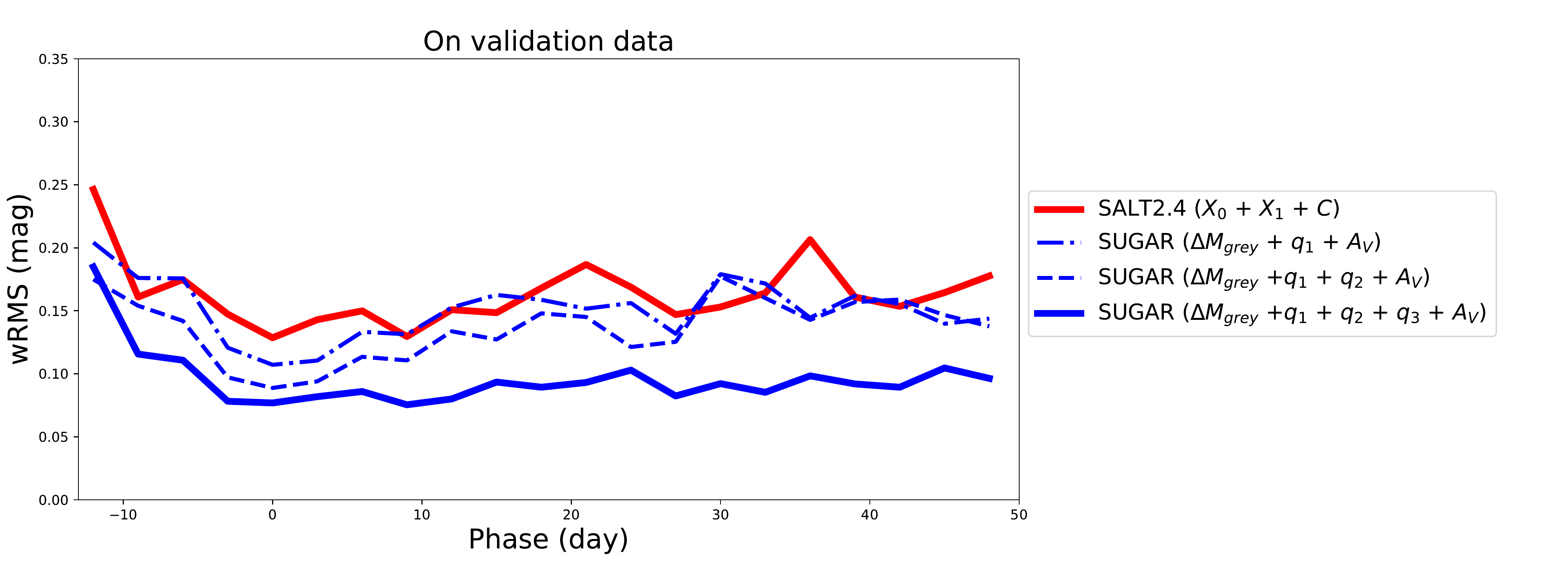}
	\caption{\small  wRMS of the residuals as a function of phase for 
	the full \sugar model (blue line), for the \sugar model only corrected by 
	$q_1$ and $A_V$ (blue dashed dotted line), for the \sugar model only correct by $q_1$, $q_2$ 
	and $A_V$ (blue dashed lines), and for the SALT2 model (red line). \textbf{Top}: Training data. \textbf{Bottom}: Validation data.}
	\label{Residual_model_plot_time}
\end{figure*}

\indent SN-training-4 is only one example among the  \NSnTrainingFilterEMFA \sne on which \sugar was 
trained. To have a statistical description of the precision of the models, we derive the dispersion of 
the residuals obtained for all \sne as a function of wavelength and phase. 
In Figure~\ref{Residual_model_plot}, we present the wRMS of 
this dispersion: for each wavelength, we compute the wRMS across all phases and weight it by the spectral variance. 
To show how the accuracy of the model evolves when going from a classic \cite{Tripp98} relation (when 
only $q_1$ and $A_V$ are included) to the full model, the wRMS for SUGAR is computed three times with progressive 
inclusion of $q_1$, $q_2$, and $q_3$. We compute the residual dispersion for both the training and the validation sample.
What was observed for SN-training-4 is confirmed in this statistical analysis across 
all \sne: at any wavelength, \sugar gives a better description than SALT2. The strongest improvement is in the
IR, but is also significant in the UV. All the spectral details are significantly improved, such as the \CaHK, \Sic, the 
\OI, and the \CaIR. The best performance of the \sugar model is obtained in the region around 6500 \AA, where 
the dispersion gets as low as 0.05 magnitude. It should be kept in mind that the spectral dispersion includes 
noise variance that may be high in some spectra and is therefore not representative of the dispersion 
expected in photometry after integration over broad band filters. The phase evolution of the wRMS is presented in Figure~
\ref{Residual_model_plot_time}: the dispersion in each phase bin is calculated across all wavelengths. 
This includes wavelengths for which SALT2 is not adapted: this explains relatively high values of wRMS for this model 
compared to SUGAR.

The successive inclusion of $q_1$, $q_2$ and $q_3$ can help us understand 
the origin of \sugar{}'s improvement over SALT2. As the training samples for both models are different, one may wonder what 
SALT2 would have given if trained on \snf data. This situation can be studied when including only $q_1$: $\text{SUGAR}|_{q_1}$ is 
slightly worse than SALT2 between \Sia and \Sic features, and between \Sic and \OI. 
These regions cover the wavelength range where SALT2 was trained and optimized, while $\text{SUGAR}|_{q_1}$ is only the 
by-product of the 3 component model and is not by itself optimal for a single component. Earlier phase are more difficult to reproduce in both
models: this could be an effect of higher variability beyond stretch at those phases. For later phases, 
$\text{SUGAR}|_{q_1}$ does not exhibit the same increase in wRMS as SALT2. This is also an effect of the lack of 
coverage for IR wavelength for SALT2. One could therefore expect SALT2 to behave at least as well as $\text{SUGAR}|_{q_1}$
 in UV and IR when trained on \snf data. As expected, the inclusion of additional components 
improves the accuracy of \sugar: $q_2$ improves the description of the spectral features because it is correlated 
with (\CaHK, \Sic and \CaIR) while one has to wait for the inclusion of $q_3$ to reach the full \sugar precision. 

The improvement brought by \sugar is thus enabled by two factors: the spectral coverage and the spectral 
resolution that allows the inclusion of the new components, $q_2$, which improves the color description, 
and $q_3$, which helps reduce the dispersion away from maximum light. Moreover, the improvement added by 
$q_2$ and $q_3$ are also seen in the validation sample, which confirms that these components are not due to overtraining.
  
\subsubsection{Relation between SALT2 and \sugar parameters}

\begin{figure*}
	\centering
	\includegraphics[scale=0.52]{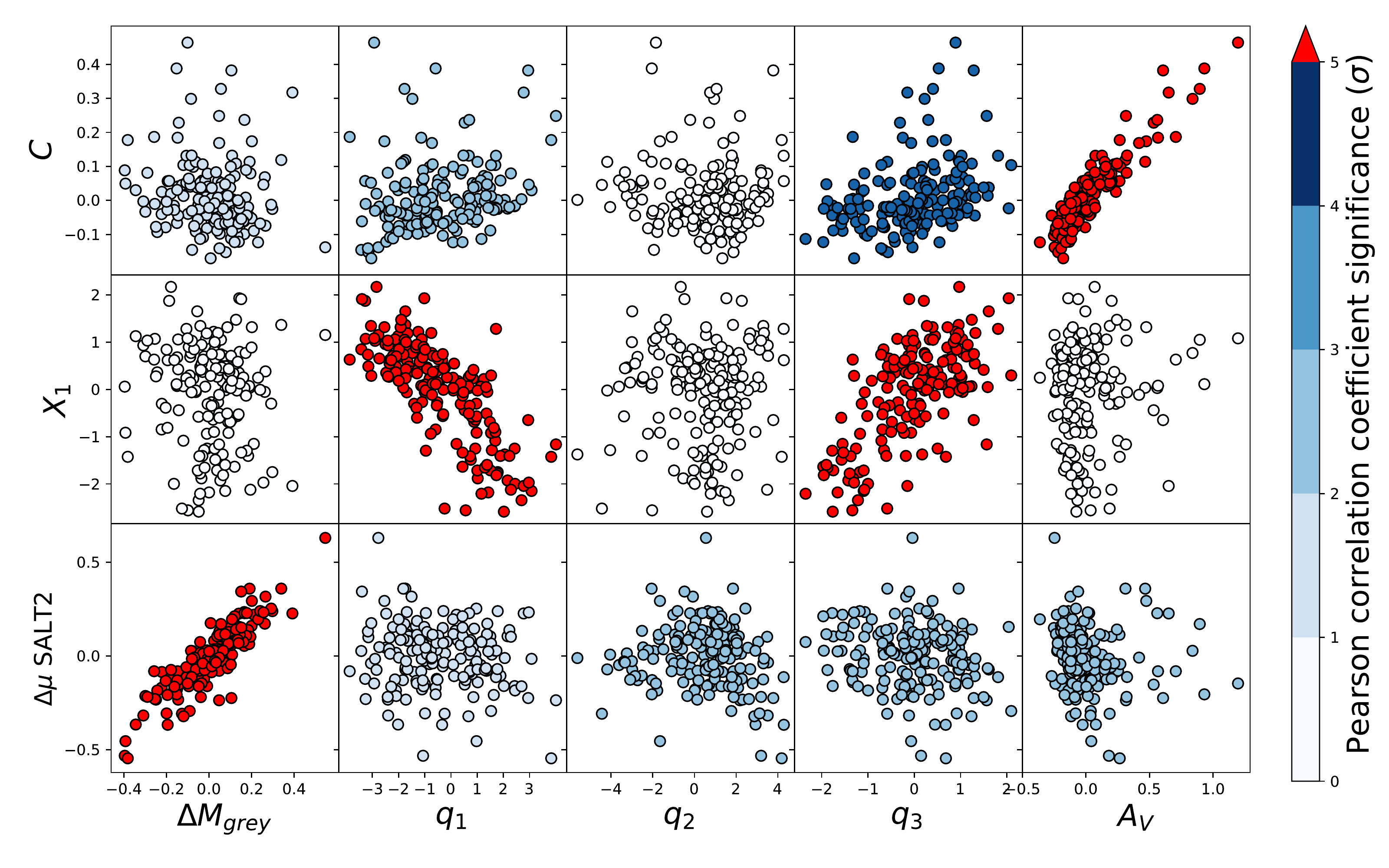}
	\caption{\small SALT2 parameters estimated in photometry as a function of 
	the \sugar parameters estimated in spectroscopy (on training and validation sample). The color code estimates 
	the correlation between the parameters SALT2 and SUGAR.}
	\label{salt_param_vs_sugar_param}
\end{figure*}

In Section~\ref{emfa_section} we presented the correlations between the SALT2 parameters and the first five 
factors found using factor analysis. After training the full \sugar model and reconstructing the 
parameters of individual supernovae as described in Section~\ref{sugarfit}, we can study the correlations 
between the SALT2 parameters ($X_0$, $X_1$, $C$) and the SUGAR parameters ($\Delta M_{grey}$, $q_1$ , $q_2$, $q_3$, 
$A_V$). The direct comparison of $X_0$ and $\Delta M_{grey}$ is irrelevant due to the different prescription for lifting the 
degeneracies of both models. We recognize $\Delta M_{grey}$ as the Hubble diagram residuals, a quantity that 
we can also compute from SALT2 parameters (and will denote as $\Delta \mu^{corr}$ ).
The results of this comparison between SALT2 and SUGAR parameters are presented in Figure~
\ref{salt_param_vs_sugar_param}. The lessons from the correlations of SALT2 parameters and the original
factors still hold: $X_1$ is strongly correlated with $q_1$ and $q_3$; $q_2$ has no significant correlation with any of SALT2 
parameters, except with  $\Delta \mu$ . The small correlation of $q_2$ and $q_3$ with $\Delta \mu$ shows that the 
inclusion of this factor can help improve the standardization. The SALT2 color $C$ is strongly correlated with $A_V$, 
as would be expected since the purpose of these two parameters is to take into account the reddening of \sne. 
Contrary to the results from Section~\ref{emfa_section} with the factors alone, $C$ now exhibits
a $4.5 \sigma$ correlation with $q_3$. This correlation might come from the slight fluctuation in color with 
respect to phase visible in Figure~\ref{alpha3_LC} and the inclusion of the validation sample. 
The slight correlation of $C$ and $\Delta M_{grey}$ comes
from the correlation of $\Delta M_{grey}$ and $A_V$. Indeed, $\Delta M_{grey}$ and $A_V$ were determined 
separately during the training, without fixing potential degeneracy between them. This explains why 
they are correlated. Finally, {$\Delta \mu$} 
has a strong correlation with $\Delta M_{grey}$. This indicates that the inclusion of the additional factors in 
\sugar is unable to catch the major source of magnitude variability of \sne, even if it provides a much improved description of 
spectral details.

\section{Discussion}
\label{discussionsugarsection} 

\subsection{Grey dispersion and dispersion matrix fitting}
\label{discussion_grey_disp_matrix}

For the model at maximum light, we did not fit a grey term but did include a dispersion matrix while estimating the extinction curve. However, for the full spectral time series, we fit for a grey offset and used the extinction curve determined at maximum light. One might wonder why we did not do a simultaneous fit of the full spectral time series, the grey offset and the extinction curve. Our decision was motivated by the following logic.\\
\indent In order to perform a simultaneous fit, one important thing is to have a good estimate of the dispersion matrix, 
which is essential for the estimation of the extinction curve, as discussed in \cite{Scolnic14}. In addition, 
this matrix would have the advantage of taking into account temporal correlations of residuals, in addition to the color 
correlations of the residuals, which would improve the model developed at maximum light.\\
\indent However, from a numerical point of view, this would break the sparse algebra approximations that significantly speed 
up the training. Indeed, the sparse algebra (due to Equation~\ref{Big_matrice_for_SUGAR_from_GP}) makes it possible to 
improve speed efficiency of minimization of Equation~\ref{chi2_sugar_0} from {$\mathcal{O}\left((N_{\lambda} N_{t} N_{\text{param}}  )^3\right)$} to {$\mathcal{O}\left(N_{\lambda} \times (N_{t} N_{\text{param}}  )^3\right)$}, where 
$N_{\lambda}$ is the number of wavelength bin,  $N_{t}$ the number of bin in terms of time, and $N_{\text{param}}$ is the 
number of parameters to fit for a given wavelength and epoch. In SUGAR training, $N_{\lambda}\sim190$, $N_t\sim30$, and 
$N_{\text{param}}=5$ (average spectrum + 3 factors + extinction curve), consequently, not doing sparse algebra will 
significatively slow down the speed of the training algorithm by  {$\mathcal{O}\left(N_{\lambda}^2 \right)$}. \\
\indent Moreover, there would be problems with numerical stability for the estimation of the dispersion matrix due to its 
size; in the case of the simultaneous fit: it would by $\sim 20000 \times 20000$. Indeed, it would be necessary to make sure that the 
dispersion matrix is positive definite, which would involve doing a Singular Value Decomposition that will again slow down the 
speed of the training algorithm.
In addition, it is evident that the grey offset is degenerate with the dispersion matrix because the dispersion matrix can fully 
capture a grey dispersion. Therefore we deliberately
did not include it explicitly in the fit of the extinction curve within SUGAR in Section~\ref{extinction_law_section}. 
There is however a degeneracy between $A_{\lambda_0}$ and the grey offset, and part of the latter is
captured by this parameter.
 \\
\indent The SUGAR model presented here is already a significant improvement 
over the SALT2 model and provides insights into understanding \snia variability.

\subsection{Test of adding an additional component}
\label{discussion_number_of_component}

\begin{figure*}
	\centering
	\includegraphics[scale=0.45]{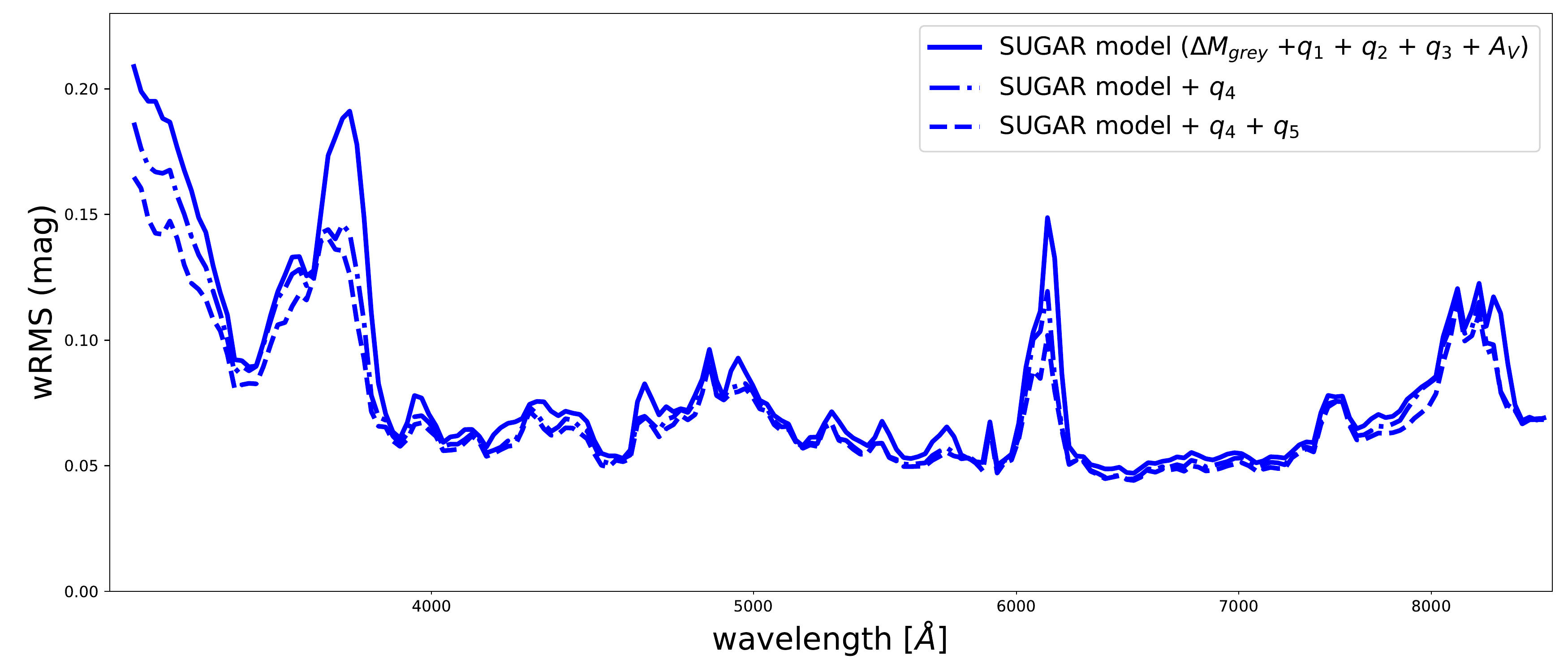}
	\caption{\small wRMS of the residuals as a function of the wavelength for 
	the full \sugar model (blue line), for the \sugar model with the addition of the factor $q_4$ 
	(blue dashed dotted line), and for the \sugar model with the addition of the factor $q_4$ and $q_5$
	 (blue dashed lines).}
	\label{Residual_model_more_components}
\end{figure*}

In the Section~\ref{emfa_section}, we discussed the number of factors needed
to describe the final \sugar SED, and we concluded that this cannot be determined 
only from the factor description of spectral features at maximum light.
Indeed, even if they are strongly related, the main goal 
is to know the number of components needed to describe the full 
SED and not the number of components needed to describe the spectral 
features space at maximum light. One way to check if the choice of three factors used here is optimal is 
to retrain the \sugar model with more than three components and observe how
 the spectral residuals evolve with this change. \\
 \indent In the following, we ran the training of \sugar twice, each time adding an additional component, i.e. 
 we reproduced Sec.~\ref{sugar_fitting_model} with $q_1$-$q_4$ and $q_1$-$q_5$ components.
 In both cases the value of $R_V=\RvValue$ found with three factors is fixed in order to focus only 
on intrinsic parameters. In Fig.~\ref{Residual_model_more_components} we compare the spectral residuals 
of the \sugar model to those of \sugar trained with the additional factors. \\
\indent As expected, the addition of the two new components from factor analysis does 
not improve the description of \sne as significantly as the addition of $q_2$ or $q_3$.
The factor $q_4$ slightly improves the description within the \CaHK area (0.05~mag) and 
the \Sic (0.03 mag), but does not improve the SED outside these areas. 
The factor $q_5$ does not significantly improve the description of the SED. This is confirmation that our 
choice of using three factors provides a good description of the SED.

\section{Conclusion}
\label{sugarconclu}

In this paper we have presented a new spectra-temporal empirical model of \sne,  
named SUGAR. This model significantly improves the spectral description of 
\sne compared to the current state of the art by going beyond the classical stretch and color 
parametrization.\\
\indent In Section~\ref{descriptiondatset}, we presented the SNfactory spectrophotometric dataset that was used to train 
the model. In Section~\ref{deriveddata}, we presented the intermediary data that were used to train the full \sugar 
model. In a first step, we selected a set of 13 spectral indicators near maximum light in B-band that are 
composed of pseudo-equivalent widths and minima of P-Cygni profiles. Those spectral indicators 
were chosen to describe the intrinsic part of the \sugar model because they are easy to define and are independent 
of host-galaxy dust extinction. Then, we defined a new basis where the 13 spectral indicators are uncorrelated. 
For this we developed a factor analysis algorithm that is more robust in the presence of errors than PCA algorithms. Three 
factors seems to be effective enough to describe the spectral indicators space at maximum light. The first factor 
describes the coherent variation of the  pseudo-equivalent widths, mainly of the silicon and calcium lines. Like those lines, this 
factor is strongly correlated with stretch. The second factor is mainly correlated with the velocities and shows a very weak link 
with pseudo-equivalent widths, except for the \CaHK and \SW. The third 
factor shows a slight correlation with the stretch parameter. Once the factors have been defined, we established 
a model of SED at maximum light based on these three first factors from spectral features, using 
the same underlying method as in \cite{Chotard11}. This allows us to separate the intrinsic behavior from color variation due to dust.
Finally, we find that the color curve obtained is compatible with a \cite{Cardelli89} extinction curve with \mbox{$R_V$ 
of $\RvValue$.} 
We then developed an interpolation method using Gaussian process in order to train the \sugar model on a uniform and fixed time grid. 
The correlation length obtained varies between 5 and 12 days depending on  
spectral regions, which justifies the use of a 3-day time step for the grid. The uncertainty from the Gaussian Processes 
is generally underestimated by a factor of 1.3 and more investigation is needed to understand why this occurs.\\
\indent The training process for the \sugar model is described in Section~\ref{sugarmodeltrainingsection}, and an interpretation of 
each new component is provided. Both factors $q_1$ and $q_3$ resemble a stretch effect, but 
$q_3$ has less impact around maximum light and in color space than $q_1$. The effects of $q_2$ are strongest 
in the areas of spectral features, and mainly evident in the infrared as compared to broad band photometry.\\
\indent After calculating the model, we showed that, instead of going through the calculation of spectral 
indicators, we can work directly with the spectral time series to recover the three factors and extinction parameter. By studying 
model residuals as a function of wavelength, it is shown that \sugar improves the spectral description 
0.1 to 0.4 mag  with respect to SALT2. This is valid for both training and validation data sets, 
which confirms that there was no overtraining resulting from the addition of $q_2$ and $q_3$.
This shows that three parameters, defined at a given phase (i.e. 
maximum light) have predictive power at other phases.
Performance of the SUGAR model makes it an excellent candidate for use with surveys such as ZTF, LSST or WFIRST, and offers
an alternative way of going beyond stretch and color to measure distance with SNe~Ia.

\begin{acknowledgements}
We thank the technical staff of the University of Hawaii 2.2-m telescope, and Dan Birchall for observing assistance. We 
recognize the significant cultural role of Mauna Kea within the indigenous Hawaiian community, and we appreciate the
opportunity to conduct observations from this revered site. This work was supported in part by the Director, Office of 
Science, Office of High Energy Physics of the U.S. Department of Energy under Contract No. DE-AC025CH11231. Support 
in France was provided by CNRS/IN2P3, CNRS/INSU, and PNC; LPNHE acknowledges support from LABEX ILP, supported 
by French state funds managed by the ANR within the Investissements d'Avenir programme under reference ANR-11-
IDEX-0004-02. NC is grateful to the LABEX Lyon Institute of Origins (ANR-10-LABX-0066) of the University de Lyon for its 
financial support within the program "Investissements d'Avenir" (ANR-11-IDEX-0007) of the French government operated by 
the National Research Agency (ANR). Support in Germany was provided by DFG through TRR33 "The Dark Universe" and 
by DLR through grants FKZ 50OR1503 and FKZ 50OR1602. In China support was provided by Tsinghua University 985 
grant and NSFC grant No 11173017. Some results were obtained using resources and support from the National Energy 
Research Scientific Computing Center, supported by the Director, Office of Science, Office of Advanced Scientific 
Computing Research of the U.S. Department of Energy under Contract No. DE-AC02- 05CH11231. We thank the Gordon 
\& Betty Moore Foundation for their continuing support. Additional support was provided by NASA under the Astrophysics 
Data Analysis Program grant 15-ADAP15-0256 (PI:Aldering). We also thank the High Performance Research and Education 
Network (HPWREN), supported by National Science Foundation Grant Nos. 0087344 \& 0426879. This project has 
received funding from the European Research Council (ERC) under the European Union's Horizon 2020 research and 
innovation programme (grant agreement No 759194 - USNAC). PFL acknowledges support from the National Science 
Foundation grant PHY-1404070. The work of MVP (participation in SUGAR implementation in sncosmo) was supported by Russian Science Foundation grant 18-72-00159. We thank Claire-Alice H\'ebert for reviewing and giving helpful advice on this paper. 
\end{acknowledgements}

\appendix

\section{Spectral-indicator measurements}
\label{app:measurements}

As the spectral indicators are a key ingredient of our statistical analysis, we need a robust and automatic algorithm to derive 
them from the rest-frame spectra and estimate the associated statistical and systematic uncertainties. The main issue to be 
addressed is the automatic detection of feature boundaries --- usually local extrema --- as they shift both along the phase and 
from one \sn to another. Sometimes, the extremum itself cannot be uniquely defined, \eg, when there is a mix of several 
local extrema at the same position or no significant extremum such as in the blue edge of the \OI feature. The photon noise 
will also limit the accuracy when determining the wavelength of the maximum, and may induce a systematic shift due to the 
non-linear process involved in finding an extremum. In this section, we give a brief overview of the general and automatic 
method developed and implemented by \cite{Chotardphd} to measure these spectral indicators and evaluate their 
corresponding uncertainties. This method shares some similarities with previous analyses 
\citep{Folatelli04,Garavini07,Nordin11,Blondin11,Silverman12b}: it is based on extrema searches in a fixed spectral domain after 
smoothing the data. Improving on previous studies, our smoothing procedure is based on an optimal determination of the 
regularization parameters, and more noteworthy, we developed a thorough determination of the uncertainties based on a 
Monte Carlo procedure taking as input the variance of our signal.

\subsection{Method}

Most of the spectral indicators presented here are defined by the position in wavelength and flux of at least one local 
minimum or maximum, \ie, peaks or troughs of the \snia spectral features. In order to compute a precise estimate of these 
local extrema, a Savitsky-Golay \citep[SG]{Savitzky64} smoothing is applied to the original spectra before the rebinning at 
1500 km s$^{-1}$ described in Section~\ref{descriptiondatset}. Since the SG optimal window depends both on the 
underlying spectral shape and on the $S/N$ ratio, which varies across wavelength, an independent smoothing is applied to 
each of the nine spectral zones of interest defined in table~\ref{table:zones}. 

\begin{table}
\caption{Independently smoothed spectral regions, with minimal and
  maximal boudaries $\lambda_{\text{min}}$ and $\lambda_{\text{max}}$
  (in \AA).}  \centering
\begin{tabular}{lcr}
  \hline \hline
  Region & $\lambda_{\text{min}}$ - $\lambda_{\text{max}}$ & Elements \\ \hline
  1 & 3450 - 4070 &  \CaHK\                \\ 
  2 & 3850 - 4150 &  \Sia;Co~\textsc{ii}   \\ 
  3 & 4000 - 4610 &  Mg~\textsc{ii} triplet\\ 
  4 & 4350 - 5350 &  Fe~\textsc{ii} blend  \\ 
  5 & 5060 - 5700 &  S~\textsc{ii}~W       \\ 
  6 & 5500 - 6050 &  \Sib\                 \\ 
  7 & 5800 - 6400 &  \Sic\                 \\ 
  8 & 5500 - 6400 &  Si~\textsc{ii}~$\lambda$5972;6355  \\ 
  9 & 6500 - 8800 &  O~\textsc{i} triplet;Ca~\textsc{ii}~IR \\ \hline
\end{tabular}
\label{table:zones}
\end{table}

Each extremum is then selected on the smoothed spectrum as the local extremum inside a given wavelength range, which 
can be found in tables~\ref{table:zonesEWv}~and~\ref{table:zonesVs} for the equivalent widths and the feature velocities 
respectively. These wavelength ranges have been trained on a set of $\sim50$ \sne within a phase range of $\pm5$~days 
around B-band maximum light so that they match the observed diversity of our spectra. The positions in wavelength and 
flux of these extrema are then used in the spectral-indicator measurements, either directly, e.g., for the velocities, or indirectly, e.g., for the equivalent widths through the definition of their pseudo-continuum.

\begin{table}
  \caption{Definition of the wavelength regions (in \AA) where the
    extrema are expected to be found. The $b$ and $r$ exponents
    respectively represent the left (blue) and right (red) peak.} 
  \centering
\begin{tabular}{llcr}
  \hline \hline
  Region & Indicators & $\lambda_{\text{min}}^{b}$ - $\lambda_{\text{max}}^{b}$ & $\lambda_{\text{min}}^{r}$ - $\lambda_{\text{max}}^{r}$ \\
  \hline
  1 & \CaHK & 3504 - 3687 & 3830 - 3990 \\   
  2 & \Sia  & 3830 - 3990 & 4030 - 4150 \\   
  3 & \Mg   & 4030 - 4150 & 4450 - 4650 \\   
  4 & \FE   & 4450 - 4650 & 5050 - 5285 \\   
  5 & \SW   & 5050 - 5285 & 5500 - 5681 \\   
  6 & \Sib  & 5550 - 5681 & 5850 - 6015 \\   
  7 & \Sic  & 5850 - 6015 & 6250 - 6365 \\   
  8 & \OI   & 7100 - 7270 & 7720 - 8000 \\   
  9 & \CaIR & 7720 - 8000 & 8300 - 8800 \\ \hline
\end{tabular}
\label{table:zonesEWv}
\end{table}
\begin{table}
  \caption{Wavelength regions (in \AA) used to compute the feature
    velocities, together with their rest-frame wavelengths,
    $\lambda_{0}$.}  \centering
  \begin{tabular}{llccr}
    \hline \hline
    Region & Velocity & $\lambda_{\text{min}}$ - $\lambda_{\text{max}}$ & $\lambda_{0}$ \\
    \hline
    2 & v(\Sia) & 3963 - 4034 & 4131 \\
    5 & v(\SWa) & 5200 - 5350 & 5454 \\
    5 & v(\SWb) & 5351 - 5550 & 5640 \\
    7 & v(\Sic) & 6000 - 6210 & 6355 \\ \hline
  \end{tabular}
  \label{table:zonesVs}
\end{table}

\subsection{Optimal smoothing}
\label{app:smoothing}

The purpose of smoothing, also known as regularization, is to transform the original noisy data in order to get closer on 
average to the unknown original spectrum, based on some regularity hypothesis. A parameter describing how smooth the 
final function is has to be introduced, and has to be estimated, either based on physical consideration or deduced from the 
data themselves. Here we follow the latter approach and describe an optimal way of setting the smoothing parameter given a 
class of transformations. 

\subsubsection{Formalism}

Smoothing a noisy spectrum $Y$ consists of the determination of the smoothed spectrum $Y'$, 
which is a function of the initial flux 
$Y'~=~f(Y)$. $Y'$ will be an estimator of the noise-free and unknown original flux, convolved by the instrumental resolution: 
$\hat{Y}$. It is related to the observed flux by:

\begin{equation}
  \label{eq:Y}
  Y=\hat{Y}+N\eqcoma
\end{equation}

\noindent where $N$ is a realization of the spectral noise vector. The goodness of the smoothing being represented by the error 
function,

\begin{equation}
  \label{eq:Ynorm}
  {\|Y' - \hat{Y}\|}~\hat{=}~(Y'-\hat{Y})^T W (Y'-\hat{Y})\eqcoma
\end{equation}

\noindent where $W$ is the inverse of the noise covariance matrix, this requires minimization of a quantity 
which depends both on the noise properties and the signal shape. As the true spectrum, $\hat{Y}$, is unknown, 
we need to build an estimator of $\|Y'~-~\hat{Y}\|$ which is independent of $\hat{Y}$. For a linear 
regularization, one can write:

\begin{equation}
  \label{eq:Yp}
  Y'=f(Y)=B_{p}Y\eqcoma
\end{equation}

\noindent where $B_{p}$ is a smoothing matrix which only depends on the chosen smoothing technique and a smoothing
parameter $p$. Introducing Equations~\ref{eq:Y} and \ref{eq:Yp} into \ref{eq:Ynorm}, one can show that:

\begin{eqnarray}
  \|Y'-\hat{Y}\| &=& (Y'-Y+N)^T W (Y'-Y+N)\\
  &=& \|Y'-Y\| + 2 N^T W (B_{p} - I )(\hat{Y} + N) + N^T W N\eqdot
\end{eqnarray}

\noindent An estimator of the error function, $\epsilon$, that has to be minimized can be constructed by noticing that

\begin{equation}
  E \Big[ \|Y'-\hat{Y}\| \Big] = E \Big[ \|Y'- Y\| \Big] + 2\, {\mbox{Tr}}\, (B_{p}) - n\eqcoma
\end{equation}

\noindent where $E$ is the mathematical expectation value, $\mbox{Tr}$ is the Trace operator, and $n$ is the rank of the vector $Y$. As 
we have only one realization of $Y$, this translates to:

\begin{equation}
  \label{eq:epsilon}
  \epsilon = \|Y'-Y\| + 2\,
  \mbox{Tr} \,(B_{p}) - n\eqcoma
\end{equation}

\noindent which is the quantity that will be minimized with respect to $p$.

Given a class of linear smoothing methods $B_p$, we have thus defined a procedure to find the value of $p$ for which the 
smoothed spectrum best reproduces the original unknown one given an observed spectrum. This method is not exempt from possible overtraining, however 
this is mitigated by restricting the optimization to a single parameter. This was tested with Monte Carlo simulations.

\subsubsection{B matrix estimation}

Among all possibilities, we chose a Savitzky-Golay regularisation \citep[SG]{Savitzky64} as it allows the reduction of the high 
frequency noise, while keeping the original shape of the feature. This method relies on fitting a $k$-order polynomial function 
for each point $i$ in a fixed window size $p$ (with $p > k + 1$) centered on the current point and is designed to preserve the 
locus of maxima for even values of $k$. Each observed data point is replaced by its fitted value, and
the window then moves to the next data point until the spectrum is completely smoothed. As we have fixed the degree of the 
polynomial function to $k = 2$, the only parameter that has to be optimized is the window size, $p$. This transformation is 
linear, and the optimal window size $p$ and corresponding smoothing matrix $B_{p}$ are found by minimizing equation~
\ref{eq:epsilon}. The value of $p$ is estimated individually for each spectral region of a given spectrum, and the spectral 
indicator measurements are performed on the corresponding smoothed spectrum, as indicated above.

\subsection{Uncertainties}

The uncertainty on a given spectral-indicator measurement arises from the statistical noise of the data and the induced 
uncertainties on the measurement method parameters. In our approach, these two uncertainties are independently measured 
using the smoothed spectrum as a reference: the SG window size is typically large enough so that the residual noise can be 
neglected in the simulations.

A random noise matching the statistical properties of the observed data is then added in order to generate a mock spectrum, 
and the spectral indicators are measured on this new spectrum as if it were the observed one. After one thousand 
generations, we are able to derive the statistical fluctuations of the obtained values, which we quote as the statistical 
uncertainty. This is computed as the standard deviation from the value measured on the real spectrum, thus taking into account 
a potential bias. This bias typically corresponds to 10 to 20\% of the total error budget.

In order to save computation time, the initial SG window size $p$ is determined once and for all using the original spectrum. It is 
then kept constant when measuring spectral indicators on all the simulated spectra. However, the noise affecting the initial 
spectrum will induce an uncertainty in the smoothing parameter. This has been studied on a reduced set of simulations for 
which the optimal $p$ was derived for each realization of the noise, and the corresponding uncertainty was found to range 
from 15\% to 20\% depending on the spectral zone. We then propagate this uncertainty by computing the spectral-indicator 
values on the original spectrum for several values of $p$ in this $15-20\%$ range. The standard deviation of the resulting 
spectral-indicator value distribution gives an estimate of the systematic error introduced by the arbitrariness of the smoothing 
method, which is found to be $\sim20\%$ of the total error on average. These two uncertainties are quadratically added 
together to derive the final uncertainty for a given spectral indicator. This estimate takes into account the statistical noise of 
the spectrum, as well as the induced scatter in wavelength and flux of the extrema. Our measurement errors thus include all 
non-linear effects due to limited signal to noise and can be trusted for subsequent statistical analysis. 
More details could be found in \cite{Chotardphd} and \cite{Nordin11}.

\subsection{Performances}
\label{sec:robustness}

\subsubsection{Failure rate}
\label{para:failurerate}

When one of the extrema defining a spectral indicator lies on a flat and/or noisy section of the spectrum, its measurement 
has a chance of failure. In that case, the measurement is automatically rejected and a visual scan using control plots is 
performed to confirm the actual lack of an extremum. The rejection of the ``bad'' measurements is performed using a 
$3\sigma$ clipping in the measured-uncertainty space: if the uncertainty made on a given measurement is larger than $m
+3\times std$, where $m$ and $std$ are the average and standard deviation of this spectral-indicator uncertainty distribution 
(for the whole sample), the corresponding measurement is rejected. Considering all the spectral indicators measured on the 
113 input spectra in a range of phase of $\pm 2.5$~days around maximum light and presented in this paper, the global 
failure rate of the measurement procedure is less than $10^{-3}$ for the automatic selections mentioned above. If it does 
happen, the spectral indicators are set to the average value and assigned an infinite error.

\subsubsection{Quoted uncertainties}

A simple test has been performed to confirm the robustness of the method.  In our selected sample, $18$ \sneia have two or 
more spectra taken in the same night in a phase interval of $\pm5$~days around maximum light. We then computed the 
distribution of the \textit{pull},

\begin{equation}
  \label{eq:pull}
  \frac{\delta I}{\sigma_{\delta_{I}}} = \frac{I_{1} - I_{2}}{\sqrt{\sigma_{I_{1}}^{2}+\sigma_{I_{2}}^{2}}}\eqcoma
\end{equation}

\noindent for all the spectral indicators $I$ (feature velocity and absorption ratio) measured on each spectrum of a
same night (for a same \sn).  This distribution is centered around $-0.01$ with a dispersion of $1.06$ which indicates that 
our estimation of the uncertainty is valid up to a possible underestimation of the error by 6\%. This number is small enough 
so that we can trust our uncertainty estimation for the main analysis.

\section{Expectation-Maximization Factor Analysis}
\label{app:EMFA}

Dimensionality reduction in the presence of noisy data is often an overlooked problem. However, standard methods like 
Principal Component Analysis tend to fail at capturing the intrinsic variability of the data and the principal components will 
align with the direction of the noise when the latter becomes important. Factor Analysis on the other hand is a statistical 
method designed to model the covariance structure of high dimensional data using a small number of latent variables. It 
estimates both the natural variability of the sample and the noise arising from the measurements, under the assumption that 
the statistics are the same for all data records. Our case is slightly different: on one hand, the noise statistics are different for 
each measurement, but on the other hand, their variance is already known. We thus adapted the expectation-minimization 
algorithm presented in \cite{Ghahramani97theem} to accommodate for the specifics of our problem. The resulting method is 
also known as Probabilistic Principal Component Analysis. The formalism is the following: $\mathbf{x}_i$ is a vector of rank 
$l$ representing the  measurement $i$. It is linked to the factor $\mathbf{q}_i$, vector of rank $k\le l$ and the noise 
$\mathbf\eta_i$ by:

\begin{equation}
\label{eq:formalism}
\mathbf{x}_i = \mathbf{x}^0 + \mathbf{\Lambda q}_i + \mathbf\eta_i\eqcoma
\end{equation}

\noindent where $\mathbf{x}^0$ is a central value which can be further neglected without loss of generality ($\mathbf{x}^0=0$), $
\mathbf{q}_i$ is assumed to follow a normal distribution of unit variance, and $\mathbf\eta_i$ follows a multivariate 
normal distrubution of  variance $\mathbf\Psi_i$. In our case, $\mathbf\Psi_i$ is diagonal, a property that  can be used to 
speed-up computations. $\mathbf{\Lambda}$ is the matrix containing the $k$ explicative vectors that we need to 
determine. While $\mathbf{\Lambda}$ is not uniquely defined,  $\mathbf{\Lambda\Lambda}^T$ is and represents the 
intrinsic covariance of the data, that is, the one we would observe in the absence of noise.  The eigenvectors of $
\mathbf{\Lambda\Lambda}^T$ thus correspond to the $k$ first eigenvectors that principal component analysis would have 
found in the absence of noise. 

To find $\boldsymbol{\Lambda}$, instead of directly maximizing the likelihood of observing $\mathbf{x}_i$, the 
expectation-maximization algorithm introduces the latent variable $\mathbf{q}_i$ and then maximizes the expected 
likelihood over $\mathbf{q}_i$. The joint probability of $\mathbf{x}_i$ and $\mathbf{q}_i$ is the following multivariate 
normal distribution:

\begin{equation}
\label{vraisemblance_EMFA}
P\left(\left[ 
\begin{array}{c} \textbf{x}_i \\ \textbf{q}_i \end{array}
\right]\right)
=
{\cal N}\left(\left[
\begin{array}{c} 0 \\ 0 \end{array}
\right],\left[
\begin{array}{cc} \boldsymbol{\Lambda} 
\boldsymbol{\Lambda^T} + \boldsymbol{\Psi_{i}} & 
\boldsymbol{\Lambda} \\
\boldsymbol{\Lambda}^T & \textbf{I} \end{array}
\right]\right)\eqdot
\end{equation}

\noindent  The block-diagonal elements of the covariance matrix represent, respectively, the covariances  of $\textbf{x}_i$ 
and $\textbf{q}_i$, and the non block-diagonal elements represent the covariance arising from 
the relation~\ref{eq:formalism}. 
Expectation-maximization is an iterative procedure which ensures that the likelihood increases at each iteration and it has 
been shown that the convergence is faster than using a gradient method \citep{dempster1977}. Each iteration proceeds in 
two steps. The first step, called the E-step, consists of calculating the expectation of the conditional first and second moments 
of 
$\textbf{q}_i$ for a given $\boldsymbol{\Lambda}$ :\\
\noindent \textbf{E-step}:

\begin{eqnarray}
\textbf{q}_i \ \ \hat{=} \ \ E\left[\textbf{q} |\textbf{x}_i\right]
 &=& \boldsymbol{\Lambda}^T(\boldsymbol{\Psi_{i}} +
 \boldsymbol{\Lambda}\boldsymbol{\Lambda}^T)^{-1} \textbf{x}_i
\end{eqnarray}

\begin{eqnarray}
     E\left[\textbf{q} \ \textbf{q}^T|\textbf{x}_i\right] &=& 
     \textbf{I} - \boldsymbol{\Lambda}^T(\boldsymbol{\Psi}_i +
      \boldsymbol{\Lambda}\boldsymbol{\Lambda}^T)^{-1}
       \boldsymbol{\Lambda} +  \textbf{q}_i \ \textbf{q}_i^T\eqdot
\end{eqnarray}

\noindent Once these two quantities are computed, the second step, or M-step, consists of estimating the $\boldsymbol{\Lambda}$  
matrix by maximizing the likelihood expectation, which provides the condition:

\begin{equation}
	\label{EMFA_gradient_lambda}
    \overset{N}{\underset{i=0}{\sum}} \boldsymbol{\Psi}_i^{-1}
     \boldsymbol{\Lambda}\; E\left[\textbf{q} \ \textbf{q}^T |
     \textbf{x}_i\right]   =\overset{N}{\underset{i=0}{\sum}} 
     \boldsymbol{\Psi}_i^{-1} \textbf{x}_i \textbf{q}_i^T\eqdot
\end{equation}

\noindent In the case where $\boldsymbol{\Psi}_i$ is diagonal, one can  independently calculate 
each row of $\Lambda$, noted $\Lambda^{j}$, according to the relation\\
\noindent \textbf{M-step}:

\begin{equation}
    \Lambda^{j}=
    \left[\overset{N}{\underset{i=0}{\sum}} 
    \frac{\text{x}_i^j}{\psi_i^{jj}} \ \textbf{q}_i^T\right]
    \left[\overset{N}{\underset{i=0}{\sum}} \frac{1}{\psi_i^{jj}}
    \ E[\textbf{q} \ \textbf{q}^T|\textbf{x}_i] \right]^{-1}
     \quad,
\end{equation}

\noindent After the last iteration, the internal degeneracies of the description are lifted with the transformation of $\mathbf{\Lambda}$ 
into an orthogonal $\boldsymbol{\Lambda}'$ matrix which satisfies the condition:

\begin{equation}
\boldsymbol{\Lambda'} \boldsymbol{\Lambda'}^T = 
\boldsymbol{\Lambda} \boldsymbol{\Lambda}^T\eqdot
\end{equation}

\noindent The columns of $\mathbf{\Lambda'}$ are aligned with the eigenvectors of $\mathbf{\Lambda\Lambda}^T$ and the square of their norm are the respective eigenvalues.

\section{Orthogonal Distance Regression}
\label{app:ODR}

\subsection{Expectation and Maximization steps} 

Similarly to the factor analysis, $\textbf{h}_i$ and $\textbf{A}$ 
will be estimated iteratively. The first step 
consists of minimizing the $\chi^2$ with respect to $\textbf{h}_i$, which amounts
to solving the equation:

\begin{equation}
\frac{\partial \chi^2}{\partial \textbf{h}_i}=0
\end{equation}

\noindent and consequently gives:\\
\noindent \textbf{E-step}:

\begin{equation}
\label{etape_E_ODR_equation}
\textbf{h}_i=\left( \textbf{A}^T \textbf{W}_{\textbf{M}_i}\textbf{A} + \textbf{W}_{\textbf{x}_i}\right)^{-1} \left( \textbf{A}^T \textbf{W}_{\textbf{M}_i}\textbf{M}_i + \textbf{W}_{\textbf{x}_i}\textbf{x}_i\right)
\eqdot
\end{equation}

Once the E-step is performed we can estimate the matrix $\textbf{A}$ by 

\begin{equation}
\frac{\partial \chi^2}{\partial \textbf{A}}=0\eqcoma
\end{equation}

\noindent which gives :\\
\noindent \textbf{M-step}:

\begin{equation}
\label{etape_M_ODR_equation}
\tilde{\textbf{A}}=\left(\left[\overset{N}{\underset{i=0}{\sum}} \textbf{h}_i^{} \ \textbf{h}_i^T  \otimes  \textbf{W}_{\textbf{M}_i} \right]\right)^{-1}\left[\overset{N}{\underset{i=0}{\sum}} \textbf{W}_{\textbf{M}_i} \textbf{M}_i \textbf{h}_i^T\right]
\eqcoma
\end{equation}

\noindent where $\tilde{\textbf{A}}$ is the vector that contains all 
columns of $\mathbf{A}$ put end-to-end and the $\otimes$ operator is 
the Kronecker product. The estimation of the $\textbf{A}$  matrix and 
the orthogonal projections $\textbf{h}_i$ is done by 
reiterating the E-step and M-step until $\chi^2$ convergences.

\subsection{Fixing the degeneracies}

The free parameters of equations~\ref{Model_SED_at_max} and~\ref{Model_SED_SUGAR_equation}
that we find through orthogonal distance regression contain degenerate degrees of freedom. 
These must be fixed in order to present an unique an interpretable solution.

\subsubsection{Degeneracies of the extinction fit}

The term describing the extinction, $A_{\lambda_0,i}\gamma(\lambda)$, 
is not directly constrained by external observations. As a consequence, 
Equation \ref{Model_SED_at_max} is invariant
under several transformations. The first one is

\begin{eqnarray}
A_{\lambda_0,i}^{new}&=&A_{\lambda_0,i}+\sum_j h_i^j \ c^j \label{decorr_h_AV_part1}\eqcoma\\
\alpha_{\lambda}^{j \ new}&=&\alpha_{\lambda}^{j} - \gamma_{\lambda} \ c^j  \eqdot\label{decorr_h_AV_part2}
\end{eqnarray}

We must therefore fix the $c^j$. A natural choice is to impose
that the  $A_{\lambda_0,i}$ are decorrelated with the $h_i^j$. Indeed, any 
correlation would imply that the extinction term 
contains information linked to the intrinsic properties described by
$h_i^j$.  Thus after each E-step, we will impose a correlation of zero between the 
$A_{\lambda_0,i}$ and the $h_i^j$, which amounts to applying the transformations \ref{decorr_h_AV_part1}
and \ref{decorr_h_AV_part2} taking for the $c^j$

\begin{equation}
\textbf{c}= \left[\text{cov}\left(\tilde{\textbf{h}}\right)\right]^{-1} \text{cov}\left(\tilde{\textbf{h}},\textbf{A}_{\lambda_0}\right)\eqcoma
\end{equation}

\noindent where $\tilde{\textbf{h}}$ is the vector that contains the $h_i$, 
$\text{cov}\left(\tilde{\textbf{h}}\right)$ is the covariance matrix of the 
$\tilde{\textbf{h}}$, $\text{cov}\left(\tilde{\textbf{h}},\textbf{A}_{\lambda_0}\right)$
is the vector that contains the covariances between the $h_i^j$  and the
$A_{\lambda_0,i}$.\\
\indent The second degeneracy to be fixed
is the scale of $\gamma_\lambda$: 
only the product $A_{\lambda_0,i} \gamma_\lambda$ plays a role, and we then 
impose after each iteration

\begin{equation}
 \gamma_{\lambda_0}=1
\end{equation}
\noindent by rescaling accordingly $A_{\lambda_0,i}$ and $\gamma_\lambda$.\\
\indent Finally, we need a prescription for the mean value of $A_{\lambda_0,i}$:
we choose it to be centered at zero. This corresponds to the following transformation

\begin{eqnarray}
A_{\lambda_0,i}^{new}&=&A_{\lambda_0,i} - \frac{1}{N}\sum_i A_{\lambda_0,i}\eqcoma\\
M_{\lambda,0}^{new}&=&M_{\lambda,0} +  \frac{\gamma_{\lambda}}{N}\sum_i A_{\lambda_0,i}
\eqdot
\end{eqnarray}

\noindent which leaves the $\chi^2$ invariant and
amounts to placing the average spectrum at the average extinction.

\subsubsection{Degeneracies of the global fit}

The estimation of the grey parameter,  $\Delta M_{grey \ i}$, is 
affected by different degeneracies which can be fixed after the 
E or M iterations. We choose to fixe them after each E-step.
The first one arises from the invariance of equation \ref{Model_SED_SUGAR_equation}
under the transformation:

\begin{eqnarray}
\Delta M_{grey \ i}^{new}&=&\Delta M_{grey \ i}+\sum_j h_i^j \ c^j \label{decorr_h_grey_part1}\eqcoma\\
\alpha_{t,\lambda}^{j \ new}&=&\alpha_{t,\lambda}^{j} - \ c^j  \label{decorr_h_grey_part2}
\eqdot
\end{eqnarray}

\noindent We must therefore fix the $c^j$. Similarly to the prescription made for the 
extinction fit, a natural choice is to impose that the  $\Delta M_{grey \ i}$ and the $h^{j}_i$ 
are uncorrelated. Indeed, the interpretation will be made easier with 
$\boldsymbol{\alpha}^{j}$ containing all the information driven by 
$h_i^j$, including its global impact on magnitudes. 
We thus impose after each E-step that the correlation between
the $\Delta M_{grey \ i}$ and the $h_i^j$ is zero. This is obtained by 
applying the tranformations~\ref{decorr_h_grey_part1} 
and~\ref{decorr_h_grey_part2} taking for the  $c^j$:

\begin{equation}
\textbf{c}= \left[\text{cov}\left(\tilde{\textbf{h}}\right)\right]^{-1} \text{cov}\left(\tilde{\textbf{h}},\boldsymbol{\Delta}\textbf{M}_{\textbf{grey}}\right)\eqcoma
\end{equation}

\noindent where $\tilde{\textbf{h}}_i$ is the vector that contains the $h_i^j$, 
$\text{cov}\left(\tilde{\textbf{h}}\right)$ is the observed
covariance matrix computed on the set of $\tilde{\textbf{h}}_i$ vectors, 
and $ \text{cov}\left(\tilde{\textbf{h}}, \boldsymbol{\Delta}\textbf{M}_{\textbf{grey}}\right)$ 
is the vector that contains the covariances between  $h_i^j$ and $\Delta M_{grey \ i}$. 
Finally, by convention, the $\Delta M_{grey \ i}$ are centered on zeros at each step. 
This is obtained by the following transformation, which leaves the $\chi^2$ 
invariant:

\begin{eqnarray}
\Delta M_{grey \ i}^{new}&=&\Delta M_{grey \ i} - \frac{1}{N}\sum_i \Delta M_{grey \ i}\eqcoma\\
M_{t,\lambda,0}^{new}&=&M_{t,\lambda,0} +  \frac{1}{N}\sum_i \Delta M_{grey \ i}\eqdot
\end{eqnarray}

\section{Matrix dispersion estimation}
\label{app:disp_matrix}

Once the minimum of the $\chi^2$ defined in Eq.~\ref{chi2_max_0} is reached, 
the dispersion matrix is estimated. For this purpose, we use the same method as in 
\cite{Chotard11} and which is described in detail in \cite{Chotardphd}. 
The approach is to calculate $\textbf{D}$  from the observed dispersion of residuals and to subtract  
the average dispersion due to uncertainties:

\begin{equation}
 \textbf{D}  =   \frac{1}{N} \sum_i \left(\boldsymbol{\delta} \textbf{M}_{i} \boldsymbol{\delta} \textbf{M}_{i}^T -  \textbf{C}_{i}\right)
\end{equation}
\\
\noindent where $\boldsymbol{\delta} \textbf{M}_{i}$ are the residuals of the model once the minimum of the  $\chi^2$ is reached, and is defined for a given wavelength as:

\begin{equation}
  \delta M_{\lambda,i}  =   M_{\lambda,i}-M_{\lambda,0}-  \ \sum_j q_i^j \alpha_{\lambda}^{j}  - A_{\lambda_0,i} \ \gamma_{\lambda}
\end{equation}

\noindent and  $\textbf{C}_{i}$  is the covariance matrix that accounts for the total propagation of residuals error and is defined as:

\begin{equation}
\textbf{C}_{i} =   \begin{pmatrix}
    \ddots &   &0 \\ 
       &  \sigma_{\lambda i}^2&\\ 
    0  &  & \ddots
\end{pmatrix} \ + \ \left(\sigma_{cal}^2+\sigma_{z}^2\right)  \begin{pmatrix}
    1      & \cdots & 1 \\ 
    \vdots & \ddots & \vdots \\ 
    1      & \cdots & 1 
\end{pmatrix} \ + \ \boldsymbol{\alpha} \text{cov}\left(\textbf{q}_{i}\right) \boldsymbol{\alpha}^T
\end{equation}
\\
\noindent where $\boldsymbol{\alpha}$ is the matrix that contains the intrinsic vectors, and is defined as:

\begin{equation}
  \boldsymbol{\alpha}  =   \left(\boldsymbol{\alpha}^1,\boldsymbol{\alpha}^2,\boldsymbol{\alpha}^3,...\right)
\end{equation}
\\
In order to ensure that the matrix $\textbf{D}$ is positive definite, the negative eigenvalues of the matrix are set to zero. Once the matrix $\textbf{D}$ has been calculated, we add it in the expression of the equation~\ref{chi2_max_0} in order to recalculate the spectral distribution in energy and we iterate the calculations of  $\textbf{D}$ and the computation of the SED, until reaching the maximum of the Restricted Maximum Likelihood \citep{Guy10}.

\bibliographystyle{aa}
\bibliography{Biblio}

\end{document}